\documentclass[Ceremade,envcountsect]{svjourCeremade}

\usepackage{amsfonts}
\usepackage{amsmath}
\usepackage{amssymb}
\usepackage{dsfont}

\newcommand{\N}{\mathbb{N}}
\newcommand{\Z}{\mathbb{Z}}

\newcommand{\R}{\mathbb{R}}
\newcommand{\C}{\mathbb{C}}

\spnewtheorem{assumption}{Assumption}[section]{\bf}{\it }

\numberwithin{equation}{section}

\def \p{\partial}
\def \ds{\displaystyle}
\def \e{\varepsilon}
\def \one{\mathds{1}}
\def \bone{\mathbf{1}}
\def \bzero{\mathbf{0}}
\def \i{{\rm i}}
\def \vs{\vspace{0.2cm}}

\title{Stable directions for small nonlinear Dirac standing waves}

\begin{document}
\author{Nabile Boussaid}

\institute{Ceremade, 
Universit\'e Paris Dauphine, 
Place du Mar\'echal De Lattre De Tassigny 
F-75775 Paris C\'edex 16 - France\\
\email{boussaid@ceremade.dauphine.fr}}

\date{December 2005}

\maketitle

\begin{abstract}
We prove that for a Dirac operator, with no resonance at thresholds
nor eigenvalue at thresholds, the propagator satisfies propagation
and dispersive estimates.

When this linear operator has only two simple eigenvalues
sufficiently close to each other, we study an associated class of
nonlinear Dirac equations which have stationary solutions. As an
application of our decay estimates, we show that these solutions
have stable directions which are tangent to the subspaces associated
with the continuous spectrum of the Dirac operator. This result is
the analogue, in the Dirac case, of a theorem by Tsai and Yau about
the Schr\"odinger equation. To our knowledge, the present work is
the first mathematical study of the stability problem for a
nonlinear Dirac equation.
\end{abstract}

\section*{Introduction}

We study the stability of stationary solutions of a time-dependent
nonlinear Dirac equation.

Usually, a localized stationary solution of a given time-dependent
equation represents the bound state of a particle. Like Ranada
\cite{Ranada}, we call it a \emph{particle like solutions} (PLS). In
the literature, the term \emph{soliton} is also found instead of
PLS, but this additionally means that the particle keeps its form
after a collision. Many works have been devoted to the proof of the
existence of such solutions for a large variety of equations.
Although their stability is a crucial problem (in particular in
numerical computation or experiment), a smaller attention has been
deserved to this issue.

There are different definitions of stability. The first one is
commonly called the \emph{orbital stability}. It
means that the orbit of the perturbation of a PLS stays close to the
PLS or a manifold of PLS but does not necessarily converge. A
stronger notion is \emph{asymptotic stability}, which means that the
perturbation of the PLS relaxes asymptotically towards a PLS which
is not far from the perturbed PLS.

In fact in many conservative problems asymptotic stability does not
hold. But one has asymptotic stability for a restricted class of
perturbations, forming the so-called \emph{stable manifold}.

\bigskip

In this paper, we deal with the problem of stability of small PLS of
the following nonlinear Dirac equation:
\begin{equation}
\i\p_t\psi=(D_m+V)\psi+\nabla F(\psi) \tag{NLDE}\label{nld_intro}
\end{equation}
where $\nabla F$ is the gradient of $F:\C^4\mapsto \R$ for the
standard scalar product of $\R^8$. Here,~$D_m$ is the usual Dirac
operator~\cite{Thaller} acting on $L^2(\R^3,\C^4)$
\begin{equation*}
D_m=\alpha\cdot\left(-\i\nabla\right)+m\beta=
-\i\sum_{k=1}^3\alpha_k\partial_k + m\beta
\end{equation*}
where~$m\in \R_+^*$ and
~$\alpha=\left(\alpha_1,\alpha_2,\alpha_3\right)$ and~$\beta$
are~$\C^4$ hermitian matrices satisfying the following properties:
\begin{equation*}
\begin{cases}
\alpha_i\alpha_k+\alpha_k\alpha_i=2\delta_{ik}\bone_{\C^4},
&i,k\in\{1,2,3\},\\
\alpha_i\beta+\beta\alpha_i=\bzero_{\C^4},&i\in\{1,2,3\},\\
\beta^2=\bone_{\C^4}.
\end{cases}
\end{equation*}
Here we choose
\begin{gather*}
\alpha_i=\left(
\begin{array}{cc}
0&\sigma_i\\
\sigma_i&0
\end{array} \right)\quad\mbox{ and }\quad\beta=\left(\begin{array}{cc}
I_{\C^2}&0\\
0&-I_{\C^2}
\end{array} \right)\\
\mbox{ where } \sigma_1=\left(
\begin{array}{cc}
0&\; 1\\
1&\;0
\end{array} \right)\quad\mbox{ and } \quad\sigma_2=\left(
\begin{array}{cc}
0&-\i\\
\i&0
\end{array} \right)\quad\mbox{ and } \quad\sigma_3=\left(
\begin{array}{cc}
1&0\\
0&-1
\end{array} \right).
\end{gather*}
In~\eqref{nld_intro},~$V$ is the external potential field
and~$F:\C^4\mapsto\R$ is a nonlinearity such that
\begin{equation*}
\forall(\theta,z)\in\R\times\C^4,\quad F(e^{i\theta}z)=F(z).
\end{equation*}
Some additional assumptions on~$F$ and $V$ will be made in the
sequel. Stationary solutions (PLS) of \eqref{nld_intro} take the
form $\psi(t,x)=e^{-\i Et}\phi(x)$ where~$\phi$ satisfies
\begin{equation}
E\phi=(D_m+V)\phi+\nabla F(\phi). \tag{PLSE}\label{eq:stationary}
\end{equation}
We prove the existence of a manifold of small solutions to
\eqref{eq:stationary}, interpreted as particle like solutions to
\eqref{nld_intro}. Then we construct a stable manifold around this
manifold. At the origin, it is tangent to the sum of the eigenspace
associated with the first eigenvalue and the continuous spectral
subspace of~$D_m+V$. This is the analogue in the Dirac case of
\cite[Theorem 1.1, non-resonant case]{TsaiYau3}. The interpretation
is that radiations (described by the continuous spectrum) do not
destabilize too much the PLS manifold. To prove stabilization
towards the PLS manifold, we shall need linear decay estimates
associated with the continuous spectral subspace of~$D_m+V$.

To our knowledge, this is the first stability result on a nonlinear
Dirac equation.

\bigskip

The problem of stability has been extensively studied for
Schr\"odinger and Klein-Gordon equations. The methods used to treat
these cases cannot be easily adapted to our problem, due to the fact
that the Dirac operator~$D_m$ is not bounded-below, contrarily to
$-\Delta$. The non-negativity of the latter permits to use
minimization and concentration-compactness methods to prove the
existence of orbitally stable standing waves, see e.g. Cazenave and
Lions~\cite{CazenaveLions} or more recently Cid and Felmer
\cite{CidFelmer}.

In his review on nonlinear Dirac models, Ranada~\cite{Ranada} writes
that physicists first claimed that PLS (Particle Like Solutions) of
the nonlinear Dirac equation couldn't be stable since the second
derivative of the energy functional is not positive-definite.
Actually, in a very general setting (not related to the Dirac case),
Shatah and Straus~\cite{ShatahStrauss} and Grillakis, Shatah and
Straus~\cite{GrillakisShatahStrauss} proved a general orbital
stability condition even if the hessian of the energy functional is
not positive-definite. Their conditions allow only one simple
negative eigenvalue (and a kernel of dimension one also) for the
second variation. It therefore cannot be directly applied to the
Dirac case. However, it gave rise to an interesting discussion about
the application of this method to the Dirac equation in some
physical papers
\cite{StraussVazquez,AlvarezSoler,BlanchardStubbeVazquez}. Ranada
also refers to numerical experiments which seem to confirm that some
PLS are asymptotically stable in the Dirac case.

In the Schr\"odinger case, the asymptotic stability has been
extensively studied during the last decade. A fundamental work is
the one of Soffer and
Weinstein~\cite{SofferWeinstein,SofferWeinstein2}, which is devoted
to the study of a small nonlinear perturbation of a Schr\"odinger
operator having one simple eigenvalue. They proved that the
perturbed small PLS relaxes to a PLS. Later, Pillet and
Wayne~\cite{PilletWayne} proposed a different proof in the spirit of
the central manifold theorem. In all these works, asymptotic
stability is a direct consequence of propagation or dispersive
estimates on the Schr\"odinger operator. In order to be able to use
these estimates, one has to to consider the initial state (at time
$t=0$) of the perturbation as localized {\it i.e.} in $L^1$ or in
$L^2$ weighted spaces with growing weight. To avoid such an
assumption, Gustafson, Nakanishi and
Tsai~\cite{GustafsonNakanishiTsai} proposed to use Strichartz
estimates.

Generalizations have been considered for instance by Tsai and Yau
\cite{TsaiYau,TsaiYau2,TsaiYau3,TsaiYau4,Tsai}, who treated the case
of a Schr\"odinger operator having two simple eigenvalues. An
interesting phenomenon appeared: if the two eigenvalues are
sufficiently distant one from the other, then after linearization
around the excited state, one obtains a resonance. Tsai and Yau
showed that if there is no resonance, the manifold of ground state
has stable directions. In the resonant case, the manifold of ground
states is asymptotically stable, whereas the manifold of excited
states has stable and unstable directions (in case of instability,
under some conditions, one has relaxation to the ground state). For
a similar result, see also \cite{SofferWeinstein4,SofferWeinstein5}. Notice
that earlier Soffer and Weinstein~\cite{SofferWeinstein3}
studied a similar resonance phenomenon in the case of the
Klein-Gordon equation with a simple eigenvalue; they showed that it
induced ``metastability''. Another problem has been studied by
Cuccagna \cite{Cuccagna2,Cuccagna3,Cuccagna3Err}. He considered the
case of big PLS, when the linearized operator has only one
eigenvalue and obtained the asymptotical stability of the manifold
of ground states. Tsai, Yau and Cuccagna also need propagation or
dispersive estimates. The latter is proved by generalizing the work
of Yajima \cite{Yajima} on wave operator.

Interesting development are also given by Rodnianski, Schlag and
Soffer \cite{RodnianskiSchlagSoffer} who proved asymptotic stability
of an arbitrary number of weakly interacting big PLS. Schlag
\cite{Schlag} and Krieger and Schlag~\cite{KriegerSchlag} proved the
existence of stable direction for unstable big PLS. We point out
that some of the works of Schlag
\cite{ErdoganSchlag,GoldbergSchlag,RodnianskiSchlagSoffer2} or
Soffer~\cite{HunzikerSigalSoffer,JourneSofferSogge,RodnianskiSchlagSoffer2}
are dedicated to prove dispersive estimates.

We also would like to mention the works of Buslaev and Perel'mann
\cite{BuslaevPerelman,BuslaevPerelman2,BuslaevPerelman3,BuslaevPerelman4},
Buslaev and Sulem~\cite{BuslaevSulem,BuslaevSulem2} or Weder
\cite{Weder}, in the one dimensional Schr\"odinger case.

\bigskip
Here, we study a nonlinear Dirac equation as a perturbation of a
linear Dirac equation with a Dirac operator possessing only two
simple eigenvalues sufficiently close to each other. Hence, we avoid
problems of resonance after linearization around a PLS. The paper is
organized as follows.

In section \ref{Section:MainResults}, we define the important
objects and state our main results. We start with the propagation
and dispersive linear estimates which will be crucial tools for this
study. Then, we consider the nonlinear equation~\eqref{nld_intro}
and state the existence of the PLS manifold. Eventually, we present
our main theorem in which the stable manifold is constructed.

The section \ref{sec:proof_theorem_propagation} is devoted to the
proof of the propagation estimate, which uses spectral techniques.
This is a time decay estimate in weighted~$L^2$ spaces, expressing
the fact that states associated with the continuous spectrum are not
stationary. We use Mourre estimate similarly to Hunziker, Sigal and
Soffer \cite{HunzikerSigalSoffer} (for a generalization of the
method, see e.g.~\cite{BoutetdeMonvelGeorgescuSahbani}). This method
cannot be used in the neighborhood of the thresholds which needs a
specific treatment. In particular, problems can occur in the
presence of eigenvalues at thresholds or resonances at thresholds,
and we shall assume in the whole paper that we are not in this
situation. For the Schr\"odinger case, a similar problem has been
studied by Jensen and Kato \cite{JensenKato}, Jensen and
Nenciu~\cite{JensenNenciu,JensenNenciuErr}. Our arguments near the
thresholds are inspired of these works. For a related study, see the
article of Fournais and Skibsted~\cite{FournaisSkibsted} dealing
with long range perturbations of Schr\"odinger operators.

In Section \ref{sec:dispersive_estimate}, we then prove the
dispersive estimate, using the propagation estimate established in
Section~\ref{sec:proof_theorem_propagation}. For an interesting
survey on dispersive estimates for Schr\"odinger operators, see
Schlag~\cite{Schlag2}. We have not been able to generalize the
methods used in the Schr\"odinger case, in fact it seems that the
Dirac equation with a potential behaves like a Klein-Gordon equation
with a magnetic potential. This fact has already been noticed by
D'Anconna and Fanelli in \cite{D'AnconnaFanelli}, where they proved
simultaneously dispersive estimates for a massless Dirac equation
with a potential and for a wave equation with a magnetic potential.
Our method is here inspired of the work by Cuccagna and
Schirmer~\cite{CuccagnaSchirmer}.

Finally, the last sections are devoted to the proof of our main
result concerning the stability of the stationary solutions
of~\eqref{nld_intro}. We assume that the Dirac operator~$D_m+V$
have only two simple eigenvalues and that it has no eigenvalues at
thresholds nor resonances at thresholds. Note that our assumptions
exclude electric potentials, for which the theorem of Kramers states
that the eigenvalues are always degenerate, see
\cite{Parisse,BalslevHelffer}. In Section
\ref{Section:LinearizedOperator}, this permits us to construct a
manifold of PLS and then to study the spectrum of the linearized
operator. This in turn, in Section~\ref{Section:Stabilization},
 will allow us to decompose a solution of~\eqref{nld_intro}
 in three parts: the PLS part, the dispersive part
associated with the continuous spectrum and a part corresponding to
``excited states". This last part needs a particular treatment since
it is not dispersive and hence disturbs the relaxation towards the
PLS manifold.

\begin{acknowledgements} I would like to thank \'Eric S\'er\'e for
fruitful discussions and advices during the preparation of this
work. I wish to thank the referee for useful remarks and
suggestions.
\end{acknowledgements}

\section{Main results}
                                                                \label{Section:MainResults}
This section is devoted to the presentation of the model and the
statement of our main results.

\subsection{Decay estimates for a Dirac operator 
with potential}

Let us first state our results concerning the time decay of~$e^{-\i
t (D_m+V)}$ in weighted~$L^2$ spaces and Besov spaces. This kind of
estimates are called respectively propagation and dispersive
estimates. As mentioned in the introduction, these results will be
very important tools for the study of our nonlinear time-dependent
Dirac equation.

The following spaces will be needed to state the main result of this
subsection.
\begin{definition}[Weighted Sobolev space]
The weighted Sobolev space is defined by
\begin{equation*}
H_{\sigma}^{t}(\R^3,\C^4)=\left\{ f\in {\mathcal S}'(\R^3),\,
\|\langle Q\rangle^\sigma\langle P\rangle^t f\|_2 <\infty\right\}
\end{equation*}
for~$\sigma,t\in\R$. We endow it with the norm
\begin{equation*}
\|f\|_{ H_{\sigma}^{t}}= \|\langle Q\rangle^\sigma\langle P\rangle^t
f\|_2.
\end{equation*}
If~$t=0$, we write~$L^2_\sigma$ instead of~$H_{\sigma}^{0}$.
\end{definition}
We have used the usual notations~$\langle u\rangle=\sqrt{1+u^2}$,
$P=-i\nabla$, and~$Q$ is the operator of multiplication by~$x$ in
$\R^3$. For the sake of clarity, let us also recall the
\begin{definition}[Besov space]
For~$s\in\R$ and~$1\leq p,q\leq\infty$, the Besov
space~$B^s_{p,q}(\R^3,\C^4)$ is the space of all~$f\in {\mathcal
S}'(\R^3,\C^4)$ (dual of the Schwartz space) such that
\begin{equation*}
\ds \|f\|_{B^s_{p,q}}= \left(\sum_{j\in\N}2^{jsq} \|\varphi_j *
f\|_p^q\right)^{\frac{1}{q}}<+\infty
\end{equation*}
with~$\widehat{\varphi} \in {\mathcal D}(\R^n\setminus
\left\{0\right\})$ such that~$\sum_{j\in
\Z}\widehat{\varphi}(2^{-j}\xi)=1$ for all~$\xi
\in\R^3\setminus\left\{0\right\}$,
$\widehat{\varphi}_j(\xi)=\widehat{\varphi}(2^{-j} \xi)$ for all
$j\in\N^*$ and for all~$\xi \in\R^3$, and
$\widehat{\varphi_0}=1-\sum_{j\in \N^*}\widehat{\varphi}_j$. We
endow it with the norm~$f\in B^s_{p,q}(\R^3,\C^4) \mapsto
\|f\|_{B^s_{p,q}}$.
\end{definition}

In the whole chapter, we shall work within the following
\begin{assumption}
                                                                \label{assumption:1}
The potential~$V:\R^3\mapsto S_4(\C)$ {\it (self-adjoint~$4\times 4$
matrices)} is a~${\mathcal C}^\infty$ function such that there
exists~$\rho>5$ with
\begin{equation*}
\forall \alpha\in\N^3,\,\exists C>0,\,\forall x\in\R^3,\, |\p^\alpha
V|(x)\leq\frac{C}{\langle x\rangle^{\rho+|\alpha|}}.
\end{equation*}
\end{assumption}
Notice that by the Kato-Rellich Theorem , the operator
\begin{equation*}
H:=D_m+V
\end{equation*}
is essentially self-adjoint on~${\mathcal C}^\infty_0(\R^3,\C^4)$
and self-adjoint on $H^1(\R^3,\C^4)$. We also work with the
\begin{assumption}
                                                                \label{assumption:2}
The operator~$H$ presents no resonance at thresholds and no
eigenvalue at thresholds.
\end{assumption}

A resonance is an eigenvector
in~$H^{1/2}_{-\sigma}(\R^3,\C^4)\setminus H^{1/2}(\R^3,\C^4)$ for
some~$\sigma\in (1/2,\rho-1/2)$ here. Let
\begin{equation}
                                                                \label{Def:ProjectorContSpect}
{\mathbf P}_c(H)=\one_{(-\infty,-m]\cup[+m,+\infty)}(H)
\end{equation}
be the projector associated with the continuous spectrum of~$H$ and
\begin{equation}
                                                                \label{Def:ContinuousSubspace}
{\mathcal H}_c={\mathbf P}_c(H)L^2(\R^3,\C^4).
\end{equation}
We are now able to state our
\begin{theorem}[Propagation for perturbed Dirac dynamics]
                                                                \label{Thm:Propagation}
Assume that Assumptions~\ref{assumption:1} and~\ref{assumption:2}
hold and let be~$\sigma>5/2$. Then one has
\begin{equation*}
\|e^{-\i t H}{\mathbf P}_c\left(H\right)\|_{
B(L^2_{\sigma},L^2_{-\sigma})}\leq C\left\langle
t\right\rangle^{-3/2}.
\end{equation*}
\end{theorem}

The proof of this result will be given in Section
\ref{sec:proof_theorem_propagation}. We notice that it is still true
if we assume~$\rho>3$ in Assumption~\ref{assumption:1}.

Our next result is the following theorem, proved in Section
\ref{sec:dispersive_estimate}.
\begin{theorem}[Dispersion for perturbed Dirac dynamics]
                                                               \label{Thm:Dispersion}
Assume that Assumptions~\ref{assumption:1} and~\ref{assumption:2}
hold. Then for~$p\in[1,2]$,~$\theta\in[0,1]$, $s-s'\geq (2
+\theta)(\frac{2}{p}-1)$ and $q\in[1,\infty]$ there exists a constant~$C>0$ such that
\begin{equation*}
\|e^{-\i t H}{\mathbf P}_c(H)\|_{B^s_{p,q},B^{s'}_{p',q}} \leq C
\left(K(t)\right)^{\frac{2}{p}-1}
\end{equation*}
with~$\frac{1}{p}+\frac{1}{p'}=1$, and

\begin{equation*}
K(t)= \left\{
\begin{array}{ll}
\ds\left|  t\right| ^{-1+\theta/2}
& \mbox{if } |t|\in (0,1],\vspace{0.3cm}\\
\ds\left|  t\right| ^{-1-\theta/2} & \mbox{if } |t|\in [1,\infty).
\end{array}
\right.
\end{equation*}
\end{theorem}

\subsection{The stable manifold around the PLS for the 
nonlinear Dirac equation}

We now want to study the following nonlinear Dirac equation
\begin{equation}
\left\{\begin{array}{l} \i\p_t \psi=H\psi+\nabla F(\psi)\\
\psi(0,\cdot)=\psi_0.
\end{array}\right.
                                                                \label{Eq:NLD}
\end{equation}
with~$\psi\in {\mathcal C}^1(I,H^1(\R^3,\C^4))$ for some open
interval~$I$ which contains~$0$ and where we recall that~$H=D_m+V$.
The nonlinearity~$F: \C^4\mapsto \R$ is a differentiable map for the
real structure of~$\C^4$ and hence the~$\nabla$ symbol has to be
understood for the real structure of~$\C^4$. For the usual hermitian
product of~$\C^4$, one has
\begin{equation*}
DF(v)h=\Re \langle \nabla F(v),h\rangle.
\end{equation*}
We work within the following
\begin{assumption}
                                                                \label{assumption:Spectrum}
The operator~$H$ has only two simple eigenvalues
$\lambda_0<\lambda_1$, with~$\phi_0$ and~$\phi_1$ as associated
normalized eigenvectors. Moreover, the non resonant condition
\begin{equation*}
|\lambda_1-\lambda_0|<\min\{|\lambda_0+m|,\,|\lambda_0-m|\}
\end{equation*}
holds.
\end{assumption}
\begin{assumption}
                                                                \label{assumption:NonLinearity}
The function~$F:\C^4\mapsto \R$ is in~${\mathcal
C}^\infty(\R^8,\R)$,
is a homogeneous polynomial of degree $4$ ({\it i.e.} with $D^\alpha
F (z)=0$ for $|\alpha|=5$ and $D^{\beta}F(0)=0$ for $|\beta|\leq 4$)
or satisfies $F(z)=O(|z|^{5})$ as~$z\to 0$. Moreover, it has the
gauge invariance property:
\begin{equation*}
F(e^{i\theta}z)=F(z), \forall z\in\C^4,\;\forall \theta \in \R.
\end{equation*}
\end{assumption}

We will prove in Theorem~\ref{Thm:StabilizationSmallPLSNR} that some
solutions of the equation~\eqref{Eq:NLD} are global and can be
decomposed as the sum of a PLS plus a remainder part which is
vanishing. Since the PLS part may change during the evolution, we
need to track it. So we prove that around the origin, PLS form a
manifold. We have the
\begin{proposition}[PLS manifold]
                                                                \label{Prop:ManifoldPLS}
Suppose that
Assumptions~\ref{assumption:1}--\ref{assumption:NonLinearity} hold.
Then for any~$\sigma \in \R^+$, there exists~$\Omega$ a neighborhood
of~$0\in \C$, a~${\mathcal C}^{\infty}$ map
\begin{equation*}
h: \Omega \mapsto
\left\{\phi_0\right\}^{\bot}\cap{H^2(\R^3,\C^4)}\cap
L^2_\sigma(\R^3,\C^4)
\end{equation*}
and a~${\mathcal C}^{\infty}$ map~$E: \Omega \mapsto \R$ such that
$S(u)=u\phi_0+h(u)$ satisfy for all~$u\in\Omega$,
\begin{equation}
                                                                \label{Eq:StationaryStates}
H S(u)+\nabla F(S(u))=E(u)S(u),
\end{equation}
with the following properties
\begin{equation*}
\left\{\begin{array}{l}
h(e^{i\theta} u)=e^{i\theta}h(u),\quad\forall \theta \in \R,\\
h(u)=O(|u|^2),\\
E(u)=E(\left|u\right|),\\
E(u)=\lambda_0+O(|u|^2).
\end{array}\right.
\end{equation*}
\end{proposition}
\begin{proof}
This kind of results is now classical and left to the reader. For
more details, see Subsection~\ref{Subsection:TheManifoldPLS}.
\end{proof}

We are now able to write the main theorem of this paper. Its proof
is given in Section 6. To state it we need the space ${\mathcal
H}_c$ defined in \ref{Def:ContinuousSubspace}.
\begin{theorem}[Stable manifold]
                                                                \label{Thm:StabilizationSmallPLSNR}
Suppose that
Assumptions~\ref{assumption:1}--\ref{assumption:NonLinearity} hold.
Let~$s,s',\beta \in \R^*_+$ be such that~$s'\geq s+3\geq \beta+6$ and $\sigma>5/2$.
There exists~$\e_0>0$,~$R>0$,~$K>0$ and a Lipshitz map
\begin{equation*}
\Psi : B_\C(0,\e)\times \left({\mathcal H}_c\cap
B_{H^{s'}_{\sigma}}(0,R)\right)\mapsto \C
\end{equation*}
with~$\Psi(v,0)=0$,
\begin{equation*}
\left|\Psi(v,\xi)\right|\leq
K\left(\left|v\right|+\|\xi\|_{H^{s'}_{\sigma}}\right)^2,
\end{equation*}
and such that the following hold. For any initial condition of the
form
\begin{equation*}
\psi_0=S(v_0)+\xi_0+ \Psi(v_0,\xi_0)\phi_1
\end{equation*}
with~$v_0\in B_\C(0,\e)$ and~$\xi_0\in {\mathcal H}_c\cap
B_{H^{s'}_{\sigma}}(0,R)$, one has
\begin{itemize}
\item[(i)] there exists a unique global solution~$\psi$
of~\eqref{Eq:NLD} in
\begin{equation*}
{\mathcal C}^\infty\left(\R,H^{s'}(\R^3,\C^4)\cap
H^{s}_{-\sigma}(\R^3,\C^4)\cap
B^{\beta}_{\infty,2}(\R^3,\C^4)\right);
\end{equation*}
\item[(ii)] there exists~$\left(v_{\infty};\xi_{\infty};
E_{ \infty}\right)\in \C \times H^{s'}_{\sigma}\cap
{\mathcal H}_c\times \R$ with
\begin{equation*}
\left|v_{ \infty}-v_0\right|\leq K\|\xi_0\|_{H^{s'}_{\sigma}}^2,
\quad\left|E_{ \infty}\right|\leq K\|\xi_0\|_{H^{s'}_{\sigma}}^2,
\quad\left\|\xi_{ \infty}-\xi_0\right\|_{H^{s'}}\leq
K\|\xi_0\|_{H^{s'}_{\sigma}}^2
\end{equation*}
such that
\begin{equation*}
\psi(t)=e^{-\i(t E(v_{ \infty})+E_{ \infty})}S(v_{ \infty})+e^{-\i
tH}\xi_{ \infty}+\varepsilon(t),
\end{equation*}
where
\begin{equation*}
\begin{cases}
\left\|\varepsilon(t)\right\|_{H^{s'}}\leq
\ds K\|\xi_0\|_{H^{s'}_{\sigma}}\vs\\
\left\|\varepsilon(t)\right\|_{H^{s}_{-\sigma}}\leq
\ds \frac{K}{\langle t\rangle^2}\|\xi_0\|_{H^{s'}_{\sigma}}\vs\\
\left\|\varepsilon(t)\right\|_{B^{\beta}_{\infty,2}}\leq \ds
\frac{K}{\langle t\rangle^2}\|\xi_0\|_{H^{s'}_{\sigma}}.
\end{cases}
\end{equation*}
as~$t\to + \infty$.
\end{itemize}
\end{theorem}
\begin{remark}
The proof of these theorem work also if we want to obtain an
expansion of the form
\begin{equation*}
\psi(t)=e^{-\i(t E(v_{\infty})+E_{ \infty})}S({v_{ \infty}})+e^{-\i
tD_m}\widetilde{\xi_{ \infty}}+\varepsilon(t)
\end{equation*}
with the free Dirac operator. But in this case, we only have the
estimates
\begin{equation*}
\left\|\widetilde{\xi}_{ \infty}-\xi_0\right\|_{H^{s'}}\leq
K\|\xi_0\|_{H^{s'}_{\sigma}}
\end{equation*}
see the remark following Lemma \ref{Lem:OnNonLinearScattering}.
\end{remark}

\medskip

We notice that the stabilization is ``faster'' than the propagation
and the dispersion: it is of order~$\langle t\rangle^{-2}$ whereas
$e^{-\i tH}\xi_\infty$ is of order~$\langle t\rangle^{-3/2}$ by
Theorems~\ref{Thm:Propagation} and \ref{Thm:Dispersion}.
Hence the theorem states the existence of a family of initial states
which form a manifold tangent at the origin to the sum of the
eigenspace of~$H$ associated with~$\lambda_0$ and the subspace
associated with the continuous spectrum of~$H$: ${\mathcal H_c}$.
This family of initial states gives rise to solutions
of~\eqref{Eq:NLD} which asymptotically split in two parts. The first
one is a PLS:~$e^{-\i \left(tE(u_\infty)+
E_\infty\right)}S(u_\infty)$ the other is a dispersive
perturbation:~$e^{-\i tH}\xi_\infty$. Hence if one perturbs a PLS in
the direction of the continuous spectrum then this PLS relaxes to
another PLS by emitting a dispersive wave.

This phenomenon is due to the propagation and the dispersion
properties of the subspace associated with the continuous spectrum
of $H$. We don't think that such a phenomenon could take place for
perturbations in the direction of the excited states~$\phi_1$.
Indeed, on this subspace, the dynamic seems to be conservative. The
fact that we use propagation and dispersive estimates restricts the
family of perturbations to regular and localized ones.

\medskip

We now turn to the proof of our results.
\section{Proof of Theorem~\ref{Thm:Propagation}: 
propagation estimates}
                                                                \label{sec:proof_theorem_propagation}
Here we prove the propagation estimates of Theorem
\ref{Thm:Propagation}. The method used by Jensen and
Kato~\cite{JensenKato} to prove this kind of estimates for
Schr\"odinger operator works only for initial states which are
spectrally localized near the thresholds~$\pm m$. They used the
spectral density as the Fourier transform of the propagator. But the
Dirac resolvent
\begin{equation*}
R_V(\lambda\pm\i \e)=(H-\lambda\mp\i \e)^{-1}
\end{equation*}
does not decay in~$B(L^2_\sigma,L^2_{-\sigma})$ as~$|\lambda|\to
+\infty$ for any~$\sigma>0$, see~\cite{Yamada}. So we cannot use its
Fourier transform. To our knowledge, this method is the only one
that permits to treat the problem of propagation for energies near
thresholds. Hence with this method, we only prove (in the section
\ref{Section:Step1}) the
\begin{proposition}[Propagation near thresholds]
                                                                   \label{Prop:PropagationNearThresholds}
Suppose that Assumptions \ref{assumption:1} and~\ref{assumption:2}
hold and let~$\chi\in {\mathcal C}^\infty_0(\R^3,\C^4)$ be such that
its support is in a sufficiently small neighborhood of~$[-m;\;m]$.
Then one has for~$\sigma>5/2$
\begin{equation*}
\|e^{-\i t H}{\mathbf  P}_c\left(H\right) \chi\left(H\right)\|_{
B(L^2_{\sigma},L^2_{-\sigma})}\leq C\left\langle
t\right\rangle^{-3/2}.
\end{equation*}
\end{proposition}
We recall that~${\mathbf P}_c(H)$ is defined by
\eqref{Def:ProjectorContSpect}.

We also need to treat the propagation estimates for initial state
whose spectrum does not contain any threshold. We cannot use the
spectral density. So we work directly with the propagator. This is
exactly the method used by Hunziker, Sigal and Soffer in
\cite{HunzikerSigalSoffer}. But in our case, their result needs some
adaptation.
Hence we need to generalize~\cite[Theorem 1.1]{HunzikerSigalSoffer}
to the case of unbounded energy. In Section~\ref{Section:Step2}, we
prove the
\begin{proposition}[Propagation far from thresholds]
                                                                   \label{Prop:PropagationHighEnergy}
Suppose that Assumption~\ref{assumption:1} holds. Then for any
$\chi\in {\mathcal C}^\infty(\R^3,\C^4)$ bounded with support in
$\R\setminus(-m;m)$ and for any~$\sigma\geq 0$, there is $C>0$ such
that
\begin{equation*}
\|e^{-\i t H}\chi\left(H\right)\|_{
B(L^2_{\sigma},L^2_{-\sigma})}\leq C\left\langle
t\right\rangle^{-\sigma}.
\end{equation*}
\end{proposition}

\medskip

The proof of Theorem~\ref{Thm:Propagation} is then a consequence of
the above propositions
\begin{proof}[Proof of Theorem~\ref{Thm:Propagation}]

We  choose~$\chi_0\in {\mathcal C}^\infty(\R^3,\C^4)$ satisfying the
assumptions of Proposition~\ref{Prop:PropagationNearThresholds},
$\chi_\infty\in {\mathcal C}^\infty(\R^3,\C^4)$ satisfying
assumptions of Proposition \ref{Prop:PropagationHighEnergy} such
that $\chi_0+\chi_\infty=1$. Hence the continuous spectrum of~$H$ is
divided in two parts.
We obtain the inequality
\begin{multline*}
\|e^{-\i t H}{\mathbf
P}_c\left(H\right)\|_{B(L^2_{\sigma},L^2_{-\sigma})}\leq \|e^{-\i t
H}\chi_0\left(H\right){\mathbf P}_c\left(H\right)
\|_{B(L^2_{\sigma},L^2_{-\sigma})}\\
+\|e^{-\i t
H}\chi_\infty\left(H\right)\|_{B(L^2_{\sigma},L^2_{-\sigma})}.
\end{multline*}
Hence from Proposition~\ref{Prop:PropagationNearThresholds},
and~\ref{Prop:PropagationHighEnergy}, we deduce Theorem
\ref{Thm:Propagation}.
\end{proof}

It therefore remains to prove
Propositions~\ref{Prop:PropagationNearThresholds},
and~\ref{Prop:PropagationHighEnergy}.

\subsection{Step~$1$ : Propagation near thresholds}
                                                                \label{Section:Step1}
\subsubsection{Proof of Proposition 
\ref{Prop:PropagationNearThresholds}}

We now prove Proposition \ref{Prop:PropagationNearThresholds}. Let
$\chi$ be in  ${\mathcal C}_0^\infty(\R^3,\C^4)$, then the operator
$e^{-\i tH}{\mathbf P}_c\left(H\right)\chi\left(H\right)$ as a
function of~$t$ is the Fourier transform with respect to~$\lambda$
of
\begin{equation*}
\lambda\mapsto\Im
R_V^{+}(\lambda)\one_{(-\infty,-m]\cup[m,\infty)}(\lambda)\chi(\lambda),
\end{equation*}
where
\begin{equation}
                                                                \label{Def:LimitDiracResolvent}
R_V^\pm(\lambda)=\lim_{\e\to 0^+}R_V(\lambda\pm \i\e),
\end{equation}
we will prove in Section~\ref{Section:Step2} that the limit exists
in~${\mathcal B}(L^2_\sigma,L^2_{-\sigma})$. So
Proposition~\ref{Prop:PropagationNearThresholds} is a consequence of
the
\begin{proposition}
                                                                \label{Prop:ThresholdsBehavior}
Suppose that Assumptions~\ref{assumption:1} and~\ref{assumption:2}
hold. Then for~$\lambda>m$ close enough to~$m$, one has
\begin{equation*}
R_V^\pm(\lambda)=\lim_{\e\to 0^+}R_V(\lambda\pm\i\e)
\end{equation*}
exists in~${\mathcal B}({\mathcal H}_{\sigma}^{-1/2},{\mathcal
H}_{-\sigma}^{1/2})$ for~$\sigma>3/2$. It is~${\mathcal C}^l$ if
$\sigma
> 1/2+l$ and~$0<l\leq 2$ with
\begin{equation}
                                                                \label{Estimate:DiracResolvent}
\frac{d^l}{d\lambda^l}\Im
R_V^\pm(\lambda)=O(\sqrt{\lambda-m}^{1/2-l}),
\end{equation}
as~$\lambda \to m^+$.

The same holds for~$\lambda<-m~$ if~$m$ is replaced by~$-m$.
\end{proposition}
We prove it in Section~\ref{Section:Step1.1}. The idea is then to
apply to
\begin{equation}
                                                                \label{Def:TFPropagator}
\lambda\mapsto\Im
R_V^{+}(\lambda)\one_{(-\infty,-m]\cup[m,\infty)}(\lambda)\chi(\lambda).
\end{equation}
with~$k=1$ and~$\theta=1/2$, the following
\begin{lemma}[Lemma 10.2 of~\cite{JensenKato}]
                                                                \label{Lem:Lemma10.2[JensenKato]}
Suppose~$F(\lambda)=0$ for~$\lambda>a>0$,~$F^{(k+1)}\in
L^1([\delta,+\infty[)$ for any~$\delta >0$ and an integer~$k\geq 0$
and that~$F^{(k+1)}(\lambda)=O(\lambda^{\theta-2})$ near~$0$ for
some~$\theta \in (0,1)$. Assume further that~$F^{(j)}(0)=0$
for~$j\leq k-1$, then
\begin{equation*}
\widehat{F}(t)=O(t^{-k-\theta}).
\end{equation*}
The symbol~$O$ may be replaced by~$o$ throughout.
\end{lemma}
We refer to~\cite{JensenKato} for the proof of Lemma
\ref{Lem:Lemma10.2[JensenKato]}. In fact to apply this lemma
to~\eqref{Def:TFPropagator}, one should split this function in two
parts, one supported in~$\R^+$ and the other in~$\R^-$. Then one
translates the first one by~$-m$ and applies the lemma. To deal with
the other part, one works exactly in the same way after a symmetry
with respect to the origin. To end the proof of Proposition
\ref{Prop:PropagationNearThresholds}, it remains to prove
Proposition \ref{Prop:ThresholdsBehavior}. This the goal of the next
section.

\subsubsection{Behavior near thresholds of the Dirac resolvent: %
proof of Proposition~\ref{Prop:ThresholdsBehavior}}
                                                                \label{Section:Step1.1}
In this section, our aim is to prove Proposition
\ref{Prop:ThresholdsBehavior}. First of all, we notice that if the
limits~\eqref{Def:LimitDiracResolvent} exist then we have
\begin{equation*}
R_V^-(\lambda)^*=R_V^+(\lambda),
\end{equation*}
and since
\begin{equation*}
\alpha_5 (D_m + V -z)^{-1}\alpha_5=-(D_m+\alpha_5V\alpha_5+z)^{-1},
\end{equation*}
for~$\ds\alpha_5=\prod_{i=1}^3\alpha_i\beta$, one obtains
\begin{equation*}
\alpha_5
R_V^\pm(\lambda)^{-1}\alpha_5=-R_{\alpha_5V\alpha_5}^\mp(-\lambda).
\end{equation*}
So we only need to study the behavior of~$R_V^+(\lambda)$ near~$+m$.
Let us introduce
\begin{equation*}
\C_{++}=\left\{z\in \C,\,\Im z > 0,\,\Re z > 0\right\}
\end{equation*}
then the behavior for the free case ($V=0$) is given by the
\begin{proposition}[Dirac's resolvent expansion]
                                                                \label{Prop:DiracExp}
Let be~$s,s'>1/2$ with~$s+s'>2$ and~$t\in \R$. Then~$R_0(z)\in
{\mathcal B}({ H}_{s}^{t-1},{ H}_{-s'}^{t})$ is uniformly continuous
in~$\C_{++}$ and so it can be continuously extended
to~$\overline{\C_{++}}$.

Moreover, the formal series ~$z\in \C_{++}$,
\begin{equation*}
R_0(z)=\ds\sum_{j=0}^\infty(\i\sqrt{z^2-m^2})^j D_m G_j +
\sum_{j=0}^\infty z (\i\sqrt{z^2-m^2})^j G_j
\end{equation*}
with $\Im(\sqrt{z^2-m^2})>0$, is an asymptotic expansion
for~$z\rightarrow m$ in the following sense:

Let~$k\in \N$, if~$R_0(z)$ is approximated by the corespondent
finite series up to~$j=k$, the remainder is~$o(|z-m|^{k/2})$, as
$z\to m$, in the norm of~${\mathcal B}\left({
H}_{s}^{t-1},{H}_{-s'}^{t}\right)$ with~$s,\,s'>k+1/2$
(and~$s+s'>2$ if~$k=0$) and~$t\in \R$.

In the same sense, this identity can be differentiated in~$z$ any
number of times. More precisely, for~$l\in \N^*$ the~$l^{th}$
derivative in~$z$ of the said finite series is equal to
$\frac{d^l}{dz^l}R(z)$ up to an error~$o(|z-m|^{k/2-l})$, as $z\to
m$, in the norm of~${\mathcal B}\left({ H}_{s}^{t-1},{
H}_{-s'}^{t}\right)$ with~$s,\,s'>k+l+1/2$ and~$t\in \R$.
\end{proposition}
\begin{proof}
It is an adaptation of lemmas of~\cite{JensenKato}. We rewrite
\cite[Lemma 2.1]{JensenKato},~\cite[Lemma 2.2]{JensenKato} and
\cite[Lemma 2.3]{JensenKato} in the Dirac case with help of the
identity
\begin{equation*}
(D_m-z)^{-1}(D_m+z)^{-1}=(-\Delta+m^2-z^2)^{-1},
\end{equation*}
or in~$\left(\C^2\right)^2$
\begin{equation*}
\left(D_m-z\right)^ {-1}= \left(\begin{array}{cc}
\ds\frac{z+m}{-\Delta-z^2+m^2}&
\ds\frac{\sigma\cdot\nabla}{-\Delta-z^2+m^2}\\
\ds\frac{\sigma\cdot\nabla}{-\Delta-z^2+m^2}&
\ds\frac{z-m}{-\Delta-z^2+m^2}
\end{array}\right)
\end{equation*}
where~$\sigma$ are the two dimensional Pauli matrices.
\end{proof}

To obtain the behavior of the Dirac resolvent in the general case,
we would like to use the formula
\begin{equation}
R_V(z)=M(z)^{-1}R_0(z)
                                                                \label{Eq:ModBornSeries}
\end{equation}
with
\begin{equation*}
M(z)=\left(1+R_0(z)V\right).
\end{equation*}
To give a meaning to Identity~\eqref{Eq:ModBornSeries}, we have to
prove that~$M(z)$ is invertible in~${\mathcal
B}({H}^{1/2}_{-\sigma})$ for~$\sigma> 1/2$ with~$\sigma+1/2<\rho$,
where $\rho$ is introduced in assumption \ref{assumption:1}. We will
also give the asymptotic behavior of~$R_V^+(z)$ and some of its
derivatives as $\lambda \to m^+$. By means of
Proposition~\ref{Prop:DiracExp}, one has
\begin{equation*}
z\in\overline{\C_{++}}\mapsto M(z)\in {\mathcal B}\left({
H}^{1/2}_{-\sigma}\right)
\end{equation*}
is uniformly continuous for~$1/2<\sigma~$ and
$2<\sigma+\sigma'\leq\rho$ and some~$\sigma'>1/2$. We also have
\begin{equation*}
M(z)=M(m)+ A(z),
\end{equation*}
with~$A(z)$ uniformly continuous in~${\mathcal B}\left({
H}^{1/2}_{-\sigma'},{H}^{1/2}_{-\sigma}\right)$ near~$m$ in
$\C_{++}$  and tending to~$0$ as~$\lambda \to m$ for~$1/2<\sigma~$
and~$2<\sigma+\sigma'\leq \rho$ and some $\sigma'>1/2$. We now prove
the
\begin{lemma}[Threshold's eigenvector and resonance]
                                                                \label{Lem:ThresholdsKernel}
Suppose that Assumption~\ref{assumption:1} holds. Let ${\mathcal
M}(s)$ be the kernel of~$M(m)$ in~${ H}^{1/2}_{-s}$ and ${\mathcal
K}(s)$ the kernel of~$\left(H-m\right)$ in~${ H}^{1/2}_{-s}$.
Then~${\mathcal M}(s)$ and~${\mathcal K}(s)$ are finite dimensional
and do not depend on~$s\in (1/2,\rho-1/2)$. So we write~${\mathcal
M}$ and~${\mathcal K}$ and we have
\begin{equation*}
{\mathcal M}={\mathcal K}
\end{equation*}
\end{lemma}
\begin{proof}
See also~\cite[Lemma 3.1]{JensenKato}.

Let ~$u\in {\mathcal K}(s)$, then~$(D_m+V-m)u=0$ and~ $u\in
{H}_{-s}^{1/2}$, so~$Vu\in {H}_{\rho-s}^{-1/2}$ and
since~$\rho-s>1/2$,~$s>1/2$, and~$s+\rho-s>2$, we obtain, by
Proposition~\ref{Prop:DiracExp},~$\left(D_m-m\right)^{-1}(D_m-m)u
=\left(D_m-m\right)^{-1}Vu\in { H}_{-s}^{1/2}$. For any~$\phi\in
{\mathcal C}^\infty_0$,
\begin{equation*}
\langle \phi,\,\left(D_m-m\right)^{-1}(D_m-m)u\rangle=\langle
(D_m-m)\left(D_m-m\right)^{-1}\phi, u\rangle=\langle \phi, u\rangle,
\end{equation*}
we obtain~$(D_m-m)(D_m-m)^{-1}Vu=Vu$ and
$(D_m-m)(u+\left(D_m-m\right)^{-1}Vu)=0$.
Since $D_m-m$ has no kernel in $H^{1/2}_{-s}$, because there's no
harmonic function in $L^2_{-s}$, we obtain
$u+\left(D_m-m\right)^{-1}Vu=0$. Hence, we have
${\mathcal K}(s)\subset {\mathcal M}(s)$.

Conversely,~$I+\left(D_m-m\right)^{-1}V$ defines a Fredholm operator
of~${\mathcal B}({ H}_{-s}^{1/2})$. If~$u\in{\mathcal M}(s)$
then~$u\in { H}_{-s}^{1/2}$ and~$(D_m-m)^{-1}Vu\in { H}_{-s}^{1/2}$.
So we write
$0=\left(D_m-m\right)(u+\left(D_m-m\right)^{-1}Vu)=(D_m-m+V)u$ and
we obtain~${\mathcal M}(s)\subset{\mathcal M}(s)$.

Now we introduce~$I+V\left(D_m-m\right)^{-1}\in{\mathcal B}({
H}_{s}^{-1/2})$, and its kernel~${\mathcal N}(s)$ which is finite
dimensional is a Fredholm operator. We have that ${\mathcal N}(s)$
is decreasing with~$s$ and~${\mathcal M}(s)$ is increasing. Since,
by duality,~$\dim {\mathcal M}(s)=\dim{\mathcal N}(s)$, we deduce
that~${\mathcal N}(s)$ and~${\mathcal K}(s)={\mathcal M}(s)$ do
not depend on~$s$.
\end{proof}

We are now able to conclude the proof of Proposition
\ref{Prop:ThresholdsBehavior}.
\begin{proof}[Proof of Proposition~\ref{Prop:ThresholdsBehavior}]
Assumption~\ref{assumption:2} gives~${\mathcal K}=0$ and so with
Lemma \ref{Lem:ThresholdsKernel}, one obtains~${\mathcal M}=0$.
Hence~$M(m)$ is invertible since it is a Fredholm operator. We use
Von Neumann series to obtain that~$M(z)$ is invertible in~${\mathcal
B}\left({ H}^{1/2}_{-\sigma}\right)$ for~$\sigma> 1/2$,
$2<\sigma+\sigma'\leq \rho$ and some~$\sigma'>1/2$ and
\begin{equation*}
M(z)^{-1}=M(m)^{-1}\sum_{j\geq 0}(A(z)M(m)^{-1})^j.
\end{equation*}
So for~$\lambda\geq m$ close enough to~$m$,
\begin{equation*}
M^+(\lambda)^{-1}=\lim_{\e\to 0^+}M(\lambda+\i\e)^{-1}
\end{equation*}
exists in~${\mathcal B}\left({ H}^{1/2}_{-\sigma}\right)$
for~$\sigma> 1/2$ with~$2<\sigma+\sigma'\leq \rho$ and
some~$\sigma'>1/2$. We obtain that~$\ds\lim_{\e\to
0^+}R_V\left(\lambda+i\e\right)$ exists in~${\mathcal B}\left({
H}^{-1/2}_{\sigma''},{ H}^{1/2}_{-\sigma}\right)$ for~$\sigma> 1/2$
and $\sigma\geq\sigma''>1/2$ with~$\sigma+\sigma''>2$,
$2<\sigma+\sigma'\leq\rho$ and some~$\sigma'>1/2$.

Using Proposition~\ref{Prop:DiracExp}, we prove that if
$1/2+k<\sigma$, and~$\sigma'+1/2+k<\rho$ then
\begin{equation*}
\frac{d^k}{d\lambda^k}M^+(\lambda)=O(\sqrt{\lambda-m}^{1/2-k})
\end{equation*}
in ~${\mathcal B}\left({ H}^{-1/2}_{-\sigma'},{
H}^{1/2}_{-\sigma}\right)$ for~$k\in\N^*$ as~$\lambda\to m^+$. Since
we have
\begin{equation*}
\frac{d}{d\lambda}F(\lambda)^{-1}=-F(\lambda)^{-1}
\left(\frac{d}{d\lambda}F(\lambda)\right) F(\lambda)^{-1},
\end{equation*}
for matrix valued differentiable function ~$F$ with invertible
values, we obtain for~$k\in\N^*$ the estimate
\begin{equation*}
\frac{d^k}{d\lambda^k}M^+(\lambda)^{-1}=O(\sqrt{\lambda-m}^{1/2-k}),
\end{equation*}
in~${\mathcal B}\left({ H}^{1/2}_{-\sigma}\right)$
with~$1/2+k<\sigma$ and~$\sigma+1/2+k< \rho$ as~$\lambda\to m^+$. So
by Leibniz formula, we also have for~$k\in\N^*$
\begin{equation*}
\frac{d^k}{d\lambda^k}R^+_V(\lambda)=O(\sqrt{\lambda-m}^{1/2-k}),
\end{equation*}
in~${\mathcal B}\left({H}^{-1/2}_{\sigma'},{
H}^{1/2}_{-\sigma}\right)$ with~$1/2+j<\sigma$,~$1/2+k-j<\sigma'$
and~$1/2+k-j+\sigma < \rho$ for all~$j\in\left\{0,\ldots,k\right\}$
as~$\lambda\to m^+$. For the case~$k=0$, we have the formula
\begin{equation*}
R_V(z)=R_0(z)\left(1+VR_0(z)\right)^{-1}.
\end{equation*}
Since~$R^+(m)=R^-(m)$, this leads to
\begin{equation*}
\Im R_V^+(m)=0,
\end{equation*}
and so
\begin{equation*}
\Im R_V^\pm(\lambda)=O(\sqrt{\lambda-m}^{1/2}),
\end{equation*}
as~$\lambda\to m^+$ in~${\mathcal B}\left({
H}^{-1/2}_{\sigma},{H}^{1/2}_{-\sigma}\right)$ with
$3/2<\sigma$ and~$\sigma+3/2<\rho$.
Hence~\eqref{Estimate:DiracResolvent} is proved.
\end{proof}

\subsection{Step 2: Propagation far from thresholds}
                                                                \label{Section:Step2}
In this section, we prove Proposition
\ref{Prop:PropagationHighEnergy}. We prove the propositions for
$t\geq 0$. Then using~$\left(e^{-\i tH}\right)^*=e^{\i t H}$, the
result easily follows for~$t\leq 0$.

\subsubsection{Proof of Proposition 
\ref{Prop:PropagationHighEnergy}}

Let us introduce
\begin{equation*}
A=\frac{1}{2}\left\{D_m^{-1}P\cdot Q +Q\cdot P D_m^{-1}\right\}.
\end{equation*}
\cite[Lemma 3.1]{IftimoviciMantoiu} gives that~$A$ is an essentially
self-adjoint operator and the domain of its closure contains the
domain of~$\langle Q\rangle$.
Proposition \ref{Prop:PropagationHighEnergy} is then a consequence
of the
\begin{theorem}[Minimal escape velocity]
                                                                \label{Thm:MinimalEscapeVelocity}
Suppose that Assumption~\ref{assumption:1} holds. Then for any
$\chi\in{\mathcal C}_0^\infty$ bounded with support in
$(-\infty,-m)\cup(m,+\infty)$, there exists~$\theta>0$ such that for
any~$l\in\R$, for any~$v\in(0,\theta)$, and any~$a\in \R$ one has
\begin{equation*}
\forall t>0,\;\left\|\one_{A-a-vt\leq 0}e^{-itH}\chi(H)\one_{A-a\geq
0}\right\|\leq C t^{-l},
\end{equation*}
where $C$ do not depend $a$ and $t$.
\end{theorem}

The proof will be given in Section~\ref{Section:Step2.1}. Let us now
show that Theorem~\ref{Thm:MinimalEscapeVelocity} implies
Proposition~\ref{Prop:PropagationHighEnergy}.
\begin{proof}[Proof of Proposition~\ref{Prop:PropagationHighEnergy}]
We notice that for $c\geq 0$
\begin{equation*}
\ds  \langle A \rangle ^{-\alpha}= \langle A \rangle
^{-\alpha}\one_{\pm A\geq ct} +O(t^{-\alpha}),
\end{equation*}
when $t\geq 0$, this leads to
\begin{equation*}
\ds  \langle A \rangle ^{-\alpha} e^{-itH}\chi(H) \langle A
\rangle^{-\alpha} = \langle A \rangle^{-\alpha}
\one_{A\leq\frac{(\theta-\e) t}{2}} e^{-itH} \chi(H)
\one_{A\geq\frac{\theta t}{2}} \langle A \rangle^{-\alpha}
+O(t^{-\alpha}),
\end{equation*}
So if we choose~$a=-\frac{\theta t}{2}$ and~$v=\theta -\frac{\e}{2}$
in Theorem~\ref{Thm:MinimalEscapeVelocity}, we obtain
\begin{equation*}
\ds \|\langle A\rangle^{-\alpha}e^{-itH}\chi(H) \langle A \rangle
^{-\alpha}\|\leq C t^{-\min(\alpha,\;l)}.
\end{equation*}

Then we prove that~$\langle A\rangle^\alpha\langle Q
\rangle^{-\alpha}$ is bounded for any positive~$\alpha$. It is quite
immediate for integer~$\alpha$ using multi-commutator expansion
\cite[Identity (B.24)]{HunzikerSigal}. To prove it for any positive
real, we use \cite[Identity (1.2)]{SigalSoffer}. This identity
states that for a self adjoint with~$B\geq 1$ and a positive real
$\beta$, we have on domain of~$B^{[\alpha]+1}$
\begin{equation*}
B^{\beta}=\frac{\sin(\pi\{\beta\})}{\pi}
\int_0^{+\infty}\frac{w^{\{\beta\}-1}}{B+w}dw B^{[\beta]+1},
\end{equation*}
where~$\{\beta\}=\beta-[\beta]$ and~$[\beta]$ is the integer part.
With this formula for~$B=\langle A\rangle^{2k}$ for any~$k\in\N$, we
prove for any~$\beta\in]0,\,1[$ that
\begin{equation*}
\langle A\rangle^{2k\beta} \leq C \langle Q\rangle^{2k\beta}.
\end{equation*}
This ends the proof of Proposition \ref{Prop:PropagationHighEnergy}.
\end{proof}

\subsubsection{Proof of Theorem 
\ref{Thm:MinimalEscapeVelocity}}
                                                                \label{Section:Step2.1}
Our proof of Theorem \ref{Thm:MinimalEscapeVelocity} is an
adaptation of the one of \cite{HunzikerSigalSoffer}, we make some
modifications.

For any self-adjoint operator~$B$ with domain~$D(B)$, we write
$Ad_A(B)$ for the operator~$[A,B]$ with domain~$D(A)\cap D(B)$ dense
in~$D(B)$, defined by
\begin{equation*}
\ds \forall u,v \in D(A)\cap D(B),\,
\left\langle\i[A,B]u,v\right\rangle=\i(\left\langle
Bu,Av\right\rangle-\left\langle Au,Bv\right\rangle).
\end{equation*}
First of all, we have
\begin{lemma}
                                                                \label{Lem:BoundComm} Suppose that Assumption~\ref{assumption:1}
holds. Then~$Ad_A^k(H)$ is bounded and can be written as a finite
sum of terms of the form
\begin{equation*}
f(P)g(Q)h(P)
\end{equation*}
where~$f$ and~$h$ are rational fractions with coefficients in
${\mathcal M}_4(\C)$ of degree at most~$0$ with no poles, and~$g$ is
a function that satisfies Assumption~\ref{assumption:1}.
\end{lemma}
\begin{proof}
The proof is a simple calculation based on the fact that
$Ad_{P_j}(f(Q))=-\i(\nabla_j f)(Q)$.
\end{proof}
We can state the
\begin{lemma}[Mourre estimate]
                                                                \label{Lem:MourreEstimate}
If~$V$ satisfies Assumption~\ref{assumption:1}, then for any $\theta
\in (0,1)$ there exists~$\nu\geq 0$, one has
\begin{equation*}
\one_{\left|H\right|\geq m+\nu}\i[H,A]\one_{\left|H\right|\geq
m+\nu}\geq \theta\one_{\left|H\right|\geq m+\nu},
\end{equation*}
and for any $\lambda\in(-\infty,-m)\cup(+m,+\infty)$ for any
$\delta>0$ there exists $\e>0$ such that
\begin{equation*}
\one_{\left|H-\lambda\right|\leq\e}\i[H,A]\one_{\left|H-\lambda\right|\leq\e}\geq
\left(\frac{\lambda^2}{\sqrt{\lambda^2+m^2}}-\delta\right)\one_{\left|H-\lambda\right|\leq\e}.
\end{equation*}
\end{lemma}
\begin{proof}
This is a consequence of
$\i[D_m,A]=\frac{-\Delta}{-\Delta+m^2}$ and $[V,A]$ is a compact
operator in $B(L^2(\R^3,\C^4))$.
\end{proof}

We now adapt \cite[Theorem 1.1]{HunzikerSigalSoffer} to the case of
unbounded energy since here the multi-commutators $Ad_A^k(H)$ are
bounded operators.
We need the
We introduce the
\begin{definition}
We call generalized indicator function of~$\R^-$  a function of the
form
\begin{equation*}
x\mapsto e^{\frac{u(x)}{x}}\one_{\R^-}(x)
\end{equation*}
with~$u\in {\mathcal C}^\infty_0(\R)$ supported in~$[-\eta,\eta]$
(for some~$\eta>0$), nonnegative, and such that~$u(0)=1$.
\end{definition}
Note that our generalized indicator function of~$\R^-$  are of
infinite order in the sense of \cite[Section
2]{HunzikerSigalSoffer}. Using the commutators expansion presented
in~\cite[Section B]{HunzikerSigal} and the Mourre estimate of
Lemma~\ref{Lem:MourreEstimate}, we have the
\begin{lemma}
                                                                \label{lemma:LemTech}
Let~$f$ be a generalized indicator function of~$\R^-$ and $g\in
{\mathcal C}^\infty(\R)$ with support sufficiently far from
thresholds $\pm m$ or support sufficiently small in
$(-\infty,-m)\cup(+m,+\infty)$. Let be
$A_s=s^{-1}\left\{A-a\right\}$ and~$0<\e\leq 1$. Then for
any~$n\in\N$ and~$\delta>0$ there exists a~$C>0$ independent of
$a\in \R$ such that for~$s\geq 1$
\begin{multline*}
g(H)i[H,f(A_{s})]g(H)\leq s^{-1}\theta
g(H)f'(A_{s})g(H)\\
+Cs^{-1-\varepsilon}g(H)f^{1-\delta}(A_{s})g(H) +C
s^{-(2n-1-\e)}g^2(H).
\end{multline*}
\end{lemma}
\begin{proof}
See~\cite[Lemma 2.1]{HunzikerSigalSoffer}, in our case we don't need
to replace~$H$ by $b(H)H$ with~$b\in {\mathcal C}_0^\infty$. Indeed,
our commutators~$Ad_A^k(H)$ are bounded by means of
Lemma~\ref{Lem:BoundComm}. Then we replace the notion of function of
order $p$ by the one of generalized indicator function. Finally, we
use the fact that a generalized indicator function~$f$ satisfies
\begin{equation*}
\forall k\in \N,\, \forall \delta \in (0,1),\,\exists C>0,\quad
\left|f^{(k)}\right|\leq C \left|f\right|^{1-\delta}.
\end{equation*}
\end{proof}

We are now able to give the
\begin{proof}[Proof of Theorem~\ref{Thm:MinimalEscapeVelocity}]
We write $\chi$ as a finite sum of function $g_j\in {\mathcal
C}^\infty(\R)$ with support sufficiently far from thresholds $\pm m$
or support sufficiently small in $(-\infty,-m)\cup(+m,+\infty)$. If
we prove the theorem for $g_j$ instead of $\chi$ the theorem follows
by summing each estimates for $g_j$ since the sum is finite. In the
rest of the proof, we will not write the index $j$ of $g$.

We notice that if~$0<v<\theta-\eta$ and if~$F$ is a positive non
increasing~${\mathcal C}^\infty$-function which equals~$0$ on~$\R^+$
and~$1$ on~$(-\infty,-\eta)$, we have
\begin{equation*}
\one_{(A-a-vs)<0}\leq F\left(\frac{A-a}{s}-\theta\right).
\end{equation*}
Now suppose~$F$ is a generalized indicator function of~$R^-$. We
consider
\begin{equation*}
F(s^{-1}\left\{A-a-\theta t\right\})
\end{equation*}
and study the time evolution of the observable~$g(H)f(A_{ts})g(H)$,
where~$f=F^{2}$, with respect to~$e^{-\i tH}\one_{A-a>0}$. That is
to say we study
\begin{equation*}
\langle e^{-\i tH}\one_{A-a>0}\psi,g(H)f(A_{ts})g(H)e^{-\i
tH}\one_{A-a>0}\psi\rangle.
\end{equation*}
We work exactly as in the proof of~\cite[Theorem
1.1]{HunzikerSigalSoffer}. Hence using Lemma~\ref{lemma:LemTech} we
obtain for~$0\leq t\leq s$ and~$s>1$
\begin{multline*}
\langle e^{-\i tH}\one_{A-a>0}\psi,g(H)f(A_{ts})g(H)e^{-\i
tH}\one_{A-a>0}\psi \rangle
\leq Cs^{-(2n-2-\e)}\|\psi\|^2\\
+Cs^{-1-\e}\int_0^t\langle e^{-\i
tH}\one_{A-a>0}\psi,g(H)f(A_{ts})g(H)e^{-\i tH}
\one_{A-a>0}\psi\rangle^{1-\delta}||\psi||^{2\delta}.
\end{multline*}
Then using the Gronwall's lemma (see~\cite[Lemma
7.A.1]{AmreinBoutetdeMonvelGeorgescu}),
 we obtain
\begin{multline*}
\langle e^{-\i tH}\one_{A-a>0}\psi,g(H)f(A_{ts})g(H)e^{-\i
tD_m}\one_{A-a>0}\psi \rangle\\ \leq
\left\{Cs^{-\delta(2n-2-\e)}\|\psi\|^{2\delta} +\delta
Cs^{-\e}\|\psi||^{2\delta}\right\}^{1/\delta},
\end{multline*}
so if we choose a small~$\delta$ and a big~$n$, the proof is done if
we choose~$s=\max\left\{1,t\right\}$.
\end{proof}

\section{Proof of Theorem~\ref{Thm:Dispersion}: 
dispersive estimates}
                                                                \label{sec:dispersive_estimate}
Dispersive estimates for Schr\"odinger operators with electric
potentials take place in Lebesgue spaces. This fact permits to use
simple perturbation methods (like Duhamel's formula) to prove the
decay estimates for perturbed Schr\"odinger equations.
Unfortunately, we have only been able to prove dispersive estimates
for Dirac operators in Besov spaces, so it was not possible for us
to use Duhamel's formula or other perturbation method used for
Schr\"odinger operators.

We notice that in the case of a Dirac operator with scalar
potentials (matrix valued functions colinear with~$\beta$), the
square of the Dirac equation gives four coupled Klein-Gordon
equations with an electrostatic potential. This permits to use
results on the Klein-Gordon equation. For example, Yajima
\cite{Yajima} proved dispersive estimates for the Klein-Gordon
equation by using wave operators associated with Schr\"odinger
operators including an electrostatic potential. But in the general
case, by taking the square of a Dirac operator with a potential, we
obtain also a magnetic potential. Hence the method used by Yajima
does not work in our case.

To our knowledge the only one study of the dispersive estimates
associated with the  Dirac equation,  is the the work of D'Anconna
and Fanelli \cite{D'AnconnaFanelli} for the massless case. For non
zero mass we have not been able to found any reference. Even for the
free case for which dispersive estimates can be deduced from those
of Klein-Gordon equation. Here, to give a sketch of the proof for
the general case, we first prove the free case estimates (see
Section \ref{Section:FreeDispersiveEstimates}), using estimates on
oscillatory integrals of Section~\ref{Section:OscilotaryIntegrals}.
In Section~\ref{Section:DistortedPlaneWaves}, following Cuccagna and
Schirmer~\cite{CuccagnaSchirmer}, we introduce the distorted plane
waves. This permits us to tackle the proof of the general case in
Section~\ref{proof_theorem_dispersion_subsec}.

\subsection{Estimates on some oscillatory integrals}
                                                                 \label{Section:OscilotaryIntegrals}
Here, we state some stationary phase type results which will be
useful for the rest of the proof. We denote by~$S^2$ the unit sphere
of~$\R^3$.

\begin{lemma}
                                                                 \label{lemma:Key}
Let be~$f\in {\mathcal C}^{1}(S^2)$ and for any~$v\in
S^2$ and any~$k\in\R$ define
\begin{equation*}
J_v(k)=\int_{S^2}e^{\i k\{1-v\cdot \omega\}} f(\omega)\,d\omega.
\end{equation*}
Then we have
\begin{equation}
                                                                \label{Estimate:Key}
\left|J_v(k)\right|\leq \frac{C}{\langle
k\rangle}\left\{\sum_{|\alpha|\leq 1}\int_{S^2}
\frac{\left|\nabla^\alpha f(\omega)\right|}{\left|\omega-
v\right|^{|\alpha|}}\,d\omega+\sum_{|\alpha|\leq 1}\int_{S^2}
\frac{\left|\nabla^\alpha f(\omega)\right|}{\left|\omega+
v\right|^{|\alpha|}}\,d\omega\right\},
\end{equation}
where~$C$ does not depend on~$f$,~$k$ or~$v$.

If~$f$ is in~${\mathcal C}^{2}(S^2)$ with~$f(v)=f(-v)=0$, we have
\begin{equation}
                                                                \label{Estimate:Key2}
\left|J_v(k)\right|\leq \frac{C}{\langle
k\rangle^{3/2}}\left\{\sum_{|\alpha|\leq 1}\int_{S^2}
\frac{\left|\nabla^{2\alpha} f(\omega)\right|}{\left|\omega-
v\right|^{|\alpha|}}\,d\omega+\sum_{|\alpha|\leq 1}\int_{S^2}
\frac{\left|\nabla^{2\alpha} f(\omega)\right|}{\left|\omega+
v\right|^{|\alpha|}}\,d\omega\right\},
\end{equation}
where~$C$ does not depend on~$f$,~$k$ or~$v$.

If~$f$ is in~${\mathcal C}^{2}(S^2)$ and vanishes in a neighborhood
of~$v$ and~$-v$, we have
\begin{equation}
                                                                \label{Estimate:Key3}
\left|J_v(k)\right|\leq \frac{C}{\langle
k\rangle^{2}}\left\{\frac{\sum_{|\alpha|\leq 1}\int_{S^2}
\frac{\left|\nabla^{2\alpha} f(\omega)\right|}{\left|\omega-
v\right|^{|\alpha|}}\,d\omega}{{\rm dist}({\rm
supp}(f),v)}+\frac{\sum_{|\alpha|\leq 1}\int_{S^2}
\frac{\left|\nabla^{2\alpha} f(\omega)\right|}{\left|\omega+
v\right|^{|\alpha|}}\,d\omega}{{\rm dist}({\rm
supp}(f),-v)}\right\},
\end{equation}
where~$C$ does not depend on~$f$,~$k$ or~$v$.
\end{lemma}

\begin{proof}
We can suppose~$v=(0,0,1)$ since estimate~\eqref{Estimate:Key},
\eqref{Estimate:Key2} and~\eqref{Estimate:Key3} are invariant under
the action of rotations. We have
\begin{equation*}
J_v(k)=\int_0^{2\pi} \int_{0}^\pi e^{\i k\{1-\cos(\phi)\}}
f(\theta,\phi) \sin(\phi)\,d\phi d\theta,
\end{equation*}
then we make an integration by parts in~$\phi$
\begin{multline*}
J_v(k)=-\frac{\i}{k}\int_0^{2\pi}\left[e^{\i k\{1-\cos(\phi)\}}
f(\theta,\phi)\right]_0^\pi\, d\theta\\ +\frac{\i}{k}\int_0^{2\pi}
\int_0^\pi e^{\i k\{1-\cos(\phi)\}} \p_\phi f(\theta,\phi) \,d\phi
d\theta,
\end{multline*}
If we suppose that~$f$ vanishes in a neighborhood of~$v$ or~$-v$,
then we use that for any $\phi'$
\begin{equation*}
|f(\theta,\phi')|\leq \int_0^\pi \left|\p_\phi
f(\theta,\phi)\right|\,d{\phi}
\end{equation*}
to obtain~\eqref{Estimate:Key} in this case. Otherwise with help of
a smooth cut-off, we split the integral in two parts, each one has a
support far from~$v$ or~$-v$. Repeating the previous proof for each
part, we prove the estimate~\eqref{Estimate:Key} in the general
case.

If moreover we have~$f(v)=f(-v)=0$ then we have for any~$\alpha>0$
\begin{multline*}
J_v(k)=\frac{\i}{k}\int_0^{2\pi} \int_0^\alpha e^{\i
k\{1-\cos(\phi)\}} \p_\phi f(\theta,\phi) \,d\phi d\theta\\
+\frac{\i}{k}\int_0^{2\pi} \int_{\pi-\alpha}^\pi e^{\i
k\{1-\cos(\phi)\}} \p_\phi f(\theta,\phi) \,d\phi d\theta\\
+\frac{\i}{k}\int_0^{2\pi} \int_\alpha^{\pi-\alpha} e^{\i
k\{1-\cos(\phi)\}} \p_\phi f(\theta,\phi) \,d\phi d\theta.
\end{multline*}
We use an integration by parts to obtain for the second term of the
right hand side
\begin{multline*}
\int_\alpha^{\pi-\alpha} e^{\i k\{1-\cos(\phi)\}} \p_\phi
f(\theta,\phi) \,d\phi=\frac{i}{k}\left[e^{\i k\{1-\cos(\phi)\}}
\frac{\p_\phi
f(\theta,\phi)}{\sin(\phi)}\right]_\alpha^{\pi-\alpha}\\
-\frac{i}{k}\int_\alpha^{\pi-\alpha} e^{\i
k\{1-\cos(\phi)\}}\left\{\frac{\p_\phi^2
f(\theta,\phi)}{\sin(\phi)}-\frac{\cos(\phi)\p_\phi
f(\theta,\phi)}{\sin(\phi)^2}\right\} \,d\phi,
\end{multline*}
for the other terms of the right hand side direct estimations give
us
\begin{multline*}
\left|J_v(k)\right|\leq
\frac{C\alpha}{|k|}\int_0^{2\pi}\ds\sup_{\phi}\left|\p_\phi
f(\theta,\phi)\right|\,d\theta\\+\frac{C}{\alpha |k|^2}\int_0^{2\pi}
\left\{\sup_{\phi}\left|\p_\phi f(\theta,\phi)\right|\,d\theta
+\int_0^{2\pi}\int_0^\pi\left|\p_\phi^2 f(\theta,\phi)\right|\;d\phi
d\theta\right\},
\end{multline*}
choosing~$\alpha=\sqrt{|k|}^{-1}$ and working like in the proof of
the estimate~\eqref{Estimate:Key}, we obtain estimate
\eqref{Estimate:Key2}. The reader recognized the proof of the well
known Van der Corput Lemma with modification in order to give
precise estimates.

For the estimate~\eqref{Estimate:Key3}, we first split the integral
$J_v(k)$ in two hemispheres with respect to the pole~$v$ and we
choose $\alpha={\rm dist}({\rm supp}(f),v)$ or $\alpha={\rm
dist}({\rm supp}(f),-v)$.
\end{proof}

We obtain first the
\begin{proposition}
                                                                \label{proposition:Decay1}
Let~$h\in {\mathcal C}(\R)$ and~$g\in {\mathcal C}^2(\R^3)$ be such
that the integrals appearing in the following estimate are finite.
Then defining
\begin{equation*}
I(k,u)=\int_{\R^3}e^{\i k\{h(|\xi|)-\xi\cdot u\}} g(\xi)\,d\xi,
\end{equation*}
for any~$u\in\R^3$ and any~$k\in\R$, we have
\begin{equation}
                                                                \label{Estimate:Decay1-1}
\ds\left|I(k,u)\right|\leq \frac{C}{\left| k
u\right|}\max_{|\alpha|\leq 1}\left\{\int_{\R^3}
|\xi|^{|\alpha|-1}\frac{\left|\nabla^\alpha g(\xi)\right|}
{\left|\frac{u}{|u|}-\frac{\xi}{|\xi|}\right|^{|\alpha|}}\,d\xi,
\int_{\R^3} |\xi|^{|\alpha|-1}\frac{\left|\nabla^\alpha
g(\xi)\right|}
{\left|\frac{u}{|u|}+\frac{\xi}{|\xi|}\right|^{|\alpha|}}\,d\xi\right\}
\end{equation}
where~$C$ does not depend on~$h$,~$g$,~$k$ or~$u$.

If moreover~$g$ vanishes in a cone of axis~$D={\rm Span}(u)$, we
have
\begin{multline*}
\ds\left|I(k,u)\right|\leq \frac{C}{\left| k u\right|^{2}{\rm
dist}({\rm supp}(g)\cap S^2,D\cap S^2)}\times\\
\times\max_{|\alpha|\leq 1}\left\{\int_{\R^3}
|\xi|^{2|\alpha|-2}\frac{\left|\nabla^{2\alpha} g(\xi)\right|}
{\left|\frac{u}{|u|}-\frac{\xi}{|\xi|}\right|^{|\alpha|}}\,d\xi,
\int_{\R^3} |\xi|^{2|\alpha|-2}\frac{\left|\nabla^{2\alpha}
g(\xi)\right|}
{\left|\frac{u}{|u|}+\frac{\xi}{|\xi|}\right|^{|\alpha|}}\,d\xi\right\}
\end{multline*}
where~$C$ does not depend on~$h$,~$g$,~$k$ or~$u$.
\end{proposition}
\begin{proof}
We write
\begin{eqnarray*}
I(k,u)=\int_{\R^3}e^{\i k\{h(|\xi|)-\xi\cdot u\}} g(\xi)\,d\xi
=\int_{\R^+}e^{\i k\{h(\rho)-\rho |u|\}} J_{\frac{u}{|u|},\rho}(\rho
k|u|) \rho^2\, d\rho,\\
\end{eqnarray*}
where~$J_{v,\rho}(k)=\int_{S^2} e^{\i k\{1-v\cdot \omega\}}
g(\rho\omega)\,d\omega$ and we apply Lemma~\ref{lemma:Key}.
\end{proof}

We introduce a first useful variant with the
\begin{proposition}
                                                                \label{proposition:Decay1-hom}
Let~$g\in {\mathcal C}^{1+k}(\R^3)$ be such that the integrals
appearing
 in the following estimate are finite. We introduce
\begin{equation*}
F(x)=\int_{\R^3}e^{\i\{|\xi||x|-\xi\cdot x\}} g(\xi)\,d\xi
\end{equation*}
for any~$x\in\R^3$. Then for all~$\alpha \in \N^3$ such that
$|\alpha|\leq k$ we have
\begin{equation}
                                                                \label{Estimate:Decay1-hom}
\left|\nabla^\alpha F(x)\right|\leq\frac{C}{|x|^{|\alpha|+1}}
\max_{|\beta|\leq 1+|\alpha|} \left\{\int_{\R^3}|
\xi|^{|\beta|-1}\frac{|\nabla^\beta g(\xi)|} {\left|\frac{x}{|x|}-
\frac{\xi}{|\xi|}\right|}\;d\xi\right\}.
\end{equation}
If moreover~$g$ vanishes in a half cone of axis
$D^+=\left\{\rho\frac{x}{|x|},\;\rho\in\R^+\right\}$. Then for all
$\alpha \in \N^3$ such that~$|\alpha|\leq k$, we have
\begin{multline}
                                                                \label{Estimate:Decay1-hom2}
\left|\nabla^\alpha F(x)\right|\leq\frac{C}{|x|^{|\alpha|+2}{\rm
dist}({\rm supp}(g)\cap S^2,D^+\cap S^2)}\times\\
\times \max_{|\beta|\leq 2+|\alpha|} \left\{\int_{\R^3}|
\xi|^{|\beta|-2}\frac{|\nabla^\beta g(\xi)|} {\left|\frac{x}{|x|}-
\frac{\xi}{|\xi|}\right|}\;d\xi\right\}.
\end{multline}
\end{proposition}
\begin{proof}
The critical points correspond to the the semi axis spanned by~$x$.
We treat the part of the integrals which is far from critical points
by using an integration by parts with help of the operator
$L=\frac{\frac{\xi}{|\xi|}-\frac{x}{|x|}}
{|x|\left|\frac{\xi}{|\xi|}-\frac{x}{|x|}\right|^2}\cdot
\nabla_\xi$. Let be~$\phi(x,\xi)=\{|\xi||x|-\xi\cdot x\}$, we have
\begin{equation*}
F(x)=\langle L e^{\i\phi(x,\cdot)},g\rangle =\langle
e^{\i\phi(\xi,\cdot)},L^*g\rangle,
\end{equation*}
with
\begin{equation*}
L^*
=-L-\frac{2}{|x||\xi|\left|\frac{\xi}{|\xi|}-\frac{x}{|x|}\right|^2}.
\end{equation*}
This gives the bound
\begin{equation*}
\frac{C}{|x|}\max_{|\beta|\leq 1}
\left\{\int_{\R^3}|\xi|^{|\beta|-1}|\nabla^\beta
g(\xi)|\,d\xi\right\},
\end{equation*}
or after an iteration
\begin{equation*}
\frac{C}{|x|^2}\max_{|\beta|\leq 2}
\left\{\int_{\R^3}|\xi|^{|\beta|-1}|\nabla^\beta
g(\xi)|\,d\xi\right\},
\end{equation*}
we obtain Estimate~\eqref{Estimate:Decay1-hom} for~$\alpha=0$. The
method to treat the other part of the integral is exactly the one we
used in the proof of Proposition~\ref{proposition:Decay1}.

For higher order derivatives, we have
\begin{equation*}
\nabla_x e^{\i k\phi(x,\xi)}=\frac{|\xi|}{|x|} \nabla_\xi e^{\i
k\phi(x,\xi)}
\end{equation*}
and so
\begin{equation*}
\langle \nabla_x e^{\i\phi(x,\cdot)},g\rangle =-\frac{1}{|x|}
\langle e^{\i\phi(x,\cdot)}, \nabla |Q|g\rangle.
\end{equation*}
the result is then obtained by applying this trick~$|\alpha|$ times
and then repeating our proof for the case~$\alpha=0$, we obtain
Estimates~\eqref{Estimate:Decay1-hom} and
\eqref{Estimate:Decay1-hom2} for~$\nabla^\alpha F(x)$.
\end{proof}

And finally, we need the
\begin{proposition}
                                                                \label{Prop:Decay2}
Let be~$g\in {\mathcal C}^2(\R^3)$ with compact support. Then for
any~$u\in \R^3$,~$k\in\R$ and
\begin{equation*}
I(k,u)=\int_{\R^3}e^{\i k\{\sqrt{\xi^2+m^2}-\xi\cdot u\}}
g(\xi)\,d\xi,
\end{equation*}
we have
\begin{multline}
                                                                \label{Estimate:Decay2}
\left|I(k,u)\right|\leq \frac{C}{\left| k\right|^{3/2}}\max \Bigg[
\max_{|\alpha|\leq 2}\left\{\int_{\R^3}\left|\left\langle
\xi\right\rangle^{|\alpha|-1}\nabla^\alpha
g(\xi)\right|\;d\xi\right\};\\
\frac{1}{|u|\sqrt{\ds\inf_{x\in {\rm
supp}(g)}\left\{\frac{m^2}{\sqrt{x^2+m^2}^3}\right\}}}\max_{\substack{l\leq
1\\n\leq 1}}\left\{\int_{\R^3}\left|\xi\right|^{l-n-1}
\frac{\left|\p_{|\xi|}^l\p_\omega^n
g(\xi)\right|}{\left|\frac{u}{|u|}-\frac{\xi}{|\xi|}\right|}
d\xi\right\}\Bigg].
\end{multline}
\end{proposition}
\begin{proof}
We can suppose~$u=(0,0,|u|)$ since estimate~\eqref{Estimate:Decay2}
is invariant under the action of rotations.

The oscillatory integral~$I(k,u)$ is bounded and critical points of
the phase of~$I(k,u)$ are supported by the semi axis spanned by~$u$.
With help of a smooth cut-off function~$\chi~$, we split the
integral in two parts~$I(k,u)=I_1(k,u)+I_2(k,u)$, where~$I_1(k,u)$
is supported in a half cone around~$u$. We then use multiple
integrations by parts with help of the operator
\begin{equation*}
L=\frac{\frac{\xi}{\sqrt{\xi^2+m ^2}}-u}
{|k|\left|\frac{\xi}{\sqrt{\xi^2+m ^2}}-u\right|^2}\cdot \nabla_\xi.
\end{equation*}
Since~$\left(1-\chi\right)g\in {\mathcal C}^2(\R^3)$ has support far
from critical points and since for~$\lambda(\xi)=\sqrt{\xi^2+m^2}$
we have~$\left\|\nabla_\xi^\alpha \lambda(\xi)\right\|\leq
C_\alpha\lambda(\xi)^{1-|\alpha|}~$, we obtain
\begin{equation*}
\left|I_1(k,u)\right|\leq\frac{C}{|k|}\sum_{|\alpha|\leq
1}\|\lambda(Q)^{|\alpha|-1}\nabla^\alpha g\|_{L^1}
\end{equation*}
and
\begin{equation*}
\left|I_1(k,u)\right|\leq\frac{C}{k^2}\sum_{|\alpha|\leq
2}\|\lambda(Q)^{|\alpha|-1}\nabla^\alpha g\|_{L^1}.
\end{equation*}
Otherwise~$I_2(k,u)$ has support in a small cone around~$u$, and we
have
\begin{eqnarray*}
I_2(k,u)&=&\int_{\R^3}e^{\i k\{\sqrt{\xi^2+m^2}-\xi\cdot u\}}
\widetilde{g}(\xi)\,d\xi\\
&=&\int_{\R^+}e^{\i k\{\sqrt{\xi^2+m^2}-|\xi|\left|u\right|\}}
J_{\frac{u}{|u|},\rho}(\rho k|u|) \rho^2\, d\rho,
\end{eqnarray*}
with
\begin{equation*}
J_{v,\rho}(k)=\int_{S^2} e^{\i k\{1-v\cdot \omega\}}
\widetilde{g}(\rho\omega)\,d\omega,
\end{equation*}
where~$\widetilde{g}=\chi g$. We obtain after an integration by
parts
\begin{multline*}
J_{v,\rho}(k)=-\frac{\i}{k}\int_0^{2\pi}\left[e^{\i
k\{1-\cos(\phi)\}}
\widetilde{g}(\rho\omega(\theta,\phi))\right]_0^\pi\, d\theta\\
+\frac{\i}{k}\int_0^{2\pi} \int_0^\pi e^{\i k\{1-\cos(\phi)\}}
\p_\phi \widetilde{g}(\rho\omega(\theta,\phi)) \,d\phi d\theta,
\end{multline*}
Since we assumed~$\widetilde{g}$ is supported in half cone around
$u$, we have~$\widetilde{g}(\rho\omega(\theta,\pi))=0$. Hence we
obtain
\begin{multline*}
J_{v,\rho}(k)=\frac{\i}{k}\int_0^{2\pi}\int_0^\pi \p_\phi
\widetilde{g}(\rho\omega(\theta,\phi))\, d\theta\\
+\frac{\i}{k}\int_0^{2\pi} \int_0^\pi e^{\i k\{1-\cos(\phi)\}}
\p_\phi \widetilde{g}(\rho\omega(\theta,\phi)) \,d\phi d\theta,
\end{multline*}
and so
\begin{multline}
I_2(k,u)=\frac{\i}{|k|\left|u\right|} \int_{\R^+}\int_0^{2\pi}
\int_{-\pi}^\pi e^{\i k\{\sqrt{\rho^2+m^2}-\rho \left|u\right|\}}
\p_\phi \widetilde{g}(\rho\omega(\theta,\phi)) \,d\phi d\theta
\rho\, d\rho\\
+\frac{\i}{|k|\left|u\right|} \int_{\R^+}\int_0^{2\pi}
\int_{-\pi}^\pi e^{\i \{\sqrt{\rho^2+m^2}-\rho \left|u\right|
\cos(\phi)\}} \p_\phi \widetilde{g}(\rho\omega(\theta,\phi)) \,d\phi
d\theta \rho\, d\rho.
                                                             \label{Estimate:Decayintermediate}
\end{multline}
Let us now study the decay resulting from the dispersive behavior of
the radial part. To this end, we follow the proof of the well-known
Van Der Corput lemma. We study
\begin{eqnarray*}
L(k,u,\phi,\phi',\theta)&=&\int_{\R^+} e^{\i
|k|\{\sqrt{\rho^2+m^2}-\rho |u|\cos(\phi')\}} \p_\phi
\widetilde{g}(\rho\omega(\theta,\phi)) \rho\, d\rho.
\end{eqnarray*}
Notice that, in view of~\eqref{Estimate:Decayintermediate}, we are
only interested by~$L(k,u,\phi,\phi,\theta)$ and
$L(k,u,\phi,0,\theta)$. First, for any differentiable function on
$\R$ such that~$|f'|\geq 1$, we have for any~$\alpha\in\R^+$
\begin{equation}
                                                                \label{Estimate:VanDerCorput}
\lambda\left(\left\{t\in\R;\,|f(t)|\leq \alpha\right\}\right)\leq
\alpha,
\end{equation}
for $\lambda$ the Lebesgue measure. We introduce
\begin{equation*}
h(\rho)=\sqrt{\rho^2+m^2}-\rho |u|\cos(\phi'),
\end{equation*}
and we apply~\eqref{Estimate:VanDerCorput} to
\begin{equation*}
f(\rho)=\frac{\p_\rho h(\rho)}{\ds\inf_{x\in {\rm
supp}(g)}\left\{|\p_{|x|}^2 h(|x|)|\right\}}.
\end{equation*}
We notice that~$\p_\rho^2h(\rho)$ does not depend on~$u$ or~$\phi'$.
With help of a smooth cut-off function, we split the integral~$L$ in
two parts, one has support
\begin{equation*}
\left\{\rho\in\R^+;\,|f(\rho)|< \alpha\right\}
\end{equation*}
and the other is its complementary. In fact, we obtain exactly three
interval corresponding to
\begin{equation*}
\left\{\rho\in\R^+;\,f(\rho)< -\alpha\right\},\;
\left\{\rho\in\R^+;\,-\alpha\leq f(\rho)\leq
\alpha\right\},\;\left\{\rho\in\R^+;\,\alpha< f(\rho)\right\}.
\end{equation*}
In the first and third interval, we make an integration by parts and
in the second interval, we use
Estimate~\eqref{Estimate:VanDerCorput}. Hence we obtain the bound
\begin{multline*}
\left|L(k,u,\phi,\phi',\theta)\right|\leq\\
\max\left\{\alpha,\;
\frac{1}{\alpha |k| \ds\inf_{x\in {\rm
supp}(\widetilde{g})}\left\{|\p_{|x|}^2h(|x|)|\right\} }\right\}
\max_{\rho\in\R^+}\left\{\rho\left|\p_\phi
\widetilde{g}(\rho\omega(\theta,\phi))\right|\right\}\\
+\frac{1}{\alpha |k| \ds\inf_{x\in {\rm
supp}(\widetilde{g})}\left\{|\p_{|x|}^2h(|x|)|\right\}}\times\\
\times \max\left\{ \int_{\R^+}\left|\p_\phi \widetilde{g}
(\rho\omega(\theta,\phi))\right|\;d\rho,\;
\int_{\R^+}\rho\left|\p_\rho \p_\phi
\widetilde{g}(\rho\omega(\theta,\phi)) \right| \, d\rho\right\}.
\end{multline*}
We use
\begin{equation*}
\rho\left|\p_\phi \widetilde{g}(\rho\omega(\theta,\phi))\right|\leq
2\max\left\{ \int_{\R^+}\left|\p_\phi \widetilde{g}
(r\omega(\theta,\phi))\right|\;dr,\; \int_{\R^+}r\left|\p_\rho
\p_\phi \widetilde{g}(r\omega(\theta,\phi)) \right| \, dr\right\}
\end{equation*}
and then choose
\begin{equation*}
\alpha=\frac{1}{\sqrt{|k|\ds\inf_{x\in {\rm
supp}(\widetilde{g})}\left\{|\p_{|x|}^2h(|x|)|\right\}}},
\end{equation*}
and plugging the resulting estimates for~$\phi'=0$ and~$\phi'=\phi$
in~\eqref{Estimate:Decayintermediate}, we obtain estimate
\eqref{Estimate:Decay2}.
\end{proof}

\subsection{Dispersive estimates for the free case equation}
                                                                \label{Section:FreeDispersiveEstimates}

Thanks to the tools introduced in
Section~\ref{Section:OscilotaryIntegrals}, we are able to state the
\begin{theorem}[Dispersive estimates for free Dirac operator]
                                                                \label{Thm:FreeDispersiveEstimates}
For any~$p\in[1,2]$, for all~$\theta \in [0;1]$, for all~$s,s'\in
\R$, such that~$s-s'\geq (\frac{2}{p}-1)(2+\theta)$ and
$q\in[1,\infty]$, we have
\begin{equation*}
\|e^{-\i tD_m}\|_{B^{s}_{p,q},B^{s'}_{p',q}} \leq
\left(K(t)\right)^{\frac{2}{p}-1},
\end{equation*}
with
\begin{equation*}
K(t)= \left\{
\begin{array}{ll}
\ds\left|  t\right| ^{-1+\theta/2}
& \mbox{if } |t|\in [0,1]\vs,\\
\ds\left|  t\right| ^{-1-\theta/2} & \mbox{if } |t|\in [1,\infty),
\end{array}
\right.
\end{equation*}
and~$p'=\frac{p}{p-1}$.
\end{theorem}
\begin{proof}
The proof is a straightforward adaptation of the one of Brenner in
Appendix 2 of \cite{Brenner} for the Klein-Gordon equation. We give
a sketch of the proof for the reader's convenience. Note that the
proof of Theorem \ref{Thm:Dispersion}, for non zero potential, is
based on the same ideas.

We only need to prove the case~$p=1$, since the general case follows
by interpolation of the case~$p=1$ and the charge conservation which
corresponds to the case~$p=2$. Then using
$D_m=\sqrt{-\Delta+m^2}(\pi_+-\pi_-)$
with~$\pi_\pm=1\!\!1_{\R^\pm}(D_m)= \frac{1}{2}\left\{1\pm
|D_m|^{-1}D_m\right\}$, we obtain the estimates from those relative
to the relativistic Schr\"odinger operator~$\sqrt{-\Delta+m^2}$:
\begin{equation*}
\|e^{-\i t\sqrt{-\Delta +m^2}}\|_{B^{s}_{1,q},B^{s'}_{\infty,q}}
\leq K(t)
\end{equation*}
which in turn follow from
\begin{proposition}
                                                                \label{Prop:DispersionFiniteEnergy}
For any~$\chi \in {\mathcal D}(\R^3,\C^4)$, we define
$\chi_j(x)=\chi(2^{-j}|x|)$. Then for~$\theta'\in [0,1]$, we have:
\begin{enumerate}
\item if~$0\notin {\rm supp}(\chi)$,
\begin{equation}
                                                                \label{Estimate:PropagatorHighEnergy}
\|e^{-\i t\sqrt{-\Delta+m^2}}\chi_j
\left(\sqrt{-\Delta+m^2}\right)\|_{L^1,\,L^\infty} \leq
C2^{(2+\theta')j}|t|^{-(1+\theta'/2)}
\end{equation}
where~$C$ is independent of~$t$ and~$j$.
\item if~$0\in {\rm supp}(\chi)$,
\begin{equation}
                                                                \label{Estimate:PropagatorThresholdsEnergy}
\|e^{-\i t\sqrt{-\Delta+m^2}}
\chi(-\Delta+m^2)\|_{L^1,\,L^\infty}\leq C \langle t\rangle^{-3/2}
\end{equation}
where~$C$ is independent~$t$.
\end{enumerate}
\end{proposition}

We postpone the proof of Proposition
\ref{Prop:DispersionFiniteEnergy} until the end of the proof of
Theorem~\ref{Thm:FreeDispersiveEstimates}.

\medskip

We have
\begin{equation*}
\left\|e^{-\i t\sqrt{-\Delta +m^2}}\chi_j(\sqrt{-\Delta
+m^2})f\right\|_\infty\leq C2^{3j}\|f\|_1
\end{equation*}
interpolating with Estimate \eqref{Estimate:PropagatorHighEnergy} of
Proposition \ref{Prop:DispersionFiniteEnergy} for~$\theta'=0$
when~$t\leq 1$ and using
Estimate~\eqref{Estimate:PropagatorHighEnergy} with $\theta'=\theta$
for~$t\geq 1$, one obtains
\begin{equation*}
2^{js'}\left\|e^{-\i t\sqrt{-\Delta +m^2}}\chi_j(\sqrt{-\Delta
+m^2})f\right\|_\infty\leq C\kappa_j(t)2^{js}\|f\|_1
\end{equation*}
with
\begin{equation*}
\kappa_j(t)= 2^{j(2+\theta+s'-s)}\left\{
\begin{array}{ll}
|t| ^{-1+\theta/2}
& \mbox{if } |t|\leq 1,\\
|t| ^{-1-\theta/2} & \mbox{if } |t|\geq 1.
\end{array}
\right.
\end{equation*}
We use
\begin{equation*}
\ds\sup_{j\in \N}\kappa_j\leq CK(t)
\end{equation*}
if~$2+\theta\leq s-s'$ and
Estimate~\eqref{Estimate:PropagatorThresholdsEnergy} to prove
Theorem~\ref{Thm:FreeDispersiveEstimates}. Hence to conclude the
proof, we need to give the
\begin{proof}[Proof of Proposition~\ref{Prop:DispersionFiniteEnergy}]
Estimates of the same type, but for ${\mathcal
B}\left(L^p,L^{p'}\right)$ spaces with $p\in[4/3,2]$ can be found in
\cite{MarshallStraussWaingner,MarshallStraussWaingner2}. In the
present case $p=1$ a proof can be found in \cite{Brenner2}. This
proof, which covers a much more general situation, is quite
complicated. We propose here a simpler proof inspired by
\cite{CuccagnaSchirmer}. The kernel of~$e^{-\i t\sqrt{-\Delta+m^2}}
\chi_j(\sqrt{-\Delta+m^2})$ is given by~$(2\pi)^{-3/2}K_j(x-y)$
where
\begin{equation*}
K_j(x,t)=\int_{\R^3}e^{-\i t\sqrt{\xi^2+m^2}+x\cdot\xi}
\chi_j(\sqrt{\xi^2+m^2})\,d\xi
\end{equation*}
Hence, we estimate the~$L^\infty$ norm of~$K_j$.

If~$|x|/|t|\ll 2^{j-1}/\sqrt{2^{2j-2}+m^2}$ or~$|x|/|t|\gg 1$, we
use non stationary phase lemma in~$\R^3$ with help of the operator
$L=\frac{\frac{\xi}{\sqrt{\xi^2+m^2}}-x}{t\left|\frac{\xi}
{\sqrt{\xi^2+m^2}}-x\right|^2}\cdot\nabla$. Hence, in this case, we
obtain the estimate
\begin{equation*}
\left|\int_{\R^3}e^{-\i t\sqrt{\xi^2+m^2}+x\cdot\xi}
\chi_j(\sqrt{\xi^2+m^2})\,d\xi \right|\leq C_n2^{-(n-3)j}|t|^{-n},
\end{equation*}
for any~$n\in\N$. Otherwise, we apply
Proposition~\ref{proposition:Decay1} with~$h(r)=\sqrt{r^2+m^2}$,
$k=t$,~$u=x/t$ and~$g(x)= \chi_j(|x|)$. So if~$0\notin {\rm
supp}(\chi)$, to obtain
\begin{eqnarray*}
\lefteqn{\left|\int_{\R^3}e^{-\i t\sqrt{\xi^2+m^2}+x\cdot\xi}
\chi_j(\sqrt{\xi^2+m^2})\,d\xi \right|}\\
&&\leq \frac{C}{|t|}\max_{|\beta|\leq
1}\int_{\R^3}\left|\xi\right|^{|\beta|-1} \frac{\left|\nabla^\beta
\chi_j(\sqrt{\xi^2+m^2})\right|}{\left|\frac{x}{|x|}
\pm\frac{\xi}{|\xi|}\right|}d\xi\\
&&\leq \frac{C2^{2j}}{|t|}.
\end{eqnarray*}
Notice that in this case,~$|x|/|t|\geq c'>0$. If instead of
Proposition~\ref{proposition:Decay1}, we use
Proposition~\ref{Prop:Decay2} with~$g=\chi_j$,~$k=t$ and~$u=x/t$, we
prove the estimate
\begin{equation*}
\left|K_j(x,t)\right|\leq \frac{C2^{3j}}{|t|^{3/2}}.
\end{equation*}
The estimate~\eqref{Estimate:PropagatorHighEnergy} is then obtained
by interpolation. For~\eqref{Estimate:PropagatorThresholdsEnergy},
we use the classical stationary (Morse lemma) and non-stationary
phase methods (integration by parts) in~$\R^3$. For more details
about the method one can look at the end of the proof of Proposition
\ref{Prop:EssentialBound}. This ends the proof
of~\ref{Prop:DispersionFiniteEnergy}.
\end{proof}
This ends the proof of Theorem~\ref{Thm:FreeDispersiveEstimates}.
\end{proof}

\subsection{Distorted Plane Waves}
                                                                \label{Section:DistortedPlaneWaves}
Our aim is now to generalize the previous method to the perturbed
case. Let us introduce the wave operators
\begin{equation}
                                                                \label{Def:WaveOperators}
\ds W^\pm=\lim_{t\to\pm\infty} e^{\i t(D_m+V)}e^{-\i tD_m}
\end{equation}
(for the existence and the completeness :~${\rm Ran}(W^\pm)={\rm
Ran}({\mathbf P}_c(H))$ of these operator see \cite[Theorem
1.5]{GeorgescuMantoiu}). With the intertwining property
\begin{equation}                                                \label{Identity:Intertwining3}
f(H)W^\pm=W^\pm f(D_m),
\end{equation}
for any bounded borelian function $f$, and Fourier transform
${\mathcal F},$ we shall obtain for~$h(\xi)=\alpha\cdot \xi+m\beta$
\begin{equation*}
e^{-\i t H}{\mathbf P}_c(H)={W^\pm}e^{-\i t
D_m}{W^\pm}^*={W^\pm}{\mathcal F}e^{-\i t h(Q)}\left(W^\pm{\mathcal
F}\right)^*.
\end{equation*}
So we can adapt the previous method if we are able to prove some
estimates about the kernel~$\psi_V$ of~$W^\pm{\mathcal F}$. The
kernel~$\psi_V$ is called \emph{distorted plane wave}. We notice
that~$\psi_V$ is a~$4\times 4$ matrix valued function.

We will show that the previous method works with
$\psi_V\psi_V^{*}\chi_j$ in place of~$\chi_j$ with small
modifications. So we need estimates on~$\psi_V$. Generally,
distorted plane waves are studied like perturbations of free plane
waves. So we will prove estimates on the perturbative part, written
$w$ in the sequel.

\subsubsection{Definition and properties}
We need to introduce the free plane wave. Let
\begin{equation*}
h(k)=\alpha\cdot k+m\beta
\end{equation*}
for any~$k\in \R^3$,  notice that~$D_m=h(P)$. This hermitian matrix
has for eigenvectors the
\begin{equation*}
\psi_0^j(k,x)=e^{\i k\cdot x}u(k)e_j
\end{equation*}
where
\begin{equation}
                                                                \label{Def:The-u-Matrix}
u(k)=\frac{(m+\lambda(k))Id-\beta\alpha\cdot
k}{\sqrt{2\lambda(k)(m+\lambda(k))}}
\end{equation}
with~$\lambda(k)=\sqrt{k^2+m^2}$ and~$\left(e_j\right)_j$ are
vectors of the canonical basis of~$\C^4$. For more details see
\cite[Section 1.4, Section 1.F]{Thaller}.

By definition, a distorted plane wave is a solution of the PDO
equation
\begin{equation}
                                                                 \label{Eq:G-eigenfct}
\left(D_m+V\right)\psi=\pm\sqrt{k^2+m^2} \psi
\end{equation}
with for some~$j$ and any~$k\in\R^3$,~$\psi(k,x)-\psi_0^j(k,x)$
tending to zero as~$x$ goes to infinity (in some sense), see
\cite[section 5]{Agmon}.

A solution of~\eqref{Eq:G-eigenfct} is a function~$\psi(k,x)$ of two
variables here~$k$ is a 3-dimensional vector which is called the
wave vector. A free plane wave~$\psi_0^j$ satisfies the PDO equation
\eqref{Eq:G-eigenfct} in the case~$V=0$. Following~\cite{Agmon}, we
introduce two families of function
\begin{equation*}
\psi_V^j(k,x)=\psi_{0}^j(k,x)
-\left\{R_V^+(\lambda(k))V(\cdot)\psi_{0}^j(k,\cdot)\right\}(x)
\end{equation*}
for~$j\in\{1,2\}$ and
\begin{equation*}
\psi_V^j(k,x)=\psi_{0}^j(k,x)
-\left\{R_V^+(-\lambda(k))V(\cdot)\psi_{0}^j(k,\cdot)\right\}(x)
\end{equation*}
for~$j\in\{3,4\}$. The rest of the proof works also for~$R_V^-$
instead of~$R_V^+$ (the trace of the resolvent~$R^\pm_V$ was
introduced in~\eqref{Def:LimitDiracResolvent}).

In case there is no resonance at thresholds and no eigenvalue at
thresholds, Theorem~\ref{Thm:Propagation} gives us that
$R_V^+(\lambda(p))$ is in~$B(L^2_\sigma,L^2_{-\sigma})$ for any
$\sigma>5/2$, this also work if~$\sigma\geq 1$ see Proposition
\ref{proposition:LAP} below. So the previous definition make sense
if Assumption~\ref{assumption:1} holds and we have the
\begin{proposition}
Suppose that Assumptions~\ref{assumption:1} and~\ref{assumption:2}
hold. Then for any~$k\in\R^3\setminus \{0\}$,~$\psi_V^j(k,x)$
satisfies equation~\eqref{Eq:G-eigenfct}.
\end{proposition}

Distorted plane waves define a generalized Fourier transform. We
introduce $\psi_V(k,x)\in{\mathcal M}_4(\C)$ the matrix with vector
column~$\psi_V^j(k,x)$ and we define
\begin{equation*}
\ds({\mathcal F}_V f)(k)=\int_{\R^3}\psi_V(k,x)f(x)\,dx,
\end{equation*}
which is {\it a priori} defined on the Schwartz space~${\mathcal
S}(\R^3,\C^4)$ but will be extended to~$L^2$.

Distorted plane waves are also called generalized eigenfunctions,
since they correspond to ``eigenvalues'' associated with the
continuous spectrum. Indeed, we can prove the
\begin{theorem}[Eigenfunction Expansion]
The operator~${\mathcal F}_V$ defines a bounded linear map from
$L^2$ into itself. Its kernel is given by the the sum of the
eigenspaces of~$H$. Moreover it is a unitary map from~${\mathbf
P}_c(H)L^2$ onto~$L^2(\R^3)$ with
\begin{equation*}
\ds({\mathcal F}_V^*
f)(x)=\lim_{n\to\infty}\int_{K_n}\psi_V(k,x)^*f(k)\,dk,
\end{equation*}
for any~$(K_n)_{n\in\N}$ a family of compact sets with~$K_n\subset
K_{n+1}$ and~$\cup_{n\in\N} K_n=\R^3$. Finally, for any interval
$I\subset\R$, one has
\begin{equation}
                                                                \label{Identity:FunctCalc}
\ds\|\one_{I}(H)f\|^2=\int_{ \sigma(h(k))\cap I\neq \emptyset
}|{\mathcal F}_V f|^2\,dk
\end{equation}
where~$\sigma(h(k))$ is the spectrum of~$h(k)$.
\end{theorem}
\begin{proof}
The proof is an easy adaptation of the proof of~\cite[Theorem
6.2]{Agmon} (see also~\cite[Theorem XI.41]{ReedSimon3}), the main
difference is that here we insert the unitary matrix~$u$ defined in
\eqref{Def:The-u-Matrix}. Formula~\eqref{Identity:FunctCalc} is
nothing more than an adaptation of~\cite[ Formula (6.6)]{Agmon} or
\cite[ Formula 82e']{ReedSimon3}.
\end{proof}
We also have the
\begin{lemma}[Intertwining Property]
Let~$g$ be a bounded borelian function with support in
$\R\setminus(-m,\,m)$ , we have
\begin{equation}
                                                                \label{Identity:Intertwining0}
{\mathcal F}_V g(H)= (g\circ h) {\mathcal F}_V.
\end{equation}
\end{lemma}
\begin{proof}
Using~\eqref{Identity:FunctCalc}, we obtain that
\eqref{Identity:Intertwining0} is true for~$g=\one_I$ with~$I$ an
interval  of~$\R\setminus(-m,\,m)$. We then obtain it for bounded
borelian function with support in~$\R\setminus (-m,\,m)$, usual
density arguments and properties of functional calculus. More
precisely, we use the fact that a bounded sequence of borelian
functions which converges everywhere gives a sequence of bounded
operators which converge strongly.
\end{proof}

Hence we deduce that, for any~$\chi\in {\mathcal C}_0^\infty(\R)$,
the kernel of~$e^{-\i tH}\chi(H)$ is given by
\begin{equation*}
\left[e^{-\i tH}\chi(H){\mathbf P_c}(H)\right](x,y)= \int_{\R^3}
\psi_V(k,x)^*e^{-\i th(k)}\chi(h(k))\psi_V(k,y)\,dk.
\end{equation*}
which exactly means
\begin{equation}
                                                                \label{Identity:Intertwining1}
e^{-\i tH}\chi(H){\mathbf P_c}(H)=({\mathcal F}_V)^{*}e^{-\i t
h}\chi(h){\mathcal F}_V
\end{equation}

We recall that we want to prove the decay of~$e^{-\i tH}\chi(H)$ as
$t\to+\infty$ in some Besov spaces. We observe that
\begin{equation*}
e^{-\i t h(k)}\chi(h(k))=e^{-\i t\lambda(k)}\chi(\lambda(k))P_+(k)
+e^{\i t\lambda(k)}\chi(-\lambda(k))P_-(k),
\end{equation*}
where~$P_+(k)$ (resp.~$P_-(k)$) is the projector associated with the
positive (resp. negative) part of the spectrum of~$h(k)$, {\it i.e.}
\begin{equation*}
P_\pm(k)=\frac{1}{2}\left(1\pm\frac{h(k)}{\lambda(k)}\right).
\end{equation*}
Hence, in the following we study the functions
\begin{equation*}
\left(x,y\right)\in \R^3\times \R^3\mapsto\int_{\R^3} e^{\mp\i t
\lambda(k)}\left(P_\pm\psi_V(k,x)\right)^*\left(P_\pm\psi_V(k,y)
\right)\chi(h(k))\,dk.
\end{equation*}
\subsubsection{End of the proof of Theorem~\ref{Thm:Dispersion}}%
                                                                 \label{proof_theorem_dispersion_subsec}

We now prove Theorem~\ref{Thm:Dispersion} with help of three
propositions which will be proven in
Section~\ref{Section:Estimates}. These propositions give some
estimates on the perturbed part of the distorted plane wave.
Following Cuccagna and Schirmer in~\cite{CuccagnaSchirmer}, we write
$\psi_V(k,x)=e^{\i k\cdot x} (u(k)+w(k,x))$ where~$w$ is the
perturbation part which satisfies
\begin{equation}
                                                                 \label{Def:W}
w(k,x)_j=
\begin{cases}
e^{-\i k\cdot x }\left\{R^+_V(+\lambda(k))\left\{Ve^{\i k\cdot Q
}u(k)_j\right\}\right\}(x),
\mbox{ if }&j\in\{1,\,2\},\\
e^{-\i k\cdot x }\left\{R^+_V(-\lambda(k))\left\{V e^{\i k\cdot Q
}u(k)_j\right\}\right\}(x), \mbox{ if }&j\in\{3,\,4\},
\end{cases}
\end{equation}
and we now state our propositions.
\begin{proposition}
                                                                \label{proposition:EstimateOnW}
Suppose that Assumptions~\ref{assumption:1} and~\ref{assumption:2}
hold. Then there exists~$C>0$ such that for any~$k,\,x\in
\R^3\setminus\{0\}$, and any~$\beta\in \N^3$ with~$|\beta|\leq 1$,
one has
\begin{equation}
                                                                \label{Estimate:OnW}
\left|\nabla_k^\beta w(k,x)\right|\leq \frac{C}{\langle
k\rangle^{|\beta|}}\frac{\langle
x\rangle^{|\beta|}}{\left\langle|x|\left|\frac{x}{|x|}-\frac{k}{|k|}\right|
\right\rangle}.
\end{equation}
Moreover one has
\begin{equation}
                                                                \label{Estimate:OnW2}
\left|\nabla_k w(k,x)\right|\leq C\frac{\langle\min\{|x|,\;|k|\}
\rangle}{\langle
k\rangle\left\langle\min\{|x|,\;|k|\}\left|\frac{x}{|x|}-\frac{k}{|k|}\right|
\right\rangle^{2}}.
\end{equation}
\end{proposition}

We use this to prove the time decay in~$|t|^{-1}$. Unfortunately
this doesn't work for the~$|t|^{-3/2}$ decay, hence we then study
\begin{equation}
                                                                \label{Def:V}
v(k,x)_j=
\begin{cases}
e^{\i|k||x|}\left\{R^+_V(+\lambda(k))\left\{Ve^{\i k\cdot Q
}u(k)_j\right\}\right\}(x),
&\mbox{ if }j\in\{1,\,2\},\\
e^{-\i|k||x|}\left\{R^+_V(-\lambda(k))\left\{Ve^{\i k\cdot Q
}u(k)_j\right\}\right\}(x), &\mbox{ if }j\in\{3,\,4\},
\end{cases}
\end{equation}
and
\begin{equation}
                                                                \label{Def:VTilde}
\widetilde{v}(k,x)_j=
\begin{cases}
e^{\i|k||x|}\left\{\nabla_k R^+_V(+\lambda(k))\left\{Ve^{\i k\cdot Q
}u(k)_j\right\}\right\}(x),
&\mbox{ if }j\in\{1,\,2\},\\
e^{-\i|k||x|}\left\{\nabla_k  R^+_V(-\lambda(k))\left\{Ve^{\i k\cdot
Q }u(k)_j\right\}\right\}(x), &\mbox{ if }j\in\{3,\,4\}.
\end{cases}
\end{equation}
One has the
\begin{proposition}
                                                                \label{proposition:EstimateOnV}
Suppose that Assumptions~\ref{assumption:1} and~\ref{assumption:2}
hold. Then if~$\rho>3+|\beta|$ for some~$\beta\in \N^3$, there
exists~$C>0$ such that for any~$k,\,x\in \R^3\setminus\{0\}$, one
has
\begin{equation*}
|\nabla_k^\beta v(k,x)|\leq \frac{C}{\langle
|x|\left|\frac{x}{|x|}-\frac{k}{|k|}\right|\rangle}.
\end{equation*}
\end{proposition}

\begin{proposition}
                                                                \label{proposition:EstimateOnVtilde}
Suppose that Assumptions~\ref{assumption:1} and~\ref{assumption:2}
hold. Then if~$\rho>3+|\beta|$ for some~$\beta\in \N^3$, there
exists~$C>0$ such that for any~$k,\,x\in \R^3\setminus\{0\}$, one
has
\begin{equation*}
|\nabla_k^\beta \widetilde{v}(k,x)|\leq
\frac{\langle\min\{|x|,\;|k|\} \rangle}{\langle
k\rangle\left\langle\min\{|x|,\;|k|\}\left|\frac{x}{|x|}-\frac{k}{|k|}\right|
\right\rangle^{2}}.
\end{equation*}
\end{proposition}

Using
Proposition~\ref{proposition:EstimateOnW},~\ref{proposition:EstimateOnV}
and \ref{proposition:EstimateOnVtilde} (which are proved in Section
\ref{Section:Estimates} below), let us prove the following
\begin{proposition}
                                                                \label{Prop:EssentialBound}
Suppose that Assumptions~\ref{assumption:1} and~\ref{assumption:2}
hold. Then we have for~$\chi\in{\mathcal C}_0^\infty(\R)$ with
support in~$\R\setminus[-m;m]$ for any~$\theta\in[0,1]$
and~$j\in \N$,
\begin{equation}
                                                                \label{Estimate:EssentialBound}
\left\|e^{-\i t H}\chi(2^{-j}H)\right\|_{L^1\to L^\infty}\leq
\frac{C2^{(2+\theta)j}}{|t|^{1+\theta/2}},
\end{equation}
with~$C$ independent of~$t$ and~$j$.

We also have for~$\chi\in{\mathcal C}_0^\infty(\R)$, for
any~$\theta\in[0,1]$
\begin{equation}
                                                                \label{Estimate:EssentialBound2}
\left\|e^{-\i t H}\chi(H){\mathbf P}_c(H)\right\|_{L^1\to
L^\infty}\leq \frac{C}{|t|^{1+\theta/2}},
\end{equation}
with~$C$ independent of~$t$.
\end{proposition}
\begin{proof}
The proof works like the one of Proposition
\ref{Prop:DispersionFiniteEnergy} with some modifications due to the
fact that high derivatives in~$k$ of~$w(k,x)$ grow with respect to
$x$.

We need the~$L^\infty$ norm of the kernel of~$e^{-\i
tH}\chi(2^{-j}H)$. This kernel thanks
to~\eqref{Identity:Intertwining1} is given by
\begin{multline*}
I_j(t,x,y)=
\int_{\R^3} e^{-\i t\sqrt{\xi^2+m^2}}e^{-i\xi\cdot(x-y)}
\left\{P_+(\xi)(u^*(\xi)+w^*(x,\xi))\right\}\times\\
\times
\left\{P_+(\xi)(u(\xi)+w(y,\xi)\right\}\chi(2^{-j}\lambda(\xi))\,d\xi\\
+\int_{\R^3} e^{+ \i t\sqrt{\xi^2+m^2}}e^{-i\xi\cdot(x-y)}
\left\{P_-(\xi)(u^*(\xi)+w^*(x,\xi))\right\}\times\\
\times
\left\{P_-(\xi)(u(\xi)+w(y,\xi)\right\}\chi(-2^{-j}\lambda(\xi))\,d\xi.
\end{multline*}
We notice that if we expand each integrand in terms of~$u$ and~$w$,
we obtain the sum of the integrals
\begin{multline*}
I^\pm_j[z,z'](t,x,y)\\
= \int_{\R^3} e^{\mp \i
t\sqrt{\xi^2+m^2}}e^{-\i \xi\cdot(x-y)}
\left\{P_\pm(\xi)z^*(x,\xi)P_\pm(\xi)z'(y,\xi)\right\}
\chi(\pm2^{-j}\lambda(\xi))\,d\xi.
\end{multline*}
with~$z,z'\in\left\{u,w\right\}$. We notice that
$I^+_j[u,u](t,x,y)+I^-_j[u,u](t,x,y)$ is
 the kernel of~$e^{-itD_m}\chi_j(D_m)$, hence we only treat
the other integrals.

\medskip

For the~$|t|^{-1}$ decay, if~$|x-y|/|t|\ll
2^{j-1}/\sqrt{2^{2j-2}+m^2}$ or~$|x-y|/|t|\gg 1$, the phase has no
critical point. We use an integration by parts in~$\R^3$ with help
of the operator
\begin{equation*}
L=\frac{\left(\frac{\xi}{\sqrt{\xi^2+m^2}}-x\right)}{t\left|\frac{\xi}
{\sqrt{\xi^2+m^2}}-x\right|^2}\cdot\nabla_\xi.
\end{equation*}
So with the estimate \eqref{Estimate:OnW} of
Proposition~\ref{proposition:EstimateOnW} and with
\begin{equation*}
\left|\p_i\frac{\xi}{\sqrt{\xi^2+m^2}}\right|\leq \frac{C}{|\xi|},
\end{equation*}
we obtain the estimate
\begin{equation*}
\left|I^\pm_j[z,z'](t,x,y) \right|\leq C2^{2j} |t|^{-1},
\end{equation*}
with~$C$ independent of~$j$ and~$t$.

Otherwise if~$|x-y|/|t|\geq c>0$, using
first~\eqref{Estimate:Decay1-1} of
Proposition~\ref{proposition:Decay1} and then~\eqref{Estimate:OnW2}
of Proposition~\ref{proposition:EstimateOnW}, we infer
\begin{equation*}
\left|I^\pm_j[z,z'](t,x,y) \right|\leq C2^{2j} |t|^{-1},
\end{equation*}
with~$C$ independent of~$j$ and~$t$.

\medskip

For the~$|t|^{-3/2}$ decay, first if~$|x-y|/|t|\geq c>0$, we write
\begin{eqnarray*}
I^\pm_j[z,z'](t,x,y)
=\int_{\R^+}e^{\mp \i t\sqrt{\rho^2+m^2}-\i
\rho|x-y|}J_{\frac{x-y}{|x-y|}}(\rho|x-y|)\,\rho^2d\rho.
\end{eqnarray*}
where
\begin{equation*}
J_v(k)=\int_{S^2}e^{\i k(1-\omega\cdot v)}
\left\{P_\pm(\rho\omega)z^*(x,\rho\omega)
P_\pm(\rho\omega)z'(y,\rho\omega)\right\}
\chi(\pm2^{-j}\lambda(\rho\omega))\,d\omega.
\end{equation*}
We can suppose~$v=\left(0;0;1\right)$ and so
\begin{multline*}
J_v(k)=\int_0^{2\pi}\int_0^\pi e^{\i k(1-\cos(\phi))}
\Big\{P_\pm(\rho\omega(\theta,\phi))z^*(x,\rho\omega(\theta,\phi))\times\\
\times
P_\pm(\rho\omega(\theta,\phi))z'(y,\rho\omega(\theta,\phi))\Big\}
\chi(\pm2^{-j}\lambda(\rho\omega(\theta,\phi)))\,\sin(\phi)d\phi
d\theta.
\end{multline*}
An integration by parts in~$\phi$ gives
\begin{multline*}
J_v(k)=\frac{1}{\i k}\int_0^{2\pi}\big[ e^{\i k(1-\cos(\phi))}
\Big\{P_\pm(\rho\omega(\theta,\phi))z^*(x,\rho\omega(\theta,\phi))\times\\
\times
P_\pm(\rho\omega(\theta,\phi))z'(y,\rho\omega(\theta,\phi))\Big\}
\chi(\pm2^{-j}\lambda(\rho\omega(\theta,\phi)))\big]_0^\pi\,
d\theta\\
-\frac{1}{\i k} \int_0^{2\pi}\int_0^\pi e^{\i
k(1-\cos(\phi))}\p_\phi\Big\{P_\pm(\rho\omega(\theta,\cdot))z^*(x,\rho\omega
(\theta,\cdot))\times\\
\times P_\pm(\rho\omega(\theta,\cdot))z'(y,\rho\omega(\theta,\cdot))
\chi(\pm2^{-j}\lambda(\rho\omega(\theta,\cdot)))\Big\}(\phi)\,d\phi
d\theta.
\end{multline*}
The integrand of the first term can be rewritten in order to obtain
a sum of two integral in~$\phi$ over the interval~$[0,\pi]$. To this
end, we introduce a smooth cut-off function which splits~$[0,\pi]$
in two parts one is a neighborhood of~$0$ and the other a
neighborhood of~$\pi$. Then most of the terms obtained after
derivation can be treated by the method used for the~$|t|^{-1}$
decay. Only the two terms where derivatives of~$z$ and $z'$ appear
need a particular treatment. Now we have to distinguish the
case~$z=z'=w$ from the two others where~$z=u$ or~$z'=u$.
If~$z=z'=w$, the terms which need a particular treatment are bounded
by~$C|t|^{-1}$ times the supremum in~$\phi'$ of the
$L^1_{\phi,\;\theta}([0,\pi]\times[0,2\pi])$ of
\begin{multline*}
L^\pm_{j,n,m}(t,x,y,\phi,\phi')= \int_{\R^+}e^{\mp \i
t\sqrt{\rho^2+m^2}-\i \rho|x-y|\cos(\phi')}
\left\{P_\pm(\rho\omega)\p_\phi^nz^*(x,\rho\omega)\right\}
\times\\
\times\left\{P_\pm(\rho\omega(\theta,\phi))\p_\phi^m
z'(y,\rho\omega(\theta,\phi))\right\}
\chi(\pm2^{-j}\lambda(\rho\omega(\theta,\phi)))\,\rho d\rho,
\end{multline*}
with~$n,m\in \N$ such that~$n+m=1$. It is a sum of terms of the form
\begin{multline*}
\int_{\R^+}e^{ \i t\left\{\mp \sqrt{\rho^2+m^2}-
\rho\frac{|x-y|}{t}\left(\cos(\phi')-\cos(\phi)\right)
-\e_{i}\rho\frac{|x|}{t}+\e_{i'}\rho\frac{|y|}{t}\right\}}\times\\
\times \left\{P_\pm(\rho\omega)_{k,i}\left(e^{-\i\psi(k,x)}
\p_\phi^nz^*(x,\rho\omega)\right)_{i,l}\right\}\times\\
\times\left\{P_\pm(\rho\omega(\theta,\phi))_{l,k'} \left(\p_\phi^m
z'(y,\rho\omega(\theta,\phi))e^{\i\psi(k,y)} \right)_{k',i'}\right\}
\chi(\pm2^{-j}\lambda(\rho\omega(\theta,\phi)))\,\rho d\rho.
\end{multline*}
where $\phi$ is the angle between $\frac{x-y}{|x-y|}$ and the
$z-$axis and~$\psi(x,k)\in {\mathcal M}_4(\C)$ is given by
\begin{equation*}
\begin{pmatrix}
  \ds\left(|x||k|+x\cdot k\right)I_{\C^2} & \ds0_{\C^2} \\
  \ds0_{\C^2} & \ds\left(-|x||k|+x\cdot k\right) I_{\C^2}
\end{pmatrix},
\end{equation*}
and $\e_i,{\e_i}'\in\left\{\pm 1\right\}$. We introduce
\begin{equation*}
K(\rho)=\left\{P_\pm(\rho\omega)_{k,i}\left(e^{-\i\psi(k,x)}
\p_\phi^nz^*(x,\rho\omega)\right)_{i,l}\right\}
\end{equation*}
and
\begin{equation*}
\left\{P_\pm(\rho\omega(\theta,\phi))_{l,k'} \left(\p_\phi^m
z'(y,\rho\omega(\theta,\phi))e^{\i\psi(k,y)}
\right)_{k',i'}\right\},
\end{equation*}
\begin{equation*}
\phi(\rho)={\mp  \sqrt{\rho^2+m^2}-
\rho\frac{|x-y|}{t}\left(\cos(\phi')-\cos(\phi)\right)
-\e_{i}\rho\frac{|x|}{t}+\e_{i'}\rho\frac{|y|}{t}}
\end{equation*}
and
\begin{equation*}
f(\rho)=\frac{\p_\rho \phi(\rho)}{\ds\inf_{x\in {\rm
supp}(\chi_j)}\left\{|\p_\rho^2\phi(|x|,\lambda)|\right\}}.
\end{equation*}
With help of a smooth cut-off function, we split the integral in two
parts. One has support~$\left\{t\in\R;\,|f(t)|\leq \alpha\right\}$
on which we use the estimate
\begin{equation*}
\lambda\left(\left\{t\in\R;\,|f(t)|\leq \alpha\right\}\right)\leq
\alpha,
\end{equation*}
for $\lambda$ the Lebesgue measure, since $|f'|>1$. The other is its
complementary, in which we make an
 integration by parts. We obtain the estimate
\begin{multline*}
\left|J^+_j[r,r'](t,x,y)\right|\leq C\alpha\max_{\rho \in A_j}
\left\{\rho \left|K(\rho)\right|\right\} +\frac{1}{\alpha t
\ds\inf_{\rho \in A_j} \left\{|\p_\rho^2
\phi(\rho)|\right\} }\times\\
\times\max\left\{ \int_{A_j}\left\{\rho\left|(\p_\rho
K)(\rho)\right|\right\};\; \int_{A_j}\left|K(\rho)\right|\,
d\rho;\;2^{-j}\int_{A_j}\left\{\rho\left|K(\rho)\right|\right\}\right\},
\end{multline*}
where~$A_j=g^{-1}\left\{{\rm supp}(\chi_j)\right\}$ with~$g(\rho)=
\sqrt{\rho^2+m^2}$. Hence with~\eqref{Estimate:OnW2} of Proposition
\ref{proposition:EstimateOnW},
Proposition~\ref{proposition:EstimateOnVtilde} and decay of
derivatives of~$P_\pm$, we can choose~$\alpha=2^{2j}\sqrt{t}^{-1}$
and we obtain the bound of \eqref{Estimate:EssentialBound} in this
case.

For the case~$(z,z')=(u,w)$ (the case~$(z,z')=(w,u)$ is similar), we
study by the same way the integral
\begin{multline*}
\int_{\R^+}e^{ \i t\left\{\mp \sqrt{\rho^2+m^2}-
\rho\frac{|x-y|}{t}\left(\cos(\phi')-\cos(\widetilde{\phi})\right)
-\e_{i'}\rho\frac{|y|}{t}\right\}}\times\\
\times\Bigg\{\left\{P_\pm(\rho\omega)_{k,i}\left(\p_\phi^nz^*(x,\rho\omega)
\right)_{i,l}\right\}\times\\
\times\left\{P_\pm(\rho\omega(\theta,\phi))_{l,k'} \left(\p_\phi^m
z'(y,\rho\omega(\theta,\phi))e^{\i\psi(k,y)}
\right)_{k',i'}\right\}\times\\
\times \chi(\pm2^{-j}\lambda(\rho\omega(\theta,\phi)))\Bigg\}\,\rho
d\rho,
\end{multline*}
where $\widetilde{\phi}$ is the angle between $\frac{y}{|y|}$ and
the $z-$axis.

\medskip

If~$|x-y|/|t|\ll 1$, we can suppose
$|x-y|/|t|<|\xi|/(2\sqrt{\xi^2+m^2})$ for any~$\xi \in {\rm
supp}(\chi_j)$ and instead of applying the trick of the proof of
Lemma~\ref{lemma:Key} (integration by parts with respect to an
angular variables) to the integral~$I^\pm_j[z,z'](t,x,y)$, we make
an integration by parts with help of
\begin{equation*}
\frac{\p_{|\xi|}}{\frac{|\xi|}{\sqrt{\xi^2+m^2}}\pm\frac{\xi}{|\xi|}
\cdot\frac{x-y}{t}}.
\end{equation*}
The rest of the proof is the same.

\bigskip

We now turn to the proof of estimate
\eqref{Estimate:EssentialBound2}, the kernel of the operator is
given by a sum of terms of the form
\begin{multline*}
I^\pm_j[z,z'](t,x,y)= \int_{\R^3} e^{\mp \i t\sqrt{\xi^2+m^2}}e^{-\i
\xi\cdot(x-y)}\times\\
\times \left\{P_\pm(\xi)z^*(x,\xi)P_\pm(\xi)z'(y,\xi)\right\}
\chi(\pm\lambda(\xi))\,d\xi.
\end{multline*}

We first notice that Proposition~\ref{proposition:EstimateOnW}
implies that this integral is bounded. Then we split the integral in
two parts. One is supported in a small neighborhood of the critical
point of the phase, the other is its complementary. To treat this
last integral we work exactly like the case~``$\frac{|x-y|}{t}\ll
1$", just mentioned above. For the other one, we apply the Morse
lemma to reduced the study to
\begin{multline*}
\int_{\R^3} e^{\mp \i t \xi^2}
\left\{P_\pm(f(\xi))z^*(x,f(\xi))P_\pm(f(\xi))z'(y,f(\xi))\right\}
\widetilde{\chi}(f(\xi))\,d\xi=\\
\int_{S^2}\int_{\R^+} \rho e^{\mp \i t \rho^2}
\bigg\{P_\pm(f(\rho\omega))z^*(x,f(\rho\omega))\times\\
\times P_\pm(f(\rho\omega)) z'(y,f(\rho\omega))\bigg\}
\widetilde{\chi}(f(\rho\omega))\, d\rho d\omega,
\end{multline*}
where~$\widetilde{\chi}$ is the product of an indicator of a small
neighborhood of the critical point with~$\chi(\pm\lambda(\cdot))$.
Then an integration by parts in~$\rho$ and the Van Der Corput lemma
give~\eqref{Estimate:EssentialBound2} when~$\theta=1$. Since we have
that the integral is bounded the general case easily follows.
\end{proof}

We are now able to write the proof of Theorem~\ref{Thm:Dispersion},
using Proposition~\ref{Prop:EssentialBound}.
\begin{proof}[Proof of Theorem~\ref{Thm:Dispersion}] We notice that
\begin{eqnarray*}
\phi_k(D_m)\phi_j(H)
&=&D_m^{-1}\phi_k(D_m)H\phi_j(H) -D_m^{-1}\phi_k(D_m) V\phi_j(H)\\
&=&2^{-k}\widetilde{\phi}_k(D_m)
2^{j}\widetilde{\phi}_j(H)-2^{-k}\widetilde{\phi}_k(D_m)V\phi_j(H)
\end{eqnarray*}
We can also use~$H^{-1}$ since the support of~$\phi_j$ is far from
$0$
\begin{eqnarray*}
\phi_k(D_m)\phi_j(H) &=&D_m\phi_k(D_m)H^{-1}\phi_j(H)-\phi_k(H)
VH^{-1} \phi_j(H)
\end{eqnarray*}
or higher power in~$D_m^{-1}$ or~$H^{-1}$ to obtain with
\eqref{Estimate:EssentialBound2}
\begin{equation*}
\|\phi_i(D_m)e^{-\i t(H)} \phi_j(H)\phi_k(D_m)\|_{L^1,\;L^\infty}
\leq C 2^{-r'|j-i|}\frac{C2^{(2+\theta)j}}{t^{1+\theta}}2^{-r|j-k|}
\end{equation*}
for any reals~$r,\,r'$ with~$C$ independent of~$i,\,j$. Hence if
$r,\,r'>0$, we work like in the proof of
Theorem~\ref{Thm:FreeDispersiveEstimates} ({\it i.e.} like in
Appendix 2 of \cite{Brenner2}) to conclude the proof.
\end{proof}

It now remains to prove Proposition~\ref{proposition:EstimateOnW},
\ref{proposition:EstimateOnV}
and~\ref{proposition:EstimateOnVtilde}.

\subsubsection{Some estimates}
                                                                \label{Section:Estimates}
\paragraph{Estimates for~$w$}
We remind us of the definition of~$w$ in~\eqref{Def:W} and we
introduce
\begin{equation}
                                                                \label{Def:RTilde}
\widetilde{R}_V^\pm(k)=e^{-i k\cdot Q }R_V^+(\pm\lambda(k)) e^{i
k\cdot Q }.
\end{equation}
We have
\begin{lemma}
For any~$\alpha\in \N^3$, let be~$\sigma>4+|\alpha|$. Then there
exists~$C>0$ such that for any~$k,\,x\in\R^3\setminus\{\,0\}$
\begin{equation}
                                                                \label{Estimate:LemOnW}
\left|\left\{\nabla_k^\alpha\widetilde{R}_0^\pm(k)\langle
Q\rangle^{-\sigma} q\right\}(x)\right|\leq \frac{C}{\langle
k\rangle^{|\alpha|}}\frac{\langle
x\rangle^{|\alpha|}}{\left\langle|x|\left|\frac{x}{|x|}-\frac{k}{|k|}
\right| \right\rangle}\|q\|_{W^{2+|\alpha|,\,\infty}}.
\end{equation}
We also have that there exists~$C>0$ such that for any
$k,\,x\in\R^3\setminus\{\,0\}$
\begin{multline}
                                                                \label{Estimate:LemOnW2}
\left|\left\{\nabla_k^\alpha \widetilde{R}_0^\pm(k)\langle
Q\rangle^{-\sigma} q\right\}(x)\right|\\\leq C\frac{\langle
x\rangle^{\alpha-1}}{\langle k\rangle^{\alpha-1}}
\frac{\langle\min\{|x|,\;|k|\} \rangle}{\langle
k\rangle\left\langle\min\{|x|,\;|k|\}\left|\frac{x}{|x|}
-\frac{k}{|k|}\right| \right\rangle^{2}}
\|q\|_{W^{2+|\alpha|,\,\infty}}.
\end{multline}
\end{lemma}
\begin{proof}
We write
\begin{multline*}
\left(\widetilde{R}_0^\pm(k)\langle
Q\rangle^{-\sigma}q\right)(x)\\
=\int_{\R^3}
\frac{e^{\i\{\pm|k||y|+k\cdot y\}}}{4\pi|y|}\left\{\frac{\alpha\cdot
(x-y) q(x-y) }{\langle x- y\rangle^{\sigma+2}}+\frac{\alpha\cdot
\nabla q(x-y)
}{\langle x-y\rangle^{\sigma}}\right\} \,dy\\
+\left(\alpha\cdot k+m\beta \pm\lambda(k)\right)\int_{\R^3}
\frac{e^{\i\{\pm|k||y|+k\cdot y\}}}{4\pi|y|}\frac{q(x-y) }{\langle
x-y\rangle^{\sigma}} \,dy.
\end{multline*}
We restrict our study to~$\widetilde{R}_0^+(k)$ since the two cases
are similar. Hence we only need to estimate integrals of the form
\begin{equation*}
R(k)(x)=\int_{\R^3} e^{\i\{|k||y|+k\cdot y\}}\frac{u(x-y)}{|y|} \,dy
\end{equation*}
with~$u\in W_\sigma^{1+|\alpha|,\,\infty}(\R^3,\C)$.

\medskip

In a first step, a straightforward calculation shows that
\begin{equation}
                                                                \label{Estimate:OnR}
\left|\nabla_k^\alpha R(k)(x)\right| \leq C\langle
x\rangle^{|\alpha|-1}\|u\|_{L_\sigma^{\infty}}
\end{equation}
if~$\sigma>3+\max\{|\alpha|-1;0\}$. Then using the trick we used in
the proof of Proposition~\ref{proposition:Decay1-hom}, we obtain
\begin{equation*}
\nabla_k^\alpha
R(k)(x)=\frac{\i^{|\alpha|}}{|k|^{|\alpha|}}\int_{\R^3}
e^{\i\{|k||y|+k\cdot
y\}}\left\{\left(\nabla|Q|\right)^\alpha\frac{u(x-\cdot)}{|Q|}
\right\}(y) \,dy,
\end{equation*}
and so with~\eqref{Estimate:OnR}, we infer
\begin{equation}
                                                                \label{Estimate:First}
\left|\nabla_k^\alpha R(k)(x)\right|\leq \frac{C\langle
x\rangle^{|\alpha|-1}}{\langle
k\rangle^{|\alpha|}}\|u\|_{W_\sigma^{|\alpha|,\infty}},
\end{equation}
since~$\sigma>3+\max\{|\alpha|-1,0\}$.

\medskip

In a second step, we apply Estimate~\eqref{Estimate:Decay1-hom} of
Proposition~\ref{proposition:Decay1-hom} to~$R(k)(x)$, this gives
\begin{multline*}
\left|\nabla_k^\alpha R(k)(x)\right|\\
\leq\frac{C}{|k|^{|\alpha|+1}}
\max_{|\beta|\leq 1+|\alpha|}
\left\{\int_{\R^3}|y|^{|\beta|-1}\frac{1} {\left|\frac{k}{|k|}-
\frac{y}{|y|}\right|}\left|\nabla^\beta
\left\{\frac{u(x-\cdot)}{|Q|}\right\}(y)\right|\;dy\right\}.
\end{multline*}
Hence we need to estimate on integral of the form
\begin{equation*}
G(x,\omega)=\int_{\R^3}\frac{1}{|y|^n}\frac{1}{\langle
x-y\rangle^s}\frac{1}{\left|\frac{y}{|y|}-\omega\right|}\,dy
\end{equation*}
with~$\omega\in S^2$,~$-|\alpha|+1\leq n\leq 2$ and~$s>\sigma$. To
obtain appropriate estimates, we use
\begin{equation*}
\left|x-y\right|\geq\frac{1}{4}
\max\{|y|,|x|\}\left|\frac{x}{|x|}-\frac{y}{|y|}\right|
+\frac{1}{2}\left||x|-|y|\right|
\end{equation*}
to write for~$\theta,\theta'\geq 0~$ such that~$\theta+\theta'=1$,
\begin{eqnarray*}
G(x,\omega)&\leq&C\int_{\R^3}\frac{1}{|y|^n}\frac{1}{\langle
|x|-|y|\rangle^{\theta s}}\frac{1}{\left\langle
|x|\left|\frac{x}{|x|}-\frac{y}{|y|}\right|\right\rangle^{\theta's}}
\frac{1}{\left|\frac{y}{|y|}-\omega\right|}\,dy\\
&\leq&\frac{C}{|x|\left\langle|x|\left|\omega-\frac{x}{|x|}\right|
\right\rangle} \int_{\R^+}\frac{1}{r^{n-2}}\frac{1}{\langle
|x|-r\rangle^{\theta
s}}\,dr\\
&\leq&\frac{C\langle
x\rangle^{|\alpha|+1}}{|x|\left\langle|x|\left|\omega-\frac{x}{|x|}
\right|\right\rangle},
\end{eqnarray*}
if~$\theta's>2$ and~$\theta s>1+\max\{2-n;0\}$. Since~$G(0,\omega)$
is bounded, we obtain
\begin{equation*}
G(x,\omega)\leq \frac{C\langle x\rangle^{|\alpha|}}{\left\langle
|x|\left|\omega-\frac{x}{|x|}\right|\right\rangle},
\end{equation*}
Hence, we obtain with estimate~\eqref{Estimate:First}
\begin{equation}
                                                                \label{Estimate:Second}
\left|\nabla_k^\alpha R(k)(x)\right|\leq\frac{C}{\langle
k\rangle^{|\alpha|+1}} \frac{\langle
x\rangle^{|\alpha|}}{\left\langle
|x|\left|\frac{k}{|k|}-\frac{x}{|x|}\right|\right\rangle}
\|u\|_{W_\sigma^{|\alpha|+1,\infty}},
\end{equation}
which gives estimate~\eqref{Estimate:LemOnW}.

\medskip

In a third step, if~$k/|k|\neq x/|x|$, we split the integral for
$\nabla_k^\alpha R(k)(x)$ in two parts with help of a smooth cut-off
function defined in~$S^2$ the support of which is a half cone
determined by the bisector plane of the couple~$\left(k/|k|;
x/|x|\right)$. So we obtain~$\nabla_k^\alpha
R(k)(x)=R_1(k)(x)+R_2(k)(x)$ with~$R_1(k)(x)$ having a support
containing~$x/|x|$ and~$R_2(k)(x)$ having a support containing
$k/|k|$. We then apply the estimate \eqref{Estimate:Decay1-hom2} of
Proposition~\ref{proposition:Decay1-hom} to~$R_1(k)(x)$ to obtain
\begin{multline*}
\left|\nabla_k^\alpha R_1(k)(x)\right|
\leq\frac{C}{|k|^{|\alpha|+2}\left|\frac{k}{|k|}-
\frac{x}{|x|}\right|}\times\\
\times\max_{|\beta|\leq 2+|\alpha|}
\left\{\int_{\R^3}|y|^{|\beta|-2}\frac{1} {\left|\frac{k}{|k|}-
\frac{y}{|y|}\right|}\left|\nabla^\beta
\left\{\frac{u(x-\cdot)}{|Q|}\right\}(y)\right|\;dy\right\}.
\end{multline*}
This gives the estimate
\begin{equation*}
\left|\nabla_k^\alpha R_1(k)(x)\right|\leq\frac{C}{|k|^{|\alpha|+2}}
\frac{\quad\langle
x\rangle^{|\alpha|-1}}{\left|\frac{k}{|k|}-\frac{x}{|x|}\right|^2}
\|u\|_{W_\sigma^{|\alpha|+2,\infty}},
\end{equation*}
since~$\sigma>2+|\alpha|$. Using~\eqref{Estimate:Second}, we infer
\begin{equation*}
\left|\nabla_k^\alpha R_1(k)(x)\right|\leq\frac{C}{\langle
k\rangle^{|\alpha|+1}} \frac{\quad\langle
x\rangle^{|\alpha|}}{\langle\sqrt{|k||x|}\left|\frac{k}{|k|}
-\frac{x}{|x|}\right|\rangle^2}
\|u\|_{W_\sigma^{|\alpha|+2,\infty}},
\end{equation*}
or, using~\eqref{Estimate:First},
\begin{equation*}
\left|\nabla_k^\alpha R_1(k)(x)\right|\leq\frac{C}{\langle
k\rangle^{|\alpha|}} \frac{\quad\langle
x\rangle^{|\alpha|-1}}{\langle|k|\left|\frac{k}{|k|}-\frac{x}{|x|}
\right|\rangle^2} \|u\|_{W_\sigma^{|\alpha|+2,\infty}}.
\end{equation*}
For~$R_2(x)(k)$, we use the
inequality $\left|x-y\right|\geq\frac{|x|\left|\frac{x}{|x|}-
\frac{y}{|y|}\right|}{2}$ to obtain
\begin{equation*}
R_2(k)(x)\leq \frac{C\langle x\rangle^{|\alpha|-1}}{\langle
k\rangle^{|\alpha|}\langle
|x|\left|\frac{k}{|k|}-\frac{x}{|x|}\right|\rangle^{s}}
\|u\|_{W_\sigma^{|\alpha|,\infty}},
\end{equation*}
since~$\sigma>3+s+\max\{|\alpha|-1,0\}$. So now we easily deduce
estimate~\eqref{Estimate:LemOnW2}.
\end{proof}

For the sequel, we need the following
\begin{lemma}
                                                                \label{lemma:Commutator}
Let be~$s\in\R$ and~$\phi$ a~${\mathcal C}^\infty$ function such
that there is~$\sigma>0$ with
\begin{equation*}
\forall \alpha\in \N^3,\; \left|\nabla^\alpha\phi(x)\right|\leq
\frac{C_\alpha}{\langle x\rangle^{\sigma}}.
\end{equation*}
We have that~$[\langle P\rangle^s,\phi(Q)]$ is bounded from
$H^t_{q}$ into~$H^{t'}_{q'}$ with~$q'+\sigma\geq q$ and~$t'+1\geq
t+s$.
\end{lemma}
\begin{proof}
We want  to prove that
\begin{equation}
                                                                \label{Def:CommutatorToBound}
\langle Q\rangle^q\langle P\rangle^t[\langle
P\rangle^s,\phi(Q)]\langle P\rangle^{-t'}\langle Q\rangle^{-q'}
\end{equation}
is bounded in~${\mathcal B}(L^2)$. Using the identity
\begin{equation*}
[\langle P\rangle^s,\phi(Q)]=[\langle P\rangle^{s/2},\phi(Q)]\langle
P\rangle^{s/2}+\langle P\rangle^{s/2}[\langle
P\rangle^{s/2},\phi(Q)]
\end{equation*}
we reduce the proof to the case~$|s|<1$. And with the identity
\begin{equation*}
[\langle P\rangle^s,\phi(Q)]=-\langle P\rangle^s[\langle
P\rangle^{-s},\phi(Q)]\langle P\rangle^s
\end{equation*}
we only need to study the case~$-1<s< 0$. The proof in this case is
based on the following identity for~$-1<s<0$
\begin{equation*}
\langle P\rangle^s=\left(-\Delta+1\right)^{s/2}=
\frac{-\sin(\pi\left\{\frac{s}{2}\right\})}{\pi}
\int_0^{+\infty}\frac{w^{\left\{\frac{s}{2}\right\}}}{-\Delta+1+w}dw.
\end{equation*}
So we have
\begin{equation*}
[\langle P\rangle^s,\phi(Q)]=\sum_{k=1}^m\frac{\Gamma(s/2+1)}
{\Gamma(s/2+1-k)}(-\Delta+1)^{s/2-k}Ad_{-\Delta+1}^k(\phi(Q))+R_m\\
\end{equation*}
with
\begin{equation*}
R_m=\frac{(-1)^m\sin(\pi\left\{\frac{s}{2}\right\})}{\pi}
\int_0^{+\infty}\frac{w^{\left\{\frac{s}{2}\right\}}}
{(-\Delta+1+w)^{m+1}}
Ad_{-\Delta+1}^{m+1}(\phi(Q))\frac{dw}{-\Delta+1+w}.
\end{equation*}
Then we use~$\frac{-\Delta+1}{-\Delta+1+w}=1-\frac{w}{-\Delta+1+w}$,
and we commute powers of~$\langle P\rangle$ with  operators of the
form~$\nabla^\alpha\phi(Q)$. Hence we can repeat the previous
computation until we obtain only non positive powers of~$\langle P
\rangle$ in~\eqref{Def:CommutatorToBound}. So we only need to prove
that operators of the form
\begin{equation*}
[\langle Q\rangle^q,\phi(P)]\langle Q\rangle^{-q'}
\end{equation*}
with~$q\leq q'+1$ and~$\phi$ satisfying the assumption of the lemma
are bounded in~${\mathcal B}(L^2)$, we just repeat the previous
calculation but we switch the role of~$P$ and~$Q$. This ends the
proof.
\end{proof}

We now state a particular version of the Limiting Absorption
Principle for~$H$.
\begin{proposition}
                                                                \label{proposition:LAP}
We assume that Assumptions~\ref{assumption:1} and~\ref{assumption:2}
hold. Then for any~$\sigma\geq 1$ there exists~$C>0$ such that for
any~$k\in\R^3$
\begin{equation*}
\|\widetilde{R}^\pm_V(k)\|_{{\mathcal
B}(L^{2}_\sigma,L^2_{-\sigma})} \leq C.
\end{equation*}
\end{proposition}
\begin{proof}
In fact, we just need to prove that for any~$\sigma\geq 1$ there
exists~$C>0$ such that for any~$\lambda\in\R\setminus(-m,m)$
\begin{equation*}
\|R^+_V(\lambda)\|_{{\mathcal B}(L^{2}_\sigma,L^2_{-\sigma})}\leq C.
\end{equation*}
Using Theorem~\ref{Thm:Propagation}, we have that it is true if
$\sigma>5/2$. Then we use Born expansion
\begin{equation*}
R^+_V(\lambda)=R^+_0(\lambda)-R^+_0(\lambda)VR^+_0(\lambda)
+R^+_0(\lambda)VR^+_V(\lambda)VR^+_0(\lambda)
\end{equation*}
and~\cite[Theorem 2.1(i)]{IftimoviciMantoiu} to end the proof.
\end{proof}
We are now able to give the
\begin{proof}[Proof of Proposition~\ref{proposition:EstimateOnW}]
We only give a the general idea of the proof and we leave the
details to the reader. We notice that with~$\widetilde{R}^\pm_V$
defined by \eqref{Def:RTilde}, we obtain
\begin{equation*}
w=\widetilde{R}_VVu
\end{equation*}
with an abuse of notation since we avoid to distinguish the case
where we have~$\widetilde{R}^+_V$ or~$\widetilde{R}^-_V$. We recall
the identities
\begin{equation}
\widetilde{R}_VV=\widetilde{R}_0V-\widetilde{R}_0V\widetilde{R}_VV
=\widetilde{R}_0V-\widetilde{R}_0V\widetilde{R}_0+\widetilde{R}_0V
\widetilde{R}_VV\widetilde{R}_0V.
                                                                \label{Identity:ModifiedBornSeries}
\end{equation}
Since, we have
\begin{equation*}
\widetilde{R}_V=(1+\widetilde{R}_0V)^{-1}\widetilde{R}_0,\quad
(1+\widetilde{R}_0V)^{-1}=1-\widetilde{R}_VV,
\end{equation*}
for~$|\alpha|=1$, we obtain
\begin{equation}
\nabla_k^\alpha \widetilde{R}_V=(1-\widetilde{R}_VV)\nabla_k^\alpha
\widetilde{R}_0 (1-V\widetilde{R}_V).
                                                                \label{Identity:FirstDerivativeRes}
\end{equation}
Using~\eqref{Identity:FirstDerivativeRes}, we obtain a formula where
only derivatives of~$\widetilde{R}_0$ appear (if there is
derivatives).
Then between a derivative of~$\widetilde{R}_0$ and
$\widetilde{R}_V$, we insert a~$\widetilde{R_0}$ with the identity
\eqref{Identity:ModifiedBornSeries}:
\begin{eqnarray*}
\widetilde{R}_VV\nabla_k^{\alpha}\widetilde{R}_0V&=&
\widetilde{R}_0V\nabla_k^{\alpha}\widetilde{R}_0V
-\widetilde{R}_0V\widetilde{R}_0V\nabla_k^{\alpha}\widetilde{R}_0V
+\widetilde{R}_0V\widetilde{R}_VV\widetilde{R}_0V\nabla_k^{\alpha}
\widetilde{R}_0V.
\end{eqnarray*}
This ensures that if~$\rho > 5$,~$V$ or its derivatives decays
enough to use Estimate~\eqref{Estimate:LemOnW} and Proposition
\ref{proposition:LAP}. Since these estimates need derivatives and
Sobolev's injections, we apply Lemma~\ref{lemma:Commutator} to
conclude the proof.
\end{proof}

\paragraph{Estimates for~$v$}
We remind us of the definition of~$v$ in~\eqref{Def:V} and we
introduce
\begin{equation*}
S^{\varepsilon_1,\varepsilon_2}_V(k)= e^{-\varepsilon_1\varepsilon_2
\i|k||Q| }R_V^{\varepsilon_1}(\varepsilon_2\lambda(k))e^{\i k\cdot Q
},
\end{equation*}
where~$\varepsilon_i\in\{-1,\,1\}$. With an abuse of notation, we
will write~$v=S_VVu$. We have
\begin{lemma}
                                                                \label{Lem:EstimateForV}
There exists~$C>0$, such that for any~$k\in \R^3\setminus\{0\}$ and
$\beta\in \N^3$
\begin{equation*}
\left|\left(\nabla_k^\beta
S_0^{\varepsilon_1,\varepsilon_2}(k)\langle
Q\rangle^{-\sigma}q\right)(x)\right|
\leq\frac{C}{\left\langle|x|\left|\frac{x}{|x|}-\frac{k}{|k|}\right|
\right\rangle}\|q\|_{W^{2+|\beta|,\,\infty}}
\end{equation*}
for any ~$\sigma>3+|\beta|$.
\end{lemma}
\begin{proof}
\begin{multline*}
\left(S_0^{\varepsilon_1,\varepsilon_2}(k)\langle
Q\rangle^{-\sigma}q\right)(x)=\\
\int_{\R^3}
\frac{e^{\i\varepsilon_1\varepsilon_2\{|k||x-y|-|k||x|+\varepsilon_1\varepsilon_2
k\cdot y\}}}{4\pi|x-y|}\left\{\frac{\alpha\cdot y q(y) }{\langle
y\rangle^{\sigma+2}}+\frac{\alpha\cdot \nabla q(y)
}{\langle y\rangle^{\sigma}}\right\} \,dy\\
+\left(\alpha\cdot k+m\beta \pm\lambda(k)\right)\int_{\R^3}
\frac{e^{\i\varepsilon_1\varepsilon_2\{|k||x-y|-|k||x|+\varepsilon_1\varepsilon_2
k\cdot y\}}}{4\pi|x-y|}\frac{q(y) }{\langle y\rangle^{\sigma}} \,dy.
\end{multline*}
For the sake of simplicity, we only write the proof when~$\beta=0$.
The proof for derivatives works in the same way using
$\left||x-y|-|x|\right|\leq |y|$ and~$\sigma>3+|\beta|$. But the
proof for the case~$\beta=0$, has been already done since
\begin{equation*}
\left|\left( S_0^{\varepsilon_1,\varepsilon_2}(k)\langle
Q\rangle^{-\sigma}q\right)(x)\right| =\left|\left(
R_0^{\varepsilon_1,\varepsilon_2}(k)\langle
Q\rangle^{-\sigma}q\right)(x)\right|.
\end{equation*}
\end{proof}
Hence using Proposition~\ref{proposition:LAP}, we able to write the
\begin{proof}[Proof of Proposition~\ref{proposition:EstimateOnV}]
We write with an abuse of notation
\begin{equation*}
v=S_VVu,
\end{equation*}
and we use the Born formula
\begin{equation*}
S_VV=S_0V-S_0V\widetilde{R}_VV,
\end{equation*}
together with Lemma~\ref{Lem:EstimateForV}, Propositions
\ref{lemma:Commutator} and~\ref{proposition:LAP}. The proof works
like the one for~$w$.
\end{proof}

\paragraph{Estimates for~$\widetilde{v}$}
We remind us of the definition of~$\widetilde{v}$ in
~$\eqref{Def:VTilde}$ and we introduce
\begin{equation*}
T^{\varepsilon_1,\varepsilon_2}_V(k)= e^{-\varepsilon_1\varepsilon_2
\i|k||Q|+\i k\cdot Q
}\nabla_{k}\widetilde{R}_V^{\varepsilon_1}(\varepsilon_2\lambda(k))e^{\i
k\cdot Q},
\end{equation*}
where~$\varepsilon_i\in\{-1,\,1\}$. With another abuse of notation,
here we will write~$\widetilde{v}=T_VVu$. We have
\begin{lemma}
                                                                \label{Lem:EstimateForVtilde}
There exists~$C>0$, such that for any~$k\in \R^3\setminus\{0\}$ and
$\beta\in \N^3$
\begin{equation*}
\left|\left(\nabla_k^\beta
T_0^{\varepsilon_1,\varepsilon_2}(k)\langle
Q\rangle^{-\sigma}q\right)(x)\right| \leq C
\frac{\langle\min\{|x|,\;|k|\} \rangle}{\langle
k\rangle\left\langle\min\{|x|,\;|k|\}\left|\frac{x}{|x|}-\frac{k}{|k|}
\right| \right\rangle^{2}} \|q\|_{W^{2+|\beta|,\,\infty}},
\end{equation*}
for any ~$\sigma>4+|\beta|$.
\end{lemma}
\begin{proof}
This is an obvious adaptation of the proof of Lemma
\ref{Lem:EstimateForV}, we just notice that one has
\begin{equation*}
\left|\left( T_0^{\varepsilon_1,\varepsilon_2}(k)\langle
Q\rangle^{-\sigma}q\right)(x)\right| =
\left|\left(\nabla_{k}\widetilde{R}_V^{\varepsilon_1}(\varepsilon_2
\lambda(k)) \langle Q\rangle^{-\sigma}q\right)(x)\right|.
\end{equation*}
\end{proof}
Hence, we have
\begin{proof}[Proof of Proposition~\ref{proposition:EstimateOnVtilde}]
One more time, we write with an abuse of notation
\begin{equation*}
v=T_VVu+S_VV\nabla_ku,
\end{equation*}
The second term of the right hand side could be studied exactly as
we done in proof of Proposition~\ref{proposition:EstimateOnV} and
for the first one we use the formula
\begin{equation*}
T_VV=T_0V-T_0V\widetilde{R}_VV+S_0V\nabla_k\widetilde{R}_VV,
\end{equation*}
together with Lemma~\ref{Lem:EstimateForVtilde}, Propositions
\ref{lemma:Commutator} and~\ref{proposition:LAP}. The proof works
like the one for~$w$.
\end{proof}

\section{The linearized operator}
                                                                \label{Section:LinearizedOperator}
In this section, we study the spectral properties of the linearized
operator, associated with Equation~\eqref{Eq:NLD}, around a
stationary state. This will be useful since we compare the dynamics
associated with Equation~\eqref{Eq:NLD} to the dynamic of the linear
Dirac equation associated with~$H$. This comparison is possible only
because when the PLS is small, the linearized operator is a small
perturbation of~$H$.

\subsection{The manifold of the particle like solutions}
                                                                \label{Subsection:TheManifoldPLS}
First we notice that Proposition~\ref{Prop:ManifoldPLS}, which gives
the existence of stationary states, is a consequence of
\begin{proposition}
Let~$H$ be a self adjoint operator on~$L^2(\R^3,\C^4)$ and with a
simple eigenvalue~$\lambda_0$ associated with a normalized
eigenvector~$\phi_0$. Assume that there is a neighborhood~${\mathcal
O}\subset \R$ of~$\lambda_0$ such that for all~$\lambda\in {\mathcal
O}$ the operator~$(H-\lambda)^{-1}P_0$ is in~${\mathcal
B}(L^2_\sigma(\R^3,\C^4))$ for any~$\sigma \in \R^+$, and
in~${\mathcal B}(H^l(\R^3,\C^4),H^{l+1}(\R^3,\C^4))$ for any~$l \in
\N$, where~$P_0$ is the projector into the orthogonal space
of~$\phi_0$. Let~$F\in {\mathcal C}^{k+1}(\C^4,\C^4)$ such that
$F(z)=O(|z|^3)$.

Then for any~$\sigma \in \R^+$, there exists~$\Omega$ a neighborhood
of ~$0\in \C$, a~${\mathcal C}^{k}$ map
\begin{equation*}
h: \Omega \mapsto
\left\{\phi_0\right\}^{\bot}\cap{H^2(\R^3,\C^4)}\cap
L^2_\sigma(\R^3,\C^4)
\end{equation*}
and a~${\mathcal C}^{k}$ map~$E: \Omega \mapsto \R$ such that
$S(u)=u\phi_0+h(u)$ satisfy for all~$u\in\Omega$,
\begin{equation*}
H S(u)+\nabla F(S(u))=E(u)S(u),
\end{equation*}
with the following properties
\begin{equation*}
\left\{\begin{array}{l}
h(e^{i\theta} u)=e^{i\theta}h(u),\quad\forall \theta \in \R,\\
h(u)=O(|u|^2),\\
E(u)=E(\left|u\right|),\\
E(u)=\lambda_0+O(|u|^2).
\end{array}\right.
\end{equation*}
\end{proposition}
The proof of this proposition is an obvious adaptation of the one of
\cite[Proposition 2.2]{PilletWayne}, and we don't repeat it here.
One can also obtain it by means of the Crandall-Rabinowitz theorem
but it doesn't give immediately the decomposition associated to the
spectrum of~$H=D_m+V$.

To show that~$(H-\lambda)^{-1}P_0$ is in~${\mathcal
B}(L^2_\sigma(\R^3,\C^4))$ for any~$\sigma>0$, we just need to prove
that~$\alpha \mapsto e^{\alpha \langle
Q\rangle}(H-\lambda)^{-1}P_0e^{-\alpha \langle Q\rangle}$ is of
class~${\mathcal C}^k$ near~$0$ in~${\mathcal B}(L^2(\R^3,\C^4))$
for any~$k\in \N$, this can be proved with help of of~\cite[Lemma
5.1]{Hislop}. To prove that~$(H-\lambda)^{-1}P_0$ for any~$l\in\N$
is in~${\mathcal B}(H^l(\R^3,\C^4),H^{l+1}(\R^3,\C^4))$ for any
$l\in\N$, we first notice that~$(D_m-\lambda)^{-1}$ is in~${\mathcal
B}(H^l(\R^3,\C^4),H^{l+1}(\R^3,\C^4))$ then we use wave operator,
see \ref{Def:WaveOperators} and \cite[Theorem
1.5]{GeorgescuMantoiu}, and the intertwining property, see
\ref{Identity:Intertwining3}, to conclude.

We shall need some properties of stationary solutions of
\eqref{Eq:NLD}. Following \cite{Hislop}, we have the
\begin{lemma}[exponential decay]
                                                                \label{lemma:ExpDecay}
For all~$\beta\in \N^2$,~$s\in \R^+$ and~$p,q\in[1,\infty]$. There
is~$\gamma
>0$,~$\e>0$ and~$C>0$ such that for all~$u \in B_\C(0,\e)$ one has
\begin{equation*}
\|e^{\gamma \langle Q\rangle}\p_u^\beta S(u)\|_{B^s_{p,q}} \leq C
\|S(u)\|_2,
\end{equation*}
where $\p_u^\beta=\frac{\p^{|\beta|}}{\p^{\beta_1}\Re u
\p^{\beta_2}\Im u}$.
\end{lemma}
\begin{proof}
In fact we prove that for any~$k$ in~$\N$ there is~$\gamma >0$ and
$\e>0$ and~$C>0$ such that for all~$u \in B_\C(0,\e)$ one has
\begin{equation*}
\|e^{\gamma \langle Q\rangle}\p_u^\beta S(u)\|_{H^k} \leq C
\|S(u)\|_2.
\end{equation*}
Then interpolation and the following property of Besov spaces over
$\R^3$ permit to conclude:~$B^s_{2,2}=H^s$,~$B^s_{p,r}\subset
B^{s'}_{p,q}$ if~$s'<s$,~$B^u_{r,q}\subset B^s_{p,q}$ if~$1\leq
r\leq p\leq \infty$ and~$u-n/r=s-n/p$ and~$\|uv\|_{B^s_{p,q}}\leq C
\|u\|_{B^{s}_{q,t}} \|v\|_{B^s_{r,t}}$ if
$\frac{1}{p}+\frac{s}{3}>\frac{1}{q}+\frac{1}{r}$.

We only prove the lemma for~$\beta=0$, the other cases are similar.
We have
\begin{equation*}
D_m S(u)+VS(u)+\nabla F(S(u))= E(u)S(u).
\end{equation*}
Let us introduce the $\R-$linear operator $W$ of multiplication by
the matrix valued function $x\in\R^3\mapsto -\i D\nabla
F(S(u)(x))\i+V(x)$. We obtain, with the gauge invariance of $F$, the
identity
\begin{equation*}
WS(u)=\nabla F(S(u))+VS(u).
\end{equation*}
The ``potential'' $W$  tends to zero as $x$ goes to $\infty$. In
fact, as a function of $x$, $W$ is in~$L^1\cap L^\infty$; we can
write $W=W_c+W_\delta$ where~$W_c$ is compactly supported and
$\|W_\delta\|_{L^1\cap L^\infty}\leq \delta$.

We have that~$D_m+W_\delta-E(u)$ is invertible for~$\delta$
sufficiently small and
\begin{equation*}
e^{\gamma \langle Q \rangle} S(u)= e^{\gamma \langle Q
\rangle}\left(D_m+W_\delta-E(u)\right)^{-1}e^{-\gamma \langle Q
\rangle} \{e^{\gamma \langle Q \rangle}W_c S(u)\}.
\end{equation*}
For~$\gamma~$ small,~$\left(D_m+\gamma  \frac{\alpha \cdot
Q}{\langle Q\rangle}+W_\delta-E(u)\right)$ is invertible in~$L^2$
and
\begin{eqnarray*}
e^{\gamma \langle Q \rangle} S(u)&=& \left(D_m+\gamma \frac{\alpha
\cdot Q}{\langle Q\rangle}+W_\delta-E(u)\right)^{-1} e^{\gamma
\langle Q \rangle}W_c S(u).
\end{eqnarray*}
This proves the lemma for~$k=0$ since~$e^{\gamma \langle Q
\rangle}W_c$ is bounded. Now we notice that
\begin{multline*}
|P|\left(D_m+\gamma  \frac{\alpha \cdot Q}{\langle
Q\rangle}+W_\delta-E(u)\right)^{-1}\\
=\frac{|P|}{D_m}-\frac{|P|}{D_m}\left(2\gamma  \frac{\alpha \cdot
Q}{\langle Q\rangle}+W_\delta-E(u)\right) \left(D_m+\gamma
\frac{\alpha \cdot Q}{\langle Q\rangle}+W_\delta-E(u)\right)^{-1}.
\end{multline*}
Hence we obtain
\begin{equation*}
\|e^{\gamma \langle Q \rangle}S(u)\|_{H^k}\leq C\|S(u)\|_{H^{k-1}}.
\end{equation*}
This identity proves the lemma by induction.
\end{proof}

\subsection{The spectrum of the linearized operator}
Here we study the spectrum of the linearized operator associated
with Equation \eqref{Eq:NLD} around a stationary state $S(u)$. Let
us introduce
\begin{equation*}
H(u)=H+d^2 F(S(u))-E(u)
\end{equation*}
where~$d^2F$ is the differential of~$\nabla F$. The operator~$H(u)$
is~$\R-$linear but not $\C-$linear. Replacing~$L^2(\R^3,\C^4)$ by
$L^2(\R^3,\R^4\times\R^4)$ with the inner product obtained by taking
the real part of the inner product of~$L^2(\R^3,\C^4)$, we obtain a
symmetric operator. We then complexify this real Hilbert space and
obtain~$L^2(\R^3,\C^4\times\C^4)$ with its canonical hermitian
product. This process transforms the operator~$-\i$ into
\begin{equation*}
J=\left(\begin{array}{cc}
0&-Id_{\C^4}\\
Id_{\C^4}&0
\end{array}\right).
\end{equation*}
For~$\phi\in L^2(\R^3,\R^4\times\R^4)\subset
L^2(\R^3,\C^4\times\C^4)$, we still write~$\phi$ instead of
\begin{equation*}
\left(\begin{array}{c} \Re \phi\\
 \Im \phi\end{array} \right).
\end{equation*}
The extension of~$H(u)$ over~$L^2(\R^3,\C^4\times\C^4)$ is also
written~$H(u)$ and is now a real operator.

The linearized operator associated with Equation~\eqref{Eq:NLD}
around the stationary state~$S(u)$ is given by~$JH(u)$. We shall now
study its spectrum. Differentiating~\eqref{Eq:StationaryStates}, we
have that
\begin{equation*}
{\mathcal H}_0={\rm Span}\left\{ \frac{\p}{\p \Re u}S(u),\;
\frac{\p}{\p \Im u}S(u)\right\}
\end{equation*}
is invariant under the action of~$JH(u)$. We notice (see
\cite{GustafsonNakanishiTsai}) that
\begin{equation*}
{\mathcal H}_0(u)={\rm Span}\left\{JS(u),\p_{|u|}S(u)\right\}.
\end{equation*}
Using gauge invariance and differentiating, we obtain
\begin{equation*}
JH(u)JS(u)=0\quad \mbox{ and }\quad
JH(u)\p_{|u|}S(u)=\p_{|u|}E(u)JS(u).
\end{equation*}
Hence~${\mathcal H}_0(u)$ is contained in the geometric null space
of~$JH(u)$, in fact it is exactly the geometric null space as proved
in the sequel of this subsection. First, we see that~$JH(u)$ has two
other simple eigenvalues, as stated in the following
\begin{lemma}
Let be
\begin{equation*}
S^+_1(0)=\left(\begin{array}{c} \phi_1\\-i\phi_1\end{array}\right)
\mbox{ and } S^-_1(0)=\left(\begin{array}{c}
\overline{\phi_1}\\i\overline{\phi_1}\end{array}\right).
\end{equation*}
Suppose that
Assumptions~\ref{assumption:1}--\ref{assumption:NonLinearity} hold,
then there are~$\e>0$ and four~${\mathcal C}^{\infty}$
maps~$E^\pm_1: B_\C(0,\e)\mapsto \C$ and~$k^\pm_1: B_\C(0,\e)\mapsto
\left\{S^\pm_1(0)\right\}^\bot$ such that
\begin{equation*}
JH(u)S^{\pm}_{1}(u)=E^\pm_1(u)S^\pm_1(u),
\end{equation*}
with~$\|S^{\pm}_{1}(u)\|=1$,
\begin{equation*}
S_1^\pm(u)=S_1^\pm(0)+k^\pm_1(u),
\end{equation*}
with~$E^\pm_1(u)=\pm i(\lambda_1-\lambda_0)+O(|u|^2)$ and
$k^\pm_1(0)=0$.
\end{lemma}
\begin{proof}
This can be proved in the same fashion as~\cite[Proposition
2.2]{PilletWayne} using Assumption \ref{assumption:Spectrum}.
\end{proof}

We also obtain
\begin{lemma}[exponential decay in Besov spaces]
                                                                \label{Lem:ExpDecayExcited}
Suppose that
Assumptions~\ref{assumption:1}--\ref{assumption:NonLinearity} hold,
then for any~$\beta\in \N^2~$,~$s\in \R$ and~$p,q\in[1,\infty]$
there is~$\gamma>0$,~$\e>0$ and a positive constant~$C$ such that
for all~$u \in B_\C(0,\e)$,
\begin{equation*}
\|e^{\gamma \langle Q\rangle}\p_u^\beta S^\pm_1(u)\|_{B^s_{p,q}}
\leq C \|S^\pm_1(u)\|_2,
\end{equation*}
where $\p_u^\beta=\frac{\p^{|\beta|}}{\p^{\beta_1}\Re u
\p^{\beta_2}\Im u}$.
\end{lemma}
\begin{proof}
The proof is exactly the same as the one of
Lemma~\ref{lemma:ExpDecay}.
\end{proof}
Let~${\mathcal H}_{\pm 1}(u)$ be the space spanned by~$S^\pm_1(u)$.
Let us now prove that the orthogonal space with respect to the
hermitian product associated to~$J$
\begin{equation*}
{\mathcal H}_c(u)=\left\{{\mathcal H }_0(u)\oplus {\mathcal
H}_{+1}(u)\oplus {\mathcal H}_{-1}(u)\right\}^\bot
\end{equation*}
contains no eigenvector. We notice that~${\mathcal H}_c(u)$ is
invariant under the action of~$JH(u)$. We have
\begin{lemma}[Continuous subspace property]
                                                                \label{Lem:ContinousProjector}
If Assumptions~\ref{assumption:1}--\ref{assumption:NonLinearity}
hold, let~${\mathbf P}_c(u)$ be the orthogonal projector
onto~${\mathcal H}_c(u)$. Then there exists~$\e>0$ such that
for~$u',\,u\in B_\C(0,\e)$
\begin{equation*}
\left.{\mathbf P}_c((u))\right|_{{\mathcal H}_c(u')}: {\mathcal
H}_c(u')\mapsto {\mathcal H}_c(u)
\end{equation*}
is an isomorphism from~$B^{s}_{p,q}(\R^3,\C^8)\cap {\mathcal
H}_c(u')$ into~$B^{s}_{p,q}(\R^3,\C^8)\cap {\mathcal H}_c(u)$, for
any~$s\in\R^+$ and any~$p,q\in[1,\infty]$. The inverse~$R(u',u)$ is
continuous with respect to~$u$ and~$u'$.
\end{lemma}
\begin{proof}
This proof is a straightforward adaptation of the one of~\cite[Lemma
2.2]{GustafsonNakanishiTsai}.
\end{proof}
So we have
\begin{lemma}
Under the assumptions of Proposition~\ref{Prop:ManifoldPLS}, there
exists~$\e>0$ such that for any~$u\in B_\C(0,\e)$
\begin{eqnarray*}
&\|\langle Q\rangle^{-\sigma}e^{sJH(u)}{\mathbf P}_c(u)\psi\| &\leq
C\langle s\rangle^{-\min\{\sigma,\;3/2\}} \|\langle
Q\rangle^{\sigma}\psi\|,\;\forall\psi\in
L^2_\sigma\\
&\ds\int_{\R}\|\langle Q\rangle^{-\sigma}e^{sJH(u)}{\mathbf P}_c(u)\psi\|^2\;ds &\leq
C\|\psi\|,\;\forall\psi\in
L^2.
\end{eqnarray*}
As a consequence,~${\mathcal H}_c(u)$ does not contain any
eigenvector.
\end{lemma}
\begin{proof}
For the sake of clarity, we introduce
\begin{equation*}
\zeta(u)=\left(J\frac{\p}{\p \Re u}S(u),\;J\frac{\p}{\p \Im
u}S(u),\;JS^+_1(u),\;JS^-_1(u)\right).
\end{equation*}
Writing Duhamel's formula for~$H(u)$ with respect to~$H-E(u)$, we
obtain
\begin{multline*}
e^{tJH(u)}{\mathbf P}_c(u)=e^{tJ(H-E(u))}{\mathbf P}_c(u)\\+\int_0^t
e^{(t-s)J(D_m+V-E(u))}Jd^2F(S(u))e^{sJH(u)}{\mathbf P}_c(u)\,ds.
\end{multline*}
We have
\begin{equation*}
\ds \left.{\mathbf P}_c(0)\right|_{{\mathcal H}_c(u')}^{-1}=R(u',0)=
Id_{L^2}+\sum_i\left|\alpha_i(u',0)\right\rangle\left\langle
\zeta_i(0)\right|
\end{equation*}
where the coordinates of~$\alpha_i(u',u)$ are linear combination of
the coordinates of~$\zeta(u)$, so it can be extended to
$L^2_{-\sigma}$ and we have
\begin{eqnarray*}
\lefteqn{\|\langle Q\rangle^{-\sigma} e^{-tJH(u)}{\mathbf P}_c(u)
\psi\|}\\
&\leq& \|\langle Q\rangle^{-\sigma}\left.{\mathbf
P}_c(0)\right|_{{\mathcal H}_c(u')}^{-1}\langle
Q\rangle^{\sigma}\|\Big\{\|\langle Q\rangle^{-\sigma}
{\mathbf P}_c(0)e^{-tJ(D_m+V-E(u))}{\mathbf P}_c(u)\psi\|\\
& & + \int_0^t \left\|\langle Q\rangle^{-\sigma} {\mathbf
P}_c(0)e^{-J(t-s)(D_m+V-E(u))}JD\nabla F(S(u))e^{-sJH(u)}{\mathbf
P}_c(u)\psi\right\|\,ds\Big\}\\
&\leq& C\langle t\rangle^{-\min\{\sigma,\;3/2\}}
\|\langle Q\rangle^{\sigma}\psi\|\\
& &+ C\int_0^t \langle
t-s\rangle^{-\min\{\sigma,\;3/2\}}\|\langle Q\rangle^{2\sigma}
D\nabla F(S(u))\|\|\langle
Q\rangle^{-\sigma}e^{-isH(u)}{\mathbf P}_c(u)\psi\|\,ds
\end{eqnarray*}
We then introduce
\begin{equation*}
M(t)=\sup_{s\in [0,t]}\{\langle
s\rangle^{-\min\{\sigma,\;3/2\}}\|\langle
Q\rangle^{-\sigma}e^{-sJH(u)}{\mathbf P}_c(u)\psi\|\}
\end{equation*}
and we obtain for~$|z|\leq\e$
\begin{equation*}
M(t)\leq C(\|\langle Q\rangle^{\sigma}\psi\| + \e  M(t))
\end{equation*}
which gives for~$\e$ sufficiently small
\begin{equation*}
M(t)\leq C \|\langle Q\rangle^{\sigma}\psi\|,
\end{equation*}
or
\begin{equation*}
\|\langle Q\rangle^{-\sigma}e^{-sJH(u)}{\mathbf P}_c(u)\psi\| \leq
C\langle s\rangle^{-\min\{\sigma,\;3/2\}} \|\langle
Q\rangle^{\sigma}\psi\|.
\end{equation*}
With the same method, see  Lemma
\ref{Lem:SmoothSmallNonSelfAdjointOperator}, we obtain the second estimate.

Then we obtain with the second estimate that there is no stationary state in the range
of~${\mathbf P}_c(u)$ that is to say~${\mathcal H}_c(u)$.
\end{proof}

This gives
\begin{lemma}
We have, for sufficiently small~$u\in\C$,~$E^\pm_1(u)\in \i \R$ with
$E^\pm_1(u)=-E^\mp_1(u)$ and $S_1^-(u)=\overline{S_1^+(u)}$ for the
conjugation of ~$\C^8$.
\end{lemma}
\begin{proof}
The last statement straightforwardly follows from
\begin{equation*}
JH(u)S^{\pm}_{1}(u)=E^\pm_1(u)S^\pm_1(u),
\end{equation*}
since there is no more eigenvalues than the $0$ and $E^\pm_1(u)$, we
obtain $\overline{E^\pm_1(u)}=E^\mp_1(u)$.

Then we specify the essential spectrum of~$JH(u)$. A classical study
gives that the continuous spectrum of~$JH(0)$ is given by
\begin{equation*}
\left\{\i\lambda;\; \lambda\in \R,\, |\lambda|\geq
\min\{|m-\lambda_0|,|m+\lambda_0|\}\right\}.
\end{equation*}
Using Weyl's criterion (see~\cite[Theorem XIII.14, Corollary
1]{ReedSimon4}, the adaptation is quite easy in our case), we obtain
that the essential spectrum is
\begin{equation*} \left\{\i\lambda;\; \lambda\in \R,\,
|\lambda|\geq \min\{|m-E|,|m+E|\}\right\}.
\end{equation*}
Hence~$E^\pm_1(u)$ are necessarily purely imaginary. Indeed if
$H(u)-E^\pm_1(u)J$ is not invertible then
$H(u)+\overline{E^\pm_1(u)}J$ is not invertible too. Since
$-\overline{E^\pm_1(u)}$ is not in the essential spectrum, it is
necessarily an eigenvalue in the neighborhood
of~$\pm\i\left(\lambda_1-\lambda_0\right)$. Hence this gives
$-\overline{E^\pm_1(u)}=E^\pm_1(u)$.
\end{proof}
\subsection{Decomposition of the system}
We want to decompose a solution~$\phi$ of the equation
\eqref{Eq:NLD} with respect to the spectrum of~$JH(u)$. And in fact,
we only study the resulting equations for these different parts of
the decomposition. First we isolate a part which corresponds to a
PLS. For any solution of~\eqref{Eq:NLD} over an interval of time~$I$
containing $0$, we write for~$t\in I$
\begin{equation*}
\phi(t)=e^{-\i\int_0^tE(u(s))\,ds}\left(S(u(t)) + \eta(t)\right).
\end{equation*}
In order to give an equation for~$\eta$, we introduce the following
space
\begin{equation*}
{\mathcal H}_0^\bot(u)=\left\{\eta\in L^2(\R^3,\C^8), \left\langle
J\eta, \frac{\p}{\p \Re u}S(u)\right\rangle=0,\;\left\langle J\eta,
\frac{\p}{\p \Im u}S(u)\right\rangle=0\right\}.
\end{equation*}
In fact it is the space
\begin{equation*}
{\mathcal H}_{+1}(u)\oplus {\mathcal H}_{-1}(u)\oplus {\mathcal
H}_c(u)
\end{equation*}
which is invariant under the action of~$JH(u)$ and we state the
\begin{lemma}[decomposition lemma]
Let be~$s\geq 0$ and~$p\geq 1$ there exist~$\delta > 0$ and a
${\mathcal C}^\infty$ map~$U : B_{W^{s,p}}(0,\delta)\mapsto
B_\C(0,\e)~$ which satisfies for~$\psi\in B_{W^{s,p}}(0,\delta)$
\begin{equation*}
\psi=S(u)+\eta, \mbox{ with } \eta \in {\mathcal H}_0^{\bot}(u)
\Longleftrightarrow u=U(\psi).
\end{equation*}
\end{lemma}
\begin{proof}
It is~\cite[Lemma 2.3]{GustafsonNakanishiTsai}.
\end{proof}
This lemma ensures that we can impose the orthogonality condition
\begin{equation}
                                                                \label{Eq:Orthogonality}
\eta(t)\in {\mathcal H}_0^{\bot}(u(t)).
\end{equation}
So instead of solving the Equation~\eqref{Eq:NLD} in~$\phi$, we want
to solve the equation
\begin{equation}
                                                                \label{Eq:FirstForEta}
\begin{array}{ccl}
\i\p_t \eta&=&\left\{H-E(u)\right\}\eta+ \left\{\nabla
F(S(u)+\eta)-\nabla
F(S(u))\right\}-\i dS(u)\dot{u}\\
&=&\left\{H+d^2F(S(u))-E(u)\right\}\eta+N(u,\eta)-\i dS(u)\dot{u}
\end{array}
\end{equation}
for~$\eta\in {\mathcal H}_0^{\bot}(u(t))$. Here~$d^2F$ is the
differential of~$\nabla F$ and~$dS$ the differential of~$S$ in
$\R^2$. To close the system, we need an equation for~$u$. Let us now
derive an equation for the path~$u$, by means
of~\eqref{Eq:Orthogonality}:
\begin{equation*}
\langle \eta(t),JdS(u(t))\rangle=0.
\end{equation*}
After a time derivation, we obtain
\begin{multline*}
0=\langle JH(u(t))\eta(t) + JN(u(t)),\eta(t))\\
+dS(u(t))\dot{u}(t),JdS(u(t))\rangle - \langle
\eta,Jd^2S(u(t))\dot{u}(t)\rangle.
\end{multline*}
Since~$S(u)\in J{\mathcal H}_0(u)$, we have
\begin{equation*}
\langle H(u)\eta,dS(u)\rangle=\langle \eta,H(u)dS(u)\rangle=\langle
\eta,dE(u)S(u)\rangle=0,
\end{equation*}
we obtain
\begin{equation*}
[\langle JdS(u(t)),dS(u(t))\rangle - \langle
J\eta(t),d^2S(u(t))\rangle] \dot{u}(t)=-\langle N(u(t),\eta(t))
,dS(u(t))\rangle.
\end{equation*}
So we notice that
\begin{equation*}
[\langle JdS(u(t)),dS(u(t))\rangle - \langle
J\eta(t),d^2S(u(t))\rangle] = \left(
\begin{array}{cc}
0&-1\\
1&0
\end{array}\right) + O(\left|u(t)\right|+\|\eta(t)\|_2),
\end{equation*}
which proves that~$[\langle JdS(u(t)),dS(u(t))\rangle - \langle
J\eta(t),d^2S(u(t))\rangle]$ is invertible for small~$|u(t)|$ and
$\|\eta(t)\|_2$, we therefore introduce its inverse
\begin{equation*}
A(u,\eta)=[\langle JdS(u),dS(u)\rangle - \langle
J\eta,d^2S(u)\rangle]^{-1}
\end{equation*}
and write
\begin{equation*}
\p_t u(t)=-A(u(t),\eta(t))\langle N(u(t),\eta(t)) ,dS(u(t))\rangle.
\end{equation*}
Plugging in Equation~\eqref{Eq:FirstForEta}, and similarly to the
linear case we decompose~$\eta$ with respect to the spectral
decomposition of~$H(u)=H+D\nabla F(S(u))-E(u)$
\begin{equation*}
\eta(t)=\alpha^+(t)S_1^+(u)+\alpha^-(t)S_1^-(u)+z(t)
\end{equation*}
with~$z\in {\mathcal H}_c(u)\cap L^2(\R^3,\R^8)$ and
$\alpha^-=\overline{\alpha^+}$. We obtain the system
\begin{equation*}
\left\{\begin{array}{lll}
\dot{u}&=&-A(u,\eta)\langle N(u,\eta) ,dS(u)\rangle\\
\dot{\alpha^\pm} &=&E^\pm(u)\alpha^\pm+\langle JN(u,\eta), JS_1^\pm(u)\rangle\\
&&\qquad+ \langle dS(u)A(u,\eta)\langle N(u,\eta)
,dS(u)\rangle\,JS^\pm_1(u)\rangle\\
&&\quad\qquad-\langle (d S_1^\pm(u))A(u,\eta)\langle N(u,\eta),
dS(u)\rangle,\,JS^\pm_1(u)\rangle \alpha^\pm\\
&&\quad\qquad-\langle (d S_1^\mp(u))A(u,\eta)\langle N(u,\eta),
dS(u)\rangle,\,JS^\pm_1(u)\rangle \alpha^\mp\\
\p_t z&=&JH(u)z+{\mathbf P}_c(u)JN(u,\eta)\\
&&\qquad+ {\mathbf P}_c(u)dS(u)A(u,\eta)\langle N(u,\eta) ,dS(u)\rangle\\
&&\quad\qquad-(D {\mathbf P}_c(u))A(u,\eta)\langle N(u,\eta) ,dS(u)\rangle \eta\\
\end{array}\right.,
\end{equation*}
which we will now study. We notice that this equation is defined
only for~$z$ small with real values,~$\alpha^-=\overline{\alpha^+}$
small and~$u$ small.

\section{The stabilization towards the PLS manifold}
                                                                \label{Section:Stabilization}
We now build a solution which stabilizes towards the manifold of the
stationary states. To this end, we will use Theorem
\ref{Thm:Propagation} and Theorem~\ref{Thm:Dispersion} to prove that
$z$ tends to zero in~$L^\infty$ and~$L^2_{\rm loc}$. It is possible
here since we build solutions for which we ensure that~$\alpha^+$
and~$\alpha^-$ also tend to zero. We do not think that this
convergence holds for all initial states but we do not know any
counterexample.

We also notice that we look for a real solution~$\phi=S(u)µ+\eta$,
hence~$\eta$ should be real and
therefore~$\alpha^-=\overline{\alpha^+}$.

We impose the following condition
\begin{equation*}
|\alpha|\leq \frac{C}{\langle t\rangle^{2}}.
\end{equation*}
Under the assumptions of Theorem \ref{Thm:StabilizationSmallPLSNR},
let us define for any~$\e,\,\delta>0$
\begin{multline*}
{\mathcal U}(\e,\delta) =\Bigg\{u\in{\mathcal
C}^{\infty}(\R,B_\C(0,\e)),\;\lim_{t\to+\infty}u(t)=u_\infty \mbox{
exists},\,\\
\left|u(t)-u_\infty\right|\leq\frac{\delta^2}{\langle t\rangle^{2}},
\forall t\geq 0\Bigg\}
\end{multline*}
and for any~$u\in {\mathcal U}(\e)$, let~$s,\,s',\,\beta$ be such
that~$s'\geq s+3\geq \beta+6$ and~$\sigma>5/2$, we define
\begin{multline*}
{\mathcal Z}(u,\delta)=\Bigg\{z\in {\mathcal C}^\infty(\R,
L^2(\R^3,\R^8)),\;z(t)\in {\mathcal
H}_c(u(t)),\,\\\!\max\!\Bigg[\!\sup_{v\in[0,+\infty]}\{\|z(v)\|_{H^{s'}}\},
\!\!\sup_{v\in[0,+\infty]}\{\langle v
\rangle^{3/2}\|z(v)\|_{B^\beta_{\infty,2}}\},\\
\!\!\sup_{v\in[0,+\infty]}\{\langle
v\rangle^{3/2}\{\|z(v)\|_{H^s_{-\sigma}}\}\Bigg]\!<\!\delta\Bigg\}.
\end{multline*}
Then we define the space
\begin{equation*}
\ds\Omega(\delta)=\left\{\alpha=\left(\alpha^+,\alpha^-\right)\in
{\mathcal
C}^{\infty}(\R),\;\alpha^-=\overline{\alpha^+},\;\sup_{t\in\R^+}\langle
t\rangle^{3/2}|\alpha(t)|<\delta^2\right\}.
\end{equation*}

\subsection{Step 1: Construction of~$\alpha$}

For any~$u\in {\mathcal U}(\e,\delta)$ and~$z\in {\mathcal
Z}(u,\delta)$, let us define a map~${\mathcal G}_{u,z}$ on
$\Omega(\delta)$ by
\begin{multline*}
{\mathcal G}_{u,z}(\alpha)^\pm(t)= -\int_t^\infty
e^{\int_s^tE_1^\pm(u(w))\,dw}
\Bigg\{\langle JN(u(v),\eta(v)), S_1^\pm(u(v))\rangle\\
+ \langle dS(u(v))A(u(v),\eta(v))\langle N(u(v),\eta(v))
,dS(u(v))\rangle\,S^\pm_1(u(v))\rangle\\
-\langle (d S_1^\pm(u(v)))A(u(v),\eta(v))\langle N(u(v),\eta(v))
,dS(u(v))\rangle,\,S^\pm_1(u(v))\rangle \alpha^\pm(v)\\
-\langle (d S_1^\mp(u(v)))A(u(v),\eta(v))\langle N(u(v),\eta(v))
,dS(u(v))\rangle,\,S^\pm_1(u(v))\rangle \alpha^\mp(v)\Bigg\}\,dv.
\end{multline*}
We want to show that~${\mathcal G}_{u,z}$ stabilizes
$\Omega(\delta)$ and is a contraction for the~$L^\infty$ norm. We
have the
\begin{lemma}
                                                                \label{Lem:EstimateOnN}
Let be~$\sigma\in\R$,~$s>1$  and~$p,p_1,p_2,q\in[1,\infty]$ such
that
\begin{equation*}
\frac{1}{p}+\frac{s}{3}>\frac{1}{p_1}+\frac{1}{p_2}.
\end{equation*}
Then there exists~$\e_0>0$ and~$C>0$ such that for all~$u\in
B_\C(0,\e_0)$ and~$\eta \in B_{p,q}^s(\R^3,\R^8)\cap
L^\infty(\R^3,\R^8)$, such that
\begin{equation}
                                                                \label{Estimate:OnN2-1}
\left\|\langle Q\rangle^\sigma N(u,\eta)\right\|_{B_{p,q}^s}
\leq
C\left(\left|u\right|+\left\|\eta\right\|_{B_{p_2,q}^s} \right)
\left\|\eta\right\|_{L^\infty} \left\|\langle
Q\rangle^\sigma\eta\right\|_{B_{p_1,q}^s}.
\end{equation}
\end{lemma}
\begin{proof}
We recall the definition
\begin{equation*}
N(u,\eta)=\nabla F(S(u)+\eta)-\nabla F(S(u))-d^2F(S(u)) \eta.
\end{equation*}
We have
\begin{equation*}
N(u,\eta)=\int_0^1\int_0^1
d^3F(S(u)+\theta'\theta\eta)\cdot\eta\cdot \theta\eta\,
d\theta'd\theta.
\end{equation*}
Since for~$s\in\R_+^*$,~$p,\,p_1,\,p_2,\,\in[1,\infty]$ such that
$\frac{1}{p}+\frac{s}{3}>\frac{1}{p_1}+\frac{1}{p_2}$, we have $
\ds\|uv\|_{B^s_{p,q}}\leq C \|u\|_{B^{s}_{p_1,q'}}
\|v\|_{B^s_{p_2,q'}}$. Then since $s>1$ , we use (see
\cite[Proposition 2.1]{EscobedoVega})
\begin{equation*}
\ds\|d^{3}  F(\psi)\|_{B^s_{p',q}}\leq
C\left(s,F,\|\psi\|_{\infty}\right)\|\psi\|_{B^s_{p',q}}.
\end{equation*}
Then using Lemma~\ref{lemma:ExpDecay}, we conclude the proof.
\end{proof}

Hence we have the
\begin{lemma}
There exists~$\delta_0>0$ and~$\e_0>0$ such that for any
$\delta\in(0,\delta_0)$ and~$\e\in (0,\e_0)$, for any~$u\in
{\mathcal U}(\e,\delta)$ and~$z\in {\mathcal Z}(u,\delta)$,
${\mathcal G}_{u,z}(\alpha)$ maps~$\Omega(\delta)$ into itself.
\end{lemma}
\begin{proof}
We have by means of Estimate~\eqref{Estimate:OnN2-1} with e.g.
$\sigma<-3$,~$s=0$,~$p=q=2$ and~$p_1=p_2=4$, if~$u_0\in \C$ and
$z_0\in {\mathcal H}_c(u_0)\cap H^{s'}_\sigma(\R^3,\R^8)$ are small
enough
\begin{eqnarray*}
\lefteqn{
\left|{\mathcal G}_{u,z}(\alpha)^\pm(t)\right|}\\
&\leq&C\int_t^\infty \Bigg\{\left|\langle JN(u(s),\eta(s)),
JS_1^\pm(u(s))\rangle\right|\\
&+& \left|\langle dS(u(v))A(u(v),\eta(v))\langle N(u(v),\eta(v))
,dS(u(v))\rangle\,JS^\pm_1(u(v))\rangle\right|\\
&+&\left|\langle (d S_1^\pm(u(v)))A(u(v),\eta(v))\langle
N(u(v),\eta(v)) ,dS(u(v))\rangle,\,JS^\pm_1(u(v))\rangle\alpha^\pm(v)\right|\\
&+&\left|\langle (d
S_1^\mp(u(v)))A(u(v),\eta(v))\langle
N(u(v),\eta(v)) ,dS(u(v))\rangle,\,JS^\pm_1(u(v))\rangle\alpha^\mp(v)\right|\}\,dv\\
&\leq&C\delta^2\langle t\rangle^{-2}.
\end{eqnarray*}
Hence for small~$\delta$ and small~$\e$, we have~${\mathcal
G}_{u,z}(\Omega(\delta))\subset\Omega(\delta)$.
\end{proof}

To prove that~${\mathcal G}_{u,z}$ is a contraction for the
$L^\infty$ norm, we use the
\begin{lemma}
                                                                \label{lemma:EstimateOnN1}
Let be~$\sigma\in\R$,~$s>0$ and~$p,q\in[1,\infty]$ such that~$sp>3$.
Then for any~$\e>0$ and~$M>0$ there exists~$C>0$ such that for all
$u,\,u'\in B_\C(0,\e)$ and~$\eta,\,\eta' \in
B_{p,q}^s(\R^3,\R^8)\cap L^\infty(\R^3,B_{\R^8}(0,M))$, such that
\begin{multline*}
\left\|\langle
Q\rangle^{\sigma}\left\{N(u,\eta)-N(u',\eta')\right\}\right\|_{B_{p,q}^s}
\leq
C\Bigg\{ \left(\|\langle
Q\rangle^{\sigma_1}\eta\|_{B_{p,q}^s}+\|\langle
Q\rangle^{\sigma_1}\eta'\|_{B_{p,q}^s}\right)^2\times\\
\times\left(\left|u-u'\right|+\left\|\langle
Q\rangle^{\sigma_2}\left(\eta-\eta'\right)\right\|_{B_{p,q}^s}\right)\\
+\left(\left|u\right| +\left|u'\right|+\left\|\langle
Q\rangle^{\sigma_1'}\eta\right\|_{B_{p,q}^s} +\left\|\langle
Q\rangle^{\sigma_1'}\eta'\right\|_{B_{p,q}^s}\right)\times
\\\times\left(\|\langle Q\rangle^{\sigma_2'}\eta\|_{B_{p,q}^s}+\|\langle
Q\rangle^{\sigma_2'}\eta'\|_{B_{p,q}^s}\right)\|\langle
Q\rangle^{\sigma_3'}\left(\eta-\eta'\right)\|_{B_{p,q}^s}\Bigg\},
\end{multline*}
with~$2\sigma_1+\sigma_2=\sigma_1'+\sigma_2'+\sigma_3'=\sigma$.
\end{lemma}
\begin{proof}
Since, we have
\begin{equation*}
N(u,\eta)=\int_0^1\int_0^1
d^3F(S(u)+\theta'\theta\eta)\cdot\eta\cdot \theta\eta\,
d\theta'd\theta.
\end{equation*}
we can also restrict the study to~$d^3 F(\phi)-d^3 F(\phi')$. If
$d^5 F \neq 0$, we have
\begin{multline*}
\|\langle Q\rangle^\sigma \left(d^3 F(\phi)-d^3
F(\phi')\right)\|_{B_{p,q}^s}\\\leq\int_0^1 \| d^{4}
F(\phi+t(\phi-\phi'))\|_{B_{p,q}^s}\|\langle
Q\rangle^\sigma(\phi-\phi')\|_{B_{p,q}^s}\,dt
\end{multline*}
then since~$s>0$, we use
\begin{equation*}
\ds\|d^{4}  F(\psi)\|_{B^s_{p,q}}\leq C(s,F, \|\psi\|_{B^s_{p,q}}).
\end{equation*}
Then using Lemma~\ref{lemma:ExpDecay}, we conclude the proof when
$d^5 F \neq 0$. Otherwise the proof is easily adaptable since~$d^4
F$ is a constant matrix of~${\mathcal M}_4(\C)$.
\end{proof}

We also need the
\begin{lemma}
                                                                \label{Lem:EstimateOnA}
Let be~$\sigma\in\R$,~$s>0$ and~$p,q\in[1,\infty]$. For any~$\e>0$
and~$M>0$, there exists~$C>0$ such that for all~$u,\, u'\in
B_\C(0,\e)$ and~$\eta,\, \eta' \in B_{p,q}^s(\R^3,\R^8)\cap
L^\infty(\R^3,B_{\R^8}(0,M))$, one has
\begin{equation*}
\left|A(u,\eta)-A(u',\eta')\right|\leq
C\left\{\left|u-u'\right|+\left\|\langle
Q\rangle^\sigma\left\{\eta-\eta'\right\}\right\|_{B^s_{p,q}}\right\}
\end{equation*}
\end{lemma}
\begin{proof}
We recall that
\begin{equation*}
A(u,\eta)=[\langle JdS(u),dS(u)\rangle - \langle
J\eta,d^2S(u)\rangle]^{-1}.
\end{equation*}
We have
\begin{multline*}
A(u,\eta)-A(u',\eta')=-[\langle JdS(u),dS(u)\rangle - \langle
J\eta,d^2S(u)\rangle]^{-1}\times\\
\times\left\{\langle JdS(u),dS(u)\rangle - \langle
J\eta,d^2S(u)\rangle-\langle JdS(u'),dS(u')\rangle + \langle
J\eta',d^2S(u')\rangle\right\}\times\\
\times[\langle JdS(u'),dS(u')\rangle - \langle
J\eta',d^2S(u')\rangle]^{-1}.
\end{multline*}
The lemma then follows from Lemma~\ref{lemma:ExpDecay}.
\end{proof}

Hence we have the
\begin{lemma}
                                                                \label{Lem:LipshitzPropertyOfG}
There exists~$\delta_0>0$ and~$\e_0$ such that there exists
$\kappa\in(0,1)$ such that for any~$\delta\in(0,\delta_0)$ and
$\e\in (0,\e_0)$, for any~$u,\,u'\in {\mathcal U}(\e)$ and~$z\in
{\mathcal Z}(u,\delta)$ and~$z'\in {\mathcal Z}(u',\delta)$, for any
$\alpha,\alpha'\in \Omega(\delta)$, one has
\begin{multline*}
\left\|{\mathcal G}_{u',z'}(\alpha')-{\mathcal
G}_{u,z}(\alpha)\right\|_{L^\infty(\R^+)} \\
\leq
\kappa\left(\left|u'-u\right|_{L^\infty(\R^+)}+
\left|\alpha'-\alpha\right|_{L^\infty(\R^+)}+
\left\|z'-z\right\|_{L^\infty(\R^+,B^\beta_{\infty,2})} \right).
\end{multline*}
\end{lemma}
\begin{proof}
It is a straightforward computation based on Lemma
\ref{lemma:EstimateOnN1} with e.g.~$\sigma<-6$,
$\sigma_2,\sigma'_3<-3$ and~$s=0$,~$p=q=2$, on Lemma
\ref{Lem:EstimateOnA}, on Lemma~\ref{Lem:EstimateOnN} with
e.g.~$\sigma<-3$, $p=q=2$ and~$p_1=p_2=4$ and on Lemma
\ref{Lem:ExpDecayExcited}.
\end{proof}

We now state the
\begin{lemma}
There exists~$\delta_0>0$ and~$\e_0>0$ such that for
$\delta\in(0,\delta_0)$ and $\e\in(0,\e_0)$ and any~$u\in {\mathcal
U}(\e)$ and~$z\in {\mathcal Z}(u,\delta)$, the equation
\begin{multline*}
\dot{\alpha^\pm} =E^\pm(u)\alpha^\pm+\langle JN(u,\eta),
 JS_1^\pm(u)\rangle\\+\langle dS(u)A(u,\eta)\langle N(u,\eta)
,dS(u)\rangle\,JS^\pm_1(u)\rangle\\
-\langle (d S_1^\pm(u))A(u,\eta)\langle N(u,\eta),
dS(u)\rangle,\,JS^\pm_1(u)\rangle \alpha^\pm\\
-\langle (d S_1^\mp(u))A(u,\eta)\langle N(u,\eta),
dS(u)\rangle,\,JS^\pm_1(u)\rangle \alpha^\mp,,
\end{multline*}
where~$\eta(t)=\alpha^+(t)S_1^+(u)+\alpha^-(t)S_1^-(u)+z(t)$, has a
unique solution in~$\Omega(\delta)$.
\end{lemma}
\begin{proof}
The proof is now classical since we proved that the integral
equation
\begin{equation*}
\alpha(t)={\mathcal G}_{u,z}(\alpha)(t)
\end{equation*}
can be solved by means of the fixed point theorem.
\end{proof}

\subsection{Step 2: Construction of~$z$}

Let be~$u\in {\mathcal U}(\e,\delta)$ and~$z_0\in {\mathcal
H}_c(u(0))\cap H^{s'}_{\sigma}$. Let us write
$u_\infty=\ds\lim_{t\to +\infty} u(t)$, we define~${\mathcal
T}_{u,z_0}(z)$ by
\begin{multline*}
{\mathcal T}_{u,z_0}(z)(t)=
e^{JtH(u_\infty)}z_0\\
-\int_0^te^{J(t-v)H(u_\infty)}{\mathbf P}_c(u(v))
J\left\{E(S(u(v)))-E(S(u_\infty)) \right\}z(v)\,dv\\
+\int_0^te^{J(t-v)H(u_\infty)}{\mathbf P}_c(u(v))
J\left\{d^2F(S(u(v)))-d^2F(S(u_\infty)) \right\}z(v)\,dv\\
+\int_0^te^{J(t-v)H(u_\infty)}{\mathbf P}_c(u(v))JN(u(v),
\eta(v))\,dv\\
+\int_0^te^{J(t-v)H(u_\infty)}{\mathbf
P}_c(u(v))dS(u(s))A(u(v),\eta)
\langle N(u(v),\eta(v)) ,dS(u(v))\rangle\,dv\\
-\int_0^te^{J(t-v)H(u_\infty)}(d {\mathbf
P}_c(u(v)))A(u(v),\eta(v))\langle N(u(v),\eta(v)) ,dS(u(v))\rangle
\eta(v)\,dv.
\end{multline*}

We have
\begin{lemma}
                                                                \label{Lem:StabilizationForT}
There exists~$\delta_0>0$ and~$\e_0>0$ and~$C>0$ such that for
any~$\delta\in(0,\delta_0)$, for any~$\e\in(0,\e_0)$ and for
any~$u\in {\mathcal U}(\e,\delta)$, the application~${\mathcal
T}_{u,z_0}$ maps ${\mathcal Z}(u,\delta)$ into itself if
$\left\|z_0\right\|_{H^{s'}_{\sigma}}\leq C\delta$
\end{lemma}
\begin{proof}
With Lemma~\ref{Lem:PerturbedLawConservation} and Lemma
\ref{Lem:ContinousProjector}, we obtain
\begin{eqnarray*}
\lefteqn{ \|{\mathcal T}_{u,z_0}(t)\|_{H^{s'}}}\\&\leq& C\|{\mathbf
P}_c(u_\infty){\mathcal T}_{u,z_0}(t)\|_{H^{s'}}
\\
&\leq&
C\|z_0\|_{H^{s'}}+C\int_0^t\left\|\left\{E(S(u(v)))-E(S(u_\infty))\right\}
z(v)\right\|_{H^{s'}}\,dv\\
&&+C\int_0^t\left\|\left\{d^2F(S(u(v)))-d^2F(S(u_\infty))\right\}
z(v)\right\|_{H^{s'}}\,dv\\
&&+C\int_0^t\|N(u(v),\eta(v))\|_{H^{s'}}\,dv\\
&&+C\int_0^t\|dS(u(v))A(u(v),\eta(v))
\langle N(u(v),\eta(v)) ,dS(u(v))\rangle\|_{H^{s'}}\,dv\\
&&+C\int_0^t\|(d {\mathbf P}_c(u(v)))A(u(v),\eta(v))\langle
N(u(v),\eta(v)) ,dS(u(v))\rangle \eta(v)\|_{H^{s'}}\,dv.
\end{eqnarray*}
Now, with Lemma~\ref{Lem:EstimateOnN}, we obtain
\begin{eqnarray*}
\|{\mathcal T}_{u,z_0}(t)\|_{H^{s'}}&\leq&C\|z_0\|_{H^{s'}}
+C\e\int_0^t\left|u(v)-u_\infty\right|\left\|z\right\|_{H^{s'}}\,dv\\
&&+C\int_0^t\left(\left|u(v)\right|+\left\|\eta(v)\right\|_{H^{s'}}
\right) \left\|\eta(v)\right\|_{L^\infty}
 \left\|\eta(v)\right\|_{H^{s'}}\,dv\\
&&+C\int_0^t\left(\left|u(v)\right|+\left\|\eta(v)\right\|_{H^{s'}}
\right)\left\|\eta(v)\right\|_{L^\infty}
\left\|\eta(v)\right\|_{H^{s'}}\|\eta(v)\|_{H^{s'}}\,dv,
\end{eqnarray*}
and so
\begin{eqnarray*}
\|{\mathcal T}_{u,z_0}(t)\|_{H^{s'}}&\leq&C\|z_0\|_{H^{s'}}
+C\e\delta^3 +C\left(\e+\delta \right) \delta^2+C\left(\e+\delta
\right)^2 \delta^2.
\end{eqnarray*}
Then, we also have
\begin{multline*}
{\mathcal T}_{u,z_0}(t)=e^{-\i tH+\i\int_0^tE(u(r))\;dr}z_0\\
+\int_0^t e^{-\i(t-v)H+\i\int_v^tE(u(r))\;dr}{\mathbf P}_c(u(v))Jd^2 F(S(u(v)))z(v)\,dv\\
+\int_0^te^{-\i(t-v)H+\i\int_v^tE(u(r))\;dr}{\mathbf P}_c(u(v))JN(u(v),\eta(v))\,dv\\
+\int_0^te^{-\i(t-v)H+\i\int_v^tE(u(r))\;dr}{\mathbf
P}_c(u(v))\times\\
\times dS(u(v))A(u(v),\eta(v))
\langle N(u(v),\eta(v)) ,dS(u(v))\rangle\,dv\\
-\int_0^te^{-\i(t-v)H+\i\int_v^tE(u(r))\;dr}\times\\
\times(d {\mathbf P}_c(u(v)))A(u(v),\eta(v))\langle N(u(v),\eta(v))
,dS(u(v))\rangle \eta(v)\,dv.
\end{multline*}
Hence by Lemma~\ref{Lem:ContinousProjector} and Theorem
\ref{Thm:Dispersion}, we have
\begin{eqnarray*}
\lefteqn{ \|{\mathcal T}_{u,z_0}(t)
\|_{B^\beta_{\infty,2}}}\\
&\leq& C\langle t\rangle^{-3/2}\|z_0\|_{B^{\beta+3}_{1,2}}
+C\int_0^t \langle t-v\rangle^{-3/2}
\left\| d^2 F(S(u(v)))z(v)\right\|_{B^{\beta+3}_{1,2}}\,dv\\
&&+C\int_0^t\langle t-v\rangle^{-3/2}
\left\|N(u(v),\eta(v))\right\|_{B^{\beta+3}_{1,2}}\,dv\\
&&+C\int_0^t\langle t-v\rangle^{-3/2} \left\|dS(u(v))A(u(v),\eta(v))
\langle N(u(v),\eta(v))
,dS(u(v))\rangle\right\|_{B^{\beta+3}_{1,2}}\,dv\\
&&+C\int_0^t\langle t-v\rangle^{-3/2}\times\\
&&\qquad\times\left\|(d {\mathbf P}_c(u(v)))A(u(v),\eta(v))\langle
N(u(v),\eta(v))
 ,dS(u(v))\rangle
\eta(v)\right\|_{B^{\beta+3}_{1,2}}\,dv.
\end{eqnarray*}
With Lemma~\ref{Lem:EstimateOnN}, we infer
\begin{eqnarray*}
\lefteqn{\|{\mathcal T}_{u,z_0}(t)
\|_{B^\beta_{\infty,2}}\leq}\\
&&C\langle t\rangle^{-3/2}\|z_0\|_{B^{\beta+3}_{1,2}} +C\int_0^t
\langle t-v\rangle^{-3/2}\left|u(v)\right|^2
\left\|z(v)\right\|_{H_{-\sigma}^{\beta+3}}\,dv\\
&&+C\int_0^t\langle t-v\rangle^{-3/2}
\left(\left|u(v)\right|+\left\|\eta(v)\right\|_{H^{\beta+3}} \right)
\left\|\eta(v)\right\|_{L^\infty}
\left\|\eta(v)\right\|_{H^{\beta+3}}\,dv\\
&&+C\int_0^t\langle t-v\rangle^{-3/2}
\left(\left|u(v)\right|+\left\|\eta(v)\right\|_{H^{\beta+3}} \right)
\left\|\eta(v)\right\|_{L^\infty}
\left\|\eta(v)\right\|_{H^{\beta+3}}\,dv\\
&&+C\int_0^t\langle t-v\rangle^{-3/2}
\left(\left|u(v)\right|+\left\|\eta(v)\right\|_{H^{\beta+3}} \right)
\left\|\eta(v)\right\|_{L^\infty}
\left\|\eta(v)\right\|_{H^{\beta+3}}^2\,dv.
\end{eqnarray*}
With the estimate
\begin{equation*}
\int_0^t \langle t-v\rangle^{-3/2}\langle v\rangle^{-3/2}\,dv\leq
C\langle t\rangle^{-3/2},
\end{equation*}
we infer
\begin{eqnarray*}
\langle t\rangle^{3/2}\|{\mathcal
T}_{u,z_0}(t)\|_{B^{\beta}_{\infty,2}}
&\leq&C\|z_0\|_{B^{\beta+3}_{1,2}}+C\e^2\delta +C\left(\e+\delta
\right) \delta^2+C\left(\e+\delta \right)^2 \delta^2.
\end{eqnarray*}
Then we also have
\begin{multline*}
{\mathcal T}_{u,z_0}(t) =e^{-\i tH+\i\int_0^tE(u(r))\;dr}z_0
+\int_0^t e^{-\i(t-v)H+\i\int_v^tE(u(r))\;dr}\times\\
\times{\mathbf P}_c(u(v))J\nabla F(\eta(v))\,dv\\
+\int_0^te^{-\i(t-v)H+\i\int_v^tE(u(r))\;dr}\times\\
\times{\mathbf
P}_c(u(v))J\{\nabla F(S(u(v))+\eta(v))
-\nabla F(S(u(v))-\nabla F(\eta(v))\}\,dv\\
+\int_0^te^{-\i(t-v)H+\i\int_v^tE(u(r))\;dr}\times\\
\times{\mathbf
P}_c(u(v))dS(u(v))A(u(v),\eta(v))
\langle N(u(v),\eta(v)) ,dS(u(v))\rangle\,dv\\
-\int_0^te^{-\i(t-v)H+\i\int_v^tE(u(r))\;dr}\times\\
\times(d
{\mathbf P}_c(u(v)))A(u(v),\eta(v))\langle N(u(v),\eta(v))
,dS(u(v))\rangle \eta(v)\,dv.
\end{multline*}
We now use Lemma~\ref{Lem:ContinousProjector} and Theorem
\ref{Thm:Propagation}, except for the second term of the right hand
side for which we used Theorem~\ref{Thm:Dispersion} since
$\sigma>3/2$. We also use Lemma~\ref{Lem:EstimateOnN}, except for the
third term of the right hand side for which an obvious adaptation
of the proof of Lemma~\ref{lemma:EstimateOnN1} gives
\begin{multline*}
\left\|\nabla F(S(u(v))+\eta(v)) -\nabla F(S(u(v))-\nabla
F(\eta(v))\right\|_{H^s_\sigma}\\
\leq
C\left(\left|u(v)\right|+\left\|
\eta(v)\right\|_{H^s}\right)\left|u(v)\right|\left\|
\eta(v)\right\|_{H^s_{-\sigma}}
\end{multline*}
and so we obtain
\begin{multline*}
\langle t\rangle^{3/2}\|{\mathcal
T}_{u,z_0}(t)\|_{H^{s}_{-\sigma}}\\
\leq C\|z_0\|_{H^s_{\sigma}}+C\delta^3 +C\left(\e+\delta
\right)\delta+C\left(\e+\delta \right) \delta^2+C\left(\e+\delta
\right)^2 \delta^2.
\end{multline*}
Therefore we have that~${\mathcal T}_{u,z_0}$ leaves~${\mathcal
Z}(u,\delta)$ invariant if~$\|z_0\|_{H^{s'}_{\sigma}}$,~$\delta$ and
$\e$ are small enough.
\end{proof}

An important property of~${\mathcal T}$ is given by the
\begin{lemma}
                                                                \label{Lem:LipshitzPropertyOfT}
There exists~$\delta_0>0$ and~$\e_0>0$ such that there exists
$\kappa\in(0,1)$ such that for any~$\delta\in(0,\delta_0)$, for any
$\e\in(0,\e_0)$, for any~$u,\,u'\in {\mathcal U}(\e,\delta)$, for
any~$z_0\in {\mathcal H}_c(u(0))$, for any~$z_0'\in {\mathcal
H}_c(u'(0))$, for~$z\in {\mathcal Z}(u,\delta)$ and for any~$z'\in
{\mathcal Z}(u',\delta)$, one has
\begin{multline*}
\left\|{\mathcal T}_{u',z_0'}(z')-{\mathcal
T}_{u,z_0}(z)\right\|_{L^\infty(\R^+,H^{s'})}
\\
\leq\left\|z_0-z_0'\right\|_{L^\infty(\R^+,H^{s'})} +\kappa
\left\{\left\|u-u'\right\|_{L^\infty}
+\left\|z-z'\right\|_{L^\infty(\R^+,H^{s'})}\right\}.
\end{multline*}
\end{lemma}
\begin{proof}
It is an easy consequences of straightforward estimates on the
following identity
\begin{multline*}
{\mathcal T}_{u,z_0}(t)=e^{-\i tH(u_\infty)}z_0 \\
-\int_0^t e^{-\i(t-v)H(u_\infty)}{\mathbf
P}_c(u(v))J\left(E(S(u(v)))
-E(S(u_\infty))\right)z(v)\,dv\\
+\int_0^t e^{-\i(t-v)H(u_\infty)}{\mathbf P}_c(u(v))J\left(d^2
F(S(u(v)))
-d^2 F(S(u_\infty))\right)z(v)\,dv\\
+\int_0^te^{-\i(t-v)H(u_\infty)}{\mathbf P}_c(u(v))JN(u(v),
\eta(v))\,dv\\
+\int_0^te^{-\i(t-v)H(u_\infty)}{\mathbf
P}_c(u(v))dS(u(v))A(u(v),\eta(v))
\langle N(u(v),\eta(v)) ,dS(u(v))\rangle\,dv\\
-\int_0^te^{-\i(t-v)H(u_\infty)}(d {\mathbf
P}_c(u(v)))A(u(v),\eta(v))\langle N(u(v),\eta(v)) ,dS(u(v))\rangle
\eta(v)\,dv,
\end{multline*}
based only on Lemma~\ref{lemma:EstimateOnN1},~\ref{Lem:EstimateOnA}
and~\ref{Lem:LipshitzPropertyOfG} and on the fact that
\begin{multline*}
{\mathbf P}_c(u_\infty)\left(e^{-\i sH(u_\infty)}-e^{-\i
sH(u'_\infty)}\right){\mathbf P}_c(u(v))=\\
-{\mathbf
P}_c(u_\infty)\!\!\int_0^s\!\!\left(e^{-\i(s-s')H(u_\infty)}\!\left(E(S(u'_\infty))
-E(S(u_\infty)\right)e^{-\i
s'H(u'_\infty)}\right)\,ds'{\mathbf P}_c(u(v))\\
+{\mathbf
P}_c(u_\infty)\!\!\int_0^s\!\!\Bigg(e^{-\i(s-s')H(u_\infty)}\!\Big(d^2
F(S(u'_\infty)) \\-d^2 F(S(u_\infty))\Big)e^{-\i
s'H(u'_\infty)}\Bigg)\,ds'{\mathbf P}_c(u(v))
\end{multline*}
form a family of operator in~${\mathcal
B}(H^{s'}(\R^3,\C^8),H^{s'}(\R^3,\C^8))$ (to this end we use Lemma
\ref{Lem:SmoothSmallNonSelfAdjointOperator}) such that
\begin{equation*}
\left\|{\mathbf P}_c(u_\infty)\left(e^{-\i sH(u_\infty)}-e^{-\i
sH(u'_\infty)}\right){\mathbf P}_c(u(v))\right\| \leq C\e
\left|u'_\infty-u_\infty\right|,
\end{equation*}
with~$C$ independent of~$u,u'$.
\end{proof}

\begin{lemma}
                                                                \label{Lem:GlobaLWellPosednessForZ}
There exists~$\delta>0$ and~$\e>0$ such that for any~$u\in {\mathcal
U}(\delta,\e)$ there is a solution~$z\in {\mathcal Z}(u,\delta)$ of
the equation
\begin{equation}
                                                                \label{Eq:ForZ}
\left\{\begin{array}{lll}
\p_t z&=&JH(u)z+{\mathbf P}_c(u)JN(u,\eta)\\
&&\qquad+ {\mathbf P}_c(u)dS(u)A(u,\eta)\langle N(u,\eta),
dS(u)\rangle\\
&&\quad\qquad-(D {\mathbf P}_c(u))A(u,\eta)\langle N(u,\eta),
dS(u)\rangle \eta,\\
z(0)&=&z_0,
\end{array}\right.
\end{equation}
where~$\eta(t)=\alpha^+(t)S_1^+(u)+\alpha^-(t)S_1^-(u)+z(t)$,
whenever~$z_0\in H^{s'}_{\sigma}$ is small enough.
\end{lemma}
\begin{proof}
It is a consequence of the fix point theorem applied to~${\mathcal
T}_{u,z_0}$.
\end{proof}

\begin{lemma}
                                                                \label{Lem:OnNonLinearScattering}
Under the assumptions of Lemma~\ref{Lem:GlobaLWellPosednessForZ},
for any~$u\in {\mathcal U}(\delta,\e)$ and solution~$z$ of
\eqref{Eq:ForZ}, with~$z_0\in H^{s'}_{\sigma}$ small, the following
limit
\begin{equation*}
z_\infty=\lim_{t\to\infty}e^{\i t H-\i\int_0^tE(u(r))\;dr}z(t)
\end{equation*}
exists in~$H^{s'}\cap B^\beta_{\infty,2}\cap H^s_{-\sigma}$.
Moreover, we have $z_\infty\in {\cal H}_c(0)$ and
\begin{eqnarray*}
\|e^{-\i tH+\i\int_0^tE(u(r))\;dr}z_\infty-z(t)\|_{H^{s'}}&\leq&  C\delta^2,\\
\|e^{-\i
tH+\i\int_0^tE(u(r))\;dr}z_\infty-z(t)\|_{B^\beta_{\infty,2}}
&\leq&C\frac{\delta^2}{\langle t\rangle^{2}},\\
\|e^{-\i tH+\i\int_0^tE(u(r))\;dr}z_\infty-z(t)\|_{
H^s_{-\sigma}}&\leq&
 C\frac{\delta^2}{\langle t\rangle^{2}}.
\end{eqnarray*}
\end{lemma}
\begin{proof}
Using exactly the same method as the one of Lemma
\ref{Lem:StabilizationForT}, applied to

\begin{multline*}
e^{\i tH-\i\int_0^tE(u(r))\;dr }z(t)=z_0\\
+\int_0^te^{\i v H-\i\int_0^vE(u(r))\;dr}{\mathbf P}_c(u(v))JN(u(v),
\eta(v))\,dv\\
+\int_0^te^{\i v H-\i\int_0^vE(u(r))\;dr}\times\\
\times{\mathbf P}_c(u(v))dS(u(v))A(u(v),\eta(v))
\langle N(u(v),\eta(v)) ,dS(u(v))\rangle\,dv\\
-\int_0^te^{\i v H-\i\int_0^v E(u(r))\;dr}\times\\
\times(d {\mathbf P}_c(u(v)))A(u(v),\eta(v))\langle N(u(v),\eta(v))
,dS(u(v))\rangle \eta(v)\,dv,
\end{multline*}
we prove that the limit exist by the same way we also obtain the
convergence rate. Since $e^{-itH}z_\infty$ tends to zero, $z_\infty$
necessarily belongs to ${\mathcal H}_c(0)$.
\end{proof}

\begin{remark}
The preceding proof also work with the formula
\begin{multline*}
e^{\i tD_m-\i\int_0^tE(u(r))\;dr}z(t)=z_0 +\int_0^te^{\i v
D_m-\i\int_0^vE(u(r))\;dr}{\mathbf
P}_c(u(v))Vz(v)\,dv\\
+\int_0^te^{\i v D_m-\i\int_0^vE(u(r))\;dr}{\mathbf
P}_c(u(v))JN(u(v),
\eta(v))\,dv\\
+\int_0^te^{\i v D_m-\i\int_0^vE(u(r))\;dr}\times\\
\times{\mathbf P}_c(u(v))dS(u(v))A(u(v),\eta(v))
\langle N(u(v),\eta(v)) ,dS(u(v))\rangle\,dv\\
-\int_0^te^{\i v D_m-\i\int_0^vE(u(r))\;dr}\times\\
\times(d {\mathbf P}_c(u(v)))A(u(v),\eta(v))\langle N(u(v),\eta(v))
,dS(u(v))\rangle \eta(v)\,dv.
\end{multline*}
Hence we obtain the same result with $$e^{-\i
tD_m+\i\int_0^tE(u(r))\;dr}\widetilde{z}_\infty$$ instead of
$$e^{-\i t H+\i\int_0^tE(u(r))\;dr}z_\infty.$$ But we obtain the
estimates
\begin{eqnarray*}
\|e^{-\i tD_m+\i\int_0^tE(u(r))\;dr}\widetilde{z}_\infty-z(t)\|_{H^{s'}}&\leq&  C\delta,\\
\|e^{-\i
tD_m+\i\int_0^tE(u(r))\;dr}\widetilde{z}_\infty-z(t)\|_{B^\beta_{\infty,2}}&\leq&
C\frac{\delta}{\langle t\rangle^{2}},\\
\|e^{-\i tD_m+\i\int_0^tE(u(r))\;dr}\widetilde{z}_\infty-z(t)\|_{
H^s_{-\sigma}}&\leq&
 C\frac{\delta}{\langle t\rangle^{2}}.
\end{eqnarray*}

\end{remark}
\subsection{Step 3: Construction of~$u$}

Here we want to solve the equation for~$u$. We notice that~$z$ and
$\alpha$ have been built in the previous section and are functions
of~$u$ and~$z_0\in {\mathcal H}_c(u(0))$. Let us introduce for any
$\alpha \in \Omega(\delta)$ and~$u_0\in B_\C(0,\e)$ the function on
${\mathcal U}(\e,\delta)$:
\begin{equation*}
f_{u_0}(u)(t)=u_0-\int_0^tA(u(v),\eta(v))\langle N(u(v),\eta(v))
,dS(u(v))\rangle\,dv,
\end{equation*}
where~$\eta(t)=\alpha^+(t)S_1^+(u)+\alpha^-(t)S_1^-(u)+z(t)$. We
have the
\begin{lemma}
There exists~$\delta_0>0$ and~$\e_0>0$ such that for any
$\delta\in(0,\delta_0)$, for any~$\e\in(0,\e_0)$, the function
$f_{u_0}$ maps~${\mathcal U}(\e,\delta)$ into itself if~$u_0$ and
$z_0\in H^{s'}_\sigma\cap {\mathcal H}_c(u_0)$ are small enough.
\end{lemma}
\begin{proof}
By means of Lemma~\ref{Lem:EstimateOnN}, we obtain
\begin{equation*}
\left|f_{u_0}(u)(t)\right|\leq
\left|u_0\right|+C\int_0^t\left\|N(u(v),\eta(v))\right\|_{H^s_{-\sigma}}
\leq \left|u_0\right|+C\left(\e+\delta \right) \delta^2.
\end{equation*}
Hence for~$u_0$ and~$\delta$ small~$f_{u_0}(u)(t)\in B_\C(0,\e)$.
Estimate~\eqref{Estimate:OnN2-1} also gives the existence of
$\left(f_{u_0}(u)\right)_\infty=\ds\lim_{t\to+\infty}f_{u_0}(u)(t)$
and then
\begin{equation*}
\left|\left(f_{u_0}(u)\right)_\infty-f_{u_0}(u)(t)\right|\leq
C\int_t^{+\infty}\left\|N(u(v),\eta(v))\right\|_{H^s_{-\sigma}} \leq
\frac{C}{t^2}\left(\e+\delta \right) \delta^2.
\end{equation*}
\end{proof}

The function~$f_{u_0}$ has also a Lipshitz property as stated by the
\begin{lemma}
There exists~$\delta_0>0$ and~$\e_0>0$ such that there exists
$\kappa\in(0,1)$ such that for any~$\delta\in(0,\delta_0)$, for any
$\e\in(0,\e_0)$, for any~$u,\,u'\in {\mathcal U}(\e,\delta)$, for
any~$z_0\in {\mathcal H}_c(u(0))\cap H^{s'}_\sigma$, for any
$z_0'\in {\mathcal H}_c(u'(0))\cap H^{s'}_\sigma$ small enough, for
$u_0,u_0'$ small enough, there exists~$\kappa\in(0,1)$ such that
\begin{equation*}
\left|f_{u_0}(u)-f_{u_0'}(u')\right|_{L^\infty(\R^+)}\leq
\left|u_0-u_0'\right|+
\kappa\left(\left\|u-u'\right\|_{L^\infty(\R^+)}+\|z_0-z_0'\|_{H^{s'}_\sigma}\right).
\end{equation*}
\end{lemma}
\begin{proof}
This a straightforward consequence of
Lemma~\ref{lemma:EstimateOnN1},~\ref{Lem:EstimateOnA},~\ref{Lem:LipshitzPropertyOfG}
and~\ref{Lem:LipshitzPropertyOfT}.
\end{proof}

We are now able to prove the
\begin{lemma}
There exists~$\delta_0>0$ such that for any~$\delta\in(0,\delta_0)$
such that for any~$u_0\in \C$ small and~$z_0\in {\mathcal
H}_c(u_0)\cap H^{s'}_\sigma$ small, the equation
\begin{eqnarray*}
\left\{\begin{array}{lll}
\dot{u}&=&-A(u,\eta)\langle N(u,\eta) ,dS(u)\rangle,\\
u(0)&=&u_0,
\end{array}\right.
\end{eqnarray*}
where~$\eta(t)=\alpha^+(t)S_1^+(u)+\alpha^-(t)S_1^-(u)+z(t)$, has a
unique solution in~${\mathcal U}(\delta,\e)$.
\end{lemma}
\begin{proof}
This is also a straightforward consequence of the fixed point
theorem for~$f_{u_0}$.
\end{proof}

\subsection{Step 4: End of the proof of Theorem 
\ref{Thm:StabilizationSmallPLSNR}}
We now conclude our proof with the
\begin{lemma}[Decomposition lemma 2]
Let be~$s\geq 0$ and~$p\geq 1$ there exist~$\delta > 0$ and a
${\mathcal C}^\infty$ map~$U_0 : B_{W^{s,p}}(0,\delta)\mapsto
B_\C(0,\e)~$ which satisfies for~$\psi\in B_{W^{s,p}}(0,\delta)$
\begin{equation*}
\psi=S(u)+\eta, \mbox{ with } \eta \in \{\phi_0\}^{\bot}
\Longleftrightarrow u=U_0(\psi)
\end{equation*}
\end{lemma}
\begin{proof}
In fact, we just write $\psi$ with respect to the spectral
decomposition of $H$:
\begin{equation*}
    \psi=u\phi_0+r=S(u)+\eta
\end{equation*}
where $r\in \{\phi_0\}^{\bot}$ and $\eta=r-h(u)$, with $h$
defined in Proposition \ref{Prop:ManifoldPLS}.
\end{proof}

With respect to the notation of
Theorem~\ref{Thm:StabilizationSmallPLSNR} for
$\psi_0=S(u_0)+\alpha^+(0)S_1^+(u_0)+\alpha^-(0)S_1^-(u_0)+{\mathbf
P}_c(u_0)\widetilde{z_0}$ where~$\widetilde{z_0}\in {\rm
Ran}({\mathbf P}_c)$, we introduce
\begin{equation*}
\begin{cases}
v_0=U_0(\psi_0)\\
\xi_0={\mathbf P}_c\left(\psi_0-S(v_0)\right)
\end{cases}
\end{equation*}
and
\begin{equation*}
\begin{cases}
G(u_0,\widetilde{z_0})_1=U_0(S(u_0)+\alpha^+(0)S_1^+(u_0)
+\alpha^-(0)S_1^-(u_0)+{\mathbf
P}_c(u_0)\widetilde{z_0})\\
G(u_0,\widetilde{z_0})_2={\mathbf
P}_c\Big(S(u_0)-S(G(u_0,\widetilde{z_0})_1)\\
\qquad\qquad\qquad\qquad+\alpha^+(0)S_1^+(u_0)
+\alpha^-(0)S_1^-(u_0)+{\mathbf P}_c(u_0)\widetilde{z_0}\Big)
\end{cases}
\end{equation*}
Then using~$U_0(S(u_0))=u_0$, we write
$G(u_0,\widetilde{z_0})=(u_0,\widetilde{z_0})+\widetilde{G}(u_0,\widetilde{z_0})$,
with
\begin{equation*}
\left\|\widetilde{G}(u_0,\widetilde{z_0})-\widetilde{G}(u'_0,\widetilde{z_0}')\right\|_{H^{s'}_\sigma}\leq
\kappa\left(\left|u_0-u'_0\right|
+\left\|\widetilde{z_0}-\widetilde{z_0}'\right\|_{H^{s'}_\sigma}\right)
\end{equation*}
with~$\kappa\leq 1/2$ if~$u_0,u'_0$ and
$\widetilde{z_0},\widetilde{z_0}'$ small enough. Hence in this case
$G$ is invertible with a Lipshitz inverse F. Then we choose
\begin{eqnarray*}
\lefteqn{\Psi(v_0,\xi_0)}\\
&&=\big\langle
S(F(v_0,\xi_0)_1)\\
&&\!\!+\alpha(F(v_0,\xi_0)_1)^+(0)S_1^+(F(v_0,\xi_0)_1)
+\alpha(F(v_0,\xi_0)_1)^-(0)S_1^-(F(v_0,\xi_0)_1)\\
&&+{\mathbf
P}_c(F(v_0,\xi_0)_1)F(v_0,\xi_0)_2-S(v_0),\phi_1\big\rangle
\end{eqnarray*}
and
\begin{equation*}
\xi_\infty =\left({\mathbf
P}_c(F(v_0,\xi_0)_1)F(v_0,\xi_0)_2\right)_\infty
\end{equation*}
and
\begin{equation*}
E_\infty=\int_0^\infty\left\{E(F(v_0,\xi_0)_1(v))
-E\left(\left(F(v_0,\xi_0)_1\right)_\infty\right)\right\}\;dv.
\end{equation*}
In the proof of Lemma \ref{Lem:StabilizationForT}, we see that
$\delta$ is of the same order as
$\left\|\xi_0\right\|_{H^{s'}_{\sigma}}$. The rest of the Theorem
easily follows.

\hfill$\qed$\\

\appendix
\begin{center}
{\bf\Large APPENDICES}
\end{center}

\section{The wave operator and similarity for the linearized 
operator}

Inspired by~\cite{Kato}, we use an argument of similarity to prove
the
\begin{lemma}
                                                                \label{Lem:PerturbedLawConservation}
For all~$s \in \R^+$, there exists~$C_s>0$, such that
\begin{equation*}
\forall t \in\R,\;\|e^{tJH(z)}\|_{{\mathcal L}(H^s)}\leq C_s.
\end{equation*}
\end{lemma}

We prove this lemma by using the boundedness in $H^s$ of the wave operator:
\begin{equation*}
\ds W_\pm= s-\lim_{t\rightarrow \pm \infty} e^{-t H(z)^*J}e^{-\i
t(H-E(z))}{\mathbf P}_c(H)
\end{equation*}
and the intertwining property:
\begin{equation*}
    e^{-tH(z)^*J}{\mathbf P}_c(z)^*=W^\pm e^{-\i
t(H-E(z))}{\mathbf P}_c(H)(W^\pm)^{-1} .
\end{equation*}
 This boundedness follows  from the
\begin{lemma}[Smooth and small non-selfadjoint perturbations]
                                                                \label{Lem:SmoothSmallNonSelfAdjointOperator}
Let be $\psi \in L^2$ and~$\sigma\geq1$. Then there exists~$\e>0$ and
$C>0$ such that
\begin{equation}
                                                                \label{Claim:PerturbationSmoothness}
\forall z\in B_\C(0,\e),\,
\int_0^\infty\|<Q>^{-\sigma}e^{s JH(z)}{\mathbf P}_c(z)\psi
\|_2^2\;ds\leq C \|\psi\|_2^2.
\end{equation}
\end{lemma}
\begin{proof}
By Lemma~\eqref{Lem:ContinousProjector}, we prove~${\mathbf
P}_c(z)R(z,0){\mathbf P}_c(0){\mathbf P}_c(z)={\mathbf P}_c(z)$.
Taking the adjoint with respect to the real structure, we infer
${\mathbf P}_c(z){\mathbf P}_c(0)R(z)^{*}{\mathbf P}_c(z)={\mathbf
P}_c(z)$. Then, we write
\begin{eqnarray*}
\lefteqn{\|\langle Q\rangle^{-\sigma} e^{tJH(z)}{\mathbf P}_c(z)\|}\\
&=& C\|\langle Q\rangle^{-\sigma}{\mathbf P}_c(z)e^{tJH(z)}
{\mathbf P}_c(0)R(z)^{*}{\mathbf P}_c(z)\|\\
&\leq& \|\langle Q\rangle^{-\sigma} {\mathbf P}_c(z)e^{-\i
t(H-E(z))}
{\mathbf P}_c(0)R(z)^{*}{\mathbf P}_c(z)\|\\
&&\!\!\!\!+\!\!\int_0^t\!\!\! \|\langle Q\rangle^{-\sigma} {\mathbf
P}_c(z)e^{(t-s)JH(z)}\!D\nabla F(S(z))e^{-\i s (H-E(z))}
{\mathbf P}_c(0)R(z){\mathbf P}_c(z)\|\,ds\\
&\leq& \|\langle Q\rangle^{-\sigma} {\mathbf P}_c(z)e^{-\i t
(H-E(z))}
{\mathbf P}_c(0)R(z){\mathbf P}_c(z)\|\\
&&+ C|z|^{2}\int_0^t \|\langle Q\rangle^{-\sigma} {\mathbf
P}_c(z)e^{(t-s)JH(z)}\langle Q\rangle^{-\sigma}\|\|\langle Q\rangle^{-\sigma}e^{-\i s
(H-E(z))}{\mathbf P}_c(0)\|\,ds.
\end{eqnarray*}
Using Proposition~\ref{proposition:LAP}, we obtain the claim
\eqref{Claim:PerturbationSmoothness} for~$z$ sufficiently small.
\end{proof}

This give us the existence and the boundedness of the wave operator, as stated by the following
\begin{lemma}
Let be~$W_t=e^{-tH(z)^*J}e^{-it(H-E(z))}{\mathbf P}_c(H)$. Then the
limits
\begin{equation*}
W^\pm =\lim_{t\to\pm \infty} W_t
\end{equation*}
exist in~$B(H^s)$ and their range is~${\rm Ran}\left({\mathbf
P}_c(z)\right)$. The same is true for~$W_t^*$ and
\begin{equation*}
\left(W^\pm\right)^{-1} =\lim_{t\to\pm \infty}
\left(W_t\right)^{-1} .
\end{equation*}
\end{lemma}
\begin{proof}
Let us define~$W_t=e^{-tH(z)^*J}e^{-it(H-E(z))}$. we have for~$\phi
\in H_c(z)$ and $\psi \in H_c(0)$
\begin{equation*}
\left\langle \phi, W_t \psi\right\rangle= \left\langle\phi, \psi
\right\rangle +\int_0^t\left\langle \phi, \frac{d}{ds}W_s
\psi\right\rangle\,ds,
\end{equation*}
Since we have
\begin{eqnarray*}
\left\langle \phi, \frac{d}{ds}W_s \psi\right\rangle&=&\left\langle
e^{-tJH(z)}\phi,D\nabla F(S(z))e^{-\i t(H-E(z))}\psi \right\rangle\\
& \leq & C|z|^2 \|\langle
Q\rangle^{-\sigma}e^{tJH(z)}\phi\|\|\langle Q\rangle^{-\sigma}e^{-\i
t(H-E(z))}\psi\|.
\end{eqnarray*}
which gives~$\langle \phi, \frac{d}{ds}W_s \psi\rangle \in L^1(\R)$,
so~$W_\pm$ exists and is bounded in~${\mathcal L}(H_c(0),H_c(z))$ by
the previous lemma. Since for any vector~$\phi$ in an eigenspace
of~$JH(z)$,~$W_t^*\phi$ tends weakly to zero, we obtain that the range
of~$W^\pm$ is a subspace of the range of~${\mathbf P}_c(z)$. Then the same
statements about~$\left(W_t\right)^{-1}$ follows by the same way. The
invertibility is then immediate.
\end{proof}
\begin{proof}[of Lemma \ref{Lem:PerturbedLawConservation}]
The $L^2$ bound follows from the intertwining property as explained
before Lemma \ref{Lem:SmoothSmallNonSelfAdjointOperator}.

The proof of the~$H^k$ bounds follows from commutation argument, we
apply the same scheme to
\begin{eqnarray*}
\p_i e^{-tJH(z)}{\mathbf P}_c(z)&=& [\p_i,{\mathbf P}_c(z)]
e^{-tJH(z)}{\mathbf P}_c(z)+{\mathbf P}_c(z)e^{-tJH(z)}
[\p_i,{\mathbf P}_c(z)]\\
&&+{\mathbf P}_c(z)[\p_i, e^{-tJH(z)}]{\mathbf P}_c(z)\\
& = &[\p_i,{\mathbf P}_c(z)]e^{-tJH(z)}\p_i
+{\mathbf P}_c(z)e^{-tJH(z)}[\p_i,{\mathbf P}_c(z)]\\
&&+\int_0^te^{-(t-s)JH(z)}{\mathbf P}_c(z)(\p_i D\nabla F(S))
e^{-sJH(z)}{\mathbf P}_c(z) dz.
\end{eqnarray*}
\end{proof}


\bibliographystyle{alpha}
\bibliography{biblio}

\def\cprime{$'$} \def\cprime{$'$} \def\cprime{$'$} \def\cprime{$'$}
  \def\cprime{$'$}
\begin{thebibliography}{ABdMG96}

\bibitem[ABdMG96]{AmreinBoutetdeMonvelGeorgescu}
W.~O. Amrein, A.~Boutet~de Monvel, and V.~Georgescu.
\newblock {\em {$C\sb 0$}-groups, commutator methods and spectral theory of
  {$N$}-body {H}amiltonians}, volume 135 of {\em Progress in Mathematics}.
\newblock Birkh\"auser Verlag, Basel, 1996.

\bibitem[Agm75]{Agmon}
S.~Agmon.
\newblock Spectral properties of {S}chr\"odinger operators and scattering
  theory.
\newblock {\em Ann. Scuola Norm. Sup. Pisa Cl. Sci. (4)}, 2(2):151--218, 1975.

\bibitem[AS86]{AlvarezSoler}
A.~{Alvarez} and M.~{Soler}.
\newblock {Stability of the minimum solitary wave of a nonlinear spinorial
  model}.
\newblock {\em Phys. Rev D}, 34:644--645, July 1986.

\bibitem[BdMGS96]{BoutetdeMonvelGeorgescuSahbani}
A.~Boutet~de Monvel, V.~Georgescu, and J.~Sahbani.
\newblock Boundary values of resolvent families and propagation properties.
\newblock {\em C. R. Acad. Sci. Paris S\'er. I Math.}, 322(3):289--294, 1996.

\bibitem[BH92]{BalslevHelffer}
E.~Balslev and B.~Helffer.
\newblock Limiting absorption principle and resonances for the {D}irac
  operator.
\newblock {\em Adv. in Appl. Math.}, 13(2):186--215, 1992.

\bibitem[BP92a]{BuslaevPerelman4}
V.~S. Buslaev and G.~S. Perel{\cprime}man.
\newblock Nonlinear scattering: states that are close to a soliton.
\newblock {\em Zap. Nauchn. Sem. S.-Peterburg. Otdel. Mat. Inst. Steklov.
  (POMI)}, 200(Kraev. Zadachi Mat. Fiz. Smezh. Voprosy Teor. Funktsii.
  24):38--50, 70, 187, 1992.

\bibitem[BP92b]{BuslaevPerelman2}
V.~S. Buslaev and G.~S. Perel{\cprime}man.
\newblock On nonlinear scattering of states which are close to a soliton.
\newblock {\em Ast\'erisque}, (210):6, 49--63, 1992.
\newblock M\'ethodes semi-classiques, Vol.\ 2 (Nantes, 1991).

\bibitem[BP92c]{BuslaevPerelman3}
V.~S. Buslaev and G.~S. Perel{\cprime}man.
\newblock Scattering for the nonlinear {S}chr\"odinger equation: states that
  are close to a soliton.
\newblock {\em Algebra i Analiz}, 4(6):63--102, 1992.

\bibitem[BP95]{BuslaevPerelman}
V.~S. Buslaev and G.~S. Perel{\cprime}man.
\newblock On the stability of solitary waves for nonlinear {S}chr\"odinger
  equations.
\newblock In {\em Nonlinear evolution equations}, volume 164 of {\em Amer.
  Math. Soc. Transl. Ser. 2}, pages 75--98. Amer. Math. Soc., Providence, RI,
  1995.

\bibitem[Bre77]{Brenner2}
P.~Brenner.
\newblock {$L\sb{p}$}-estimates of difference schemes for strictly hyperbolic
  systems with nonsmooth data.
\newblock {\em SIAM J. Numer. Anal.}, 14(6):1126--1144, 1977.

\bibitem[Bre85]{Brenner}
P.~Brenner.
\newblock On scattering and everywhere defined scattering operators for
  nonlinear {K}lein-{G}ordon equations.
\newblock {\em J. Differential Equations}, 56(3):310--344, 1985.

\bibitem[BS02]{BuslaevSulem2}
V.~S. Buslaev and C.~Sulem.
\newblock Asymptotic stability of solitary waves for nonlinear {S}chr\"odinger
  equations.
\newblock In {\em The legacy of the inverse scattering transform in applied
  mathematics (South Hadley, MA, 2001)}, volume 301 of {\em Contemp. Math.},
  pages 163--181. Amer. Math. Soc., Providence, RI, 2002.

\bibitem[BS03]{BuslaevSulem}
V.~S. Buslaev and C.~Sulem.
\newblock On asymptotic stability of solitary waves for nonlinear
  {S}chr\"odinger equations.
\newblock {\em Ann. Inst. H. Poincar\'e Anal. Non Lin\'eaire}, 20(3):419--475,
  2003.

\bibitem[BSV87]{BlanchardStubbeVazquez}
P.~{Blanchard}, J.~{Stubbe}, and L.~{V{\`a}zquez}.
\newblock {Stability of nonlinear spinor fields with application to the
  Gross-Neveu model}.
\newblock {\em Phys. Rev. D}, 36:2422--2428, October 1987.

\bibitem[CF01]{CidFelmer}
C.~Cid and P.~Felmer.
\newblock Orbital stability and standing waves for the nonlinear
  {S}chr\"odinger equation with potential.
\newblock {\em Rev. Math. Phys.}, 13(12):1529--1546, 2001.

\bibitem[CL82]{CazenaveLions}
T.~Cazenave and P.-L. Lions.
\newblock Orbital stability of standing waves for some nonlinear
  {S}chr\"odinger equations.
\newblock {\em Comm. Math. Phys.}, 85(4):549--561, 1982.

\bibitem[CS01]{CuccagnaSchirmer}
S.~Cuccagna and P.~P. Schirmer.
\newblock On the wave equation with a magnetic potential.
\newblock {\em Comm. Pure Appl. Math.}, 54(2):135--152, 2001.

\bibitem[Cuc01]{Cuccagna2}
S.~Cuccagna.
\newblock Stabilization of solutions to nonlinear {S}chr\"odinger equations.
\newblock {\em Comm. Pure Appl. Math.}, 54(9):1110--1145, 2001.

\bibitem[Cuc03]{Cuccagna3}
S.~Cuccagna.
\newblock On asymptotic stability of ground states of {NLS}.
\newblock {\em Rev. Math. Phys.}, 15(8):877--903, 2003.

\bibitem[Cuc05]{Cuccagna3Err}
S.~Cuccagna.
\newblock Erratum: ``{S}tabilization of solutions to nonlinear {S}chr\"odinger
  equations'' [{C}omm. {P}ure {A}ppl. {M}ath. {\bf 54} (2001), no. 9,
  1110--1145; mr1835384].
\newblock {\em Comm. Pure Appl. Math.}, 58(1):147, 2005.

\bibitem[DF]{D'AnconnaFanelli}
P.~D'Ancona and L.~Fanelli.
\newblock Decay estimates for the wave and dirac equations with a magnetic
  potential.
\newblock To appear on Comm.Pure Appl.Math.

\bibitem[ES04]{ErdoganSchlag}
M.~B. Erdo{\u{g}}an and W.~Schlag.
\newblock Dispersive estimates for {S}chr\"odinger operators in the presence of
  a resonance and/or an eigenvalue at zero energy in dimension three. {I}.
\newblock {\em Dyn. Partial Differ. Equ.}, 1(4):359--379, 2004.

\bibitem[EV97]{EscobedoVega}
M.~Escobedo and L.~Vega.
\newblock A semilinear {D}irac equation in {$H\sp s({\bf R}\sp 3)$} for
  {$s>1$}.
\newblock {\em SIAM J. Math. Anal.}, 28(2):338--362, 1997.

\bibitem[FS04]{FournaisSkibsted}
S.~Fournais and E.~Skibsted.
\newblock Zero energy asymptotics of the resolvent for a class of slowly
  decaying potentials.
\newblock {\em Math. Z.}, 248(3):593--633, 2004.

\bibitem[GM01]{GeorgescuMantoiu}
V.~Georgescu and M.~M{\u{a}}ntoiu.
\newblock On the spectral theory of singular {D}irac type {H}amiltonians.
\newblock {\em J. Operator Theory}, 46(2):289--321, 2001.

\bibitem[GNT04]{GustafsonNakanishiTsai}
S.~Gustafson, K.~Nakanishi, and T.-P. Tsai.
\newblock Asymptotic stability and completeness in the energy space for
  nonlinear {S}chr\"odinger equations with small solitary waves.
\newblock {\em Int. Math. Res. Not.}, (66):3559--3584, 2004.

\bibitem[GS04]{GoldbergSchlag}
M.~Goldberg and W.~Schlag.
\newblock A limiting absorption principle for the three-dimensional
  {S}chr\"odinger equation with {$L\sp p$} potentials.
\newblock {\em Int. Math. Res. Not.}, (75):4049--4071, 2004.

\bibitem[GSS87]{GrillakisShatahStrauss}
M.~Grillakis, J.~Shatah, and W.~Strauss.
\newblock Stability theory of solitary waves in the presence of symmetry. {I}.
\newblock {\em J. Funct. Anal.}, 74(1):160--197, 1987.

\bibitem[His00]{Hislop}
P.~D. Hislop.
\newblock Exponential decay of two-body eigenfunctions: a review.
\newblock In {\em Proceedings of the Symposium on Mathematical Physics and
  Quantum Field Theory (Berkeley, CA, 1999)}, volume~4 of {\em Electron. J.
  Differ. Equ. Conf.}, pages 265--288 (electronic), San Marcos, TX, 2000.
  Southwest Texas State Univ.

\bibitem[HS00]{HunzikerSigal}
W.~Hunziker and I.~M. Sigal.
\newblock Time-dependent scattering theory of {$N$}-body quantum systems.
\newblock {\em Rev. Math. Phys.}, 12(8):1033--1084, 2000.

\bibitem[HSS99]{HunzikerSigalSoffer}
W.~Hunziker, I.~M. Sigal, and A.~Soffer.
\newblock Minimal escape velocities.
\newblock {\em Comm. Partial Differential Equations}, 24(11-12):2279--2295,
  1999.

\bibitem[IM99]{IftimoviciMantoiu}
A.~Iftimovici and M.~M{\u{a}}ntoiu.
\newblock Limiting absorption principle at critical values for the {D}irac
  operator.
\newblock {\em Lett. Math. Phys.}, 49(3):235--243, 1999.

\bibitem[JK79]{JensenKato}
A.~Jensen and T.~Kato.
\newblock Spectral properties of {S}chr\"odinger operators and time-decay of
  the wave functions.
\newblock {\em Duke Math. J.}, 46(3):583--611, 1979.

\bibitem[JN01]{JensenNenciu}
A.~Jensen and G.~Nenciu.
\newblock A unified approach to resolvent expansions at thresholds.
\newblock {\em Rev. Math. Phys.}, 13(6):717--754, 2001.

\bibitem[JN04]{JensenNenciuErr}
A.~Jensen and G.~Nenciu.
\newblock Erratum: ``{A} unified approach to resolvent expansions at
  thresholds'' [{R}ev. {M}ath. {P}hys. {\bf 13} (2001), no. 6, 717--754; mr
  1841744].
\newblock {\em Rev. Math. Phys.}, 16(5):675--677, 2004.

\bibitem[JSS91]{JourneSofferSogge}
J.-L. Journ{\'e}, A.~Soffer, and C.~D. Sogge.
\newblock Decay estimates for {S}chr\"odinger operators.
\newblock {\em Comm. Pure Appl. Math.}, 44(5):573--604, 1991.

\bibitem[Kat66]{Kato}
T.~Kato.
\newblock Wave operators and similarity for some non-selfadjoint operators.
\newblock {\em Math. Ann.}, 162:258--279, 1965/1966.

\bibitem[KS05]{KriegerSchlag}
J.~Krieger and W.~Schlag.
\newblock Stable manifolds for all supercritical monic nls in one dimension.
\newblock preprint, 2005.

\bibitem[MSW79]{MarshallStraussWaingner2}
B.~Marshall, W.~Strauss, and S.~Wainger.
\newblock Estimates from {$L\sp{p}$} to its dual for the {K}lein-{G}ordon
  equation.
\newblock In {\em Harmonic analysis in Euclidean spaces (Proc. Sympos. Pure
  Math., Williams Coll., Williamstown, Mass., 1978), Part 2}, Proc. Sympos.
  Pure Math., XXXV, Part, pages 175--177. Amer. Math. Soc., Providence, R.I.,
  1979.

\bibitem[MSW80]{MarshallStraussWaingner}
B.~Marshall, W.~Strauss, and S.~Wainger.
\newblock {$L\sp{p}-L\sp{q}$} estimates for the {K}lein-{G}ordon equation.
\newblock {\em J. Math. Pures Appl. (9)}, 59(4):417--440, 1980.

\bibitem[Par90]{Parisse}
B.~Parisse.
\newblock R\'esonances paires pour l'op\'erateur de {D}irac.
\newblock {\em C. R. Acad. Sci. Paris S\'er. I Math.}, 310(5):265--268, 1990.

\bibitem[PW97]{PilletWayne}
C.-A. Pillet and C.~E. Wayne.
\newblock Invariant manifolds for a class of dispersive, {H}amiltonian, partial
  differential equations.
\newblock {\em J. Differential Equations}, 141(2):310--326, 1997.

\bibitem[Ran]{Ranada}
A.~F Ranada.
\newblock Classical nonlinear dirac field models of extended particles.
\newblock In {\em Quantum theory, groups, fields and particles,}, volume 198,
  pages 271--291. A. O. Barut, Amsterdam, Reidel.

\bibitem[RS78]{ReedSimon4}
M.~Reed and B.~Simon.
\newblock {\em Methods of modern mathematical physics. {IV}. {A}nalysis of
  operators}.
\newblock Academic Press [Harcourt Brace Jovanovich Publishers], New York,
  1978.

\bibitem[RS79]{ReedSimon3}
M.~Reed and B.~Simon.
\newblock {\em Methods of modern mathematical physics. {III}}.
\newblock Academic Press [Harcourt Brace Jovanovich Publishers], New York,
  1979.
\newblock Scattering theory.

\bibitem[RSS05a]{RodnianskiSchlagSoffer}
I.~Rodnianski, W.~Schlag, and A.~Soffer.
\newblock Asymptotic stability of n-soliton states of nls.
\newblock to appear in Comm. Pure and Appl. Math, 2005.

\bibitem[RSS05b]{RodnianskiSchlagSoffer2}
I.~Rodnianski, W.~Schlag, and A.~Soffer.
\newblock Dispersive analysis of charge transfer models.
\newblock {\em Comm. Pure Appl. Math.}, 58(2):149--216, 2005.

\bibitem[Sch04]{Schlag}
W.~Schlag.
\newblock Stable manifolds for an orbitally unstable nls.
\newblock preprint, 2004.

\bibitem[Sch05]{Schlag2}
W.~Schlag.
\newblock Dispersive estimates for schroedinger operators: a survey.
\newblock preprint, 2005.

\bibitem[SS85]{ShatahStrauss}
J.~Shatah and W.~Strauss.
\newblock Instability of nonlinear bound states.
\newblock {\em Comm. Math. Phys.}, 100(2):173--190, 1985.

\bibitem[SS98]{SigalSoffer}
I.M. Sigal and A.~Soffer.
\newblock Local decay and velocity bounds for quantum propagation.
\newblock preprint, 1998.

\bibitem[SV86]{StraussVazquez}
W.~A. {Strauss} and L.~{V{\'a}zquez}.
\newblock {Stability under dilations of nonlinear spinor fields}.
\newblock {\em Phys. Rev. D}, 34:641--643, July 1986.

\bibitem[SW90]{SofferWeinstein}
A.~Soffer and M.~I. Weinstein.
\newblock Multichannel nonlinear scattering for nonintegrable equations.
\newblock {\em Comm. Math. Phys.}, 133(1):119--146, 1990.

\bibitem[SW92]{SofferWeinstein2}
A.~Soffer and M.~I. Weinstein.
\newblock Multichannel nonlinear scattering for nonintegrable equations. {II}.
  {T}he case of anisotropic potentials and data.
\newblock {\em J. Differential Equations}, 98(2):376--390, 1992.

\bibitem[SW99]{SofferWeinstein3}
A.~Soffer and M.~I. Weinstein.
\newblock Resonances, radiation damping and instability in {H}amiltonian
  nonlinear wave equations.
\newblock {\em Invent. Math.}, 136(1):9--74, 1999.

\bibitem[SW04]{SofferWeinstein4}
A.~Soffer and M.~I. Weinstein.
\newblock Selection of the ground state for nonlinear {S}chr\"odinger
  equations.
\newblock {\em Rev. Math. Phys.}, 16(8):977--1071, 2004.

\bibitem[SW05]{SofferWeinstein5}
A.~{Soffer} and M.~I. {Weinstein}.
\newblock {Theory of Nonlinear Dispersive Waves and Selection of the Ground
  State}.
\newblock {\em Physical Review Letters}, 95(21):213905--+, November 2005.

\bibitem[Tha92]{Thaller}
B.~Thaller.
\newblock {\em The {D}irac equation}.
\newblock Texts and Monographs in Physics. Springer-Verlag, Berlin, 1992.

\bibitem[Tsa03]{Tsai}
T.-P. Tsai.
\newblock Asymptotic dynamics of nonlinear {S}chr\"odinger equations with many
  bound states.
\newblock {\em J. Differential Equations}, 192(1):225--282, 2003.

\bibitem[TY02a]{TsaiYau}
T.-P. Tsai and H.-T. Yau.
\newblock Asymptotic dynamics of nonlinear {S}chr\"odinger equations:
  resonance-dominated and dispersion-dominated solutions.
\newblock {\em Comm. Pure Appl. Math.}, 55(2):153--216, 2002.

\bibitem[TY02b]{TsaiYau4}
T.-P. Tsai and H.-T. Yau.
\newblock Classification of asymptotic profiles for nonlinear {S}chr\"odinger
  equations with small initial data.
\newblock {\em Adv. Theor. Math. Phys.}, 6(1):107--139, 2002.

\bibitem[TY02c]{TsaiYau2}
T.-P. Tsai and H.-T. Yau.
\newblock Relaxation of excited states in nonlinear {S}chr\"odinger equations.
\newblock {\em Int. Math. Res. Not.}, (31):1629--1673, 2002.

\bibitem[TY02d]{TsaiYau3}
T.-P. Tsai and H.-T. Yau.
\newblock Stable directions for excited states of nonlinear {S}chr\"odinger
  equations.
\newblock {\em Comm. Partial Differential Equations}, 27(11-12):2363--2402,
  2002.

\bibitem[Wed00]{Weder}
R.~Weder.
\newblock Center manifold for nonintegrable nonlinear {S}chr\"odinger equations
  on the line.
\newblock {\em Comm. Math. Phys.}, 215(2):343--356, 2000.

\bibitem[Yaj95]{Yajima}
K.~Yajima.
\newblock The {$W\sp {k,p}$}-continuity of wave operators for {S}chr\"odinger
  operators.
\newblock {\em J. Math. Soc. Japan}, 47(3):551--581, 1995.

\bibitem[Yam93]{Yamada}
O.~Yamada.
\newblock A remark on the limiting absorption method for {D}irac operators.
\newblock {\em Proc. Japan Acad. Ser. A Math. Sci.}, 69(7):243--246, 1993.

\end{thebibliography}

\end{document}